\documentclass[final,a4paper]{amsart}
\usepackage[utf8]{inputenc}
\usepackage{amsmath, amsthm} 
\usepackage{amsfonts,amssymb}
\usepackage{graphicx,epstopdf}
\usepackage{hyperref}
\usepackage{cite}
\usepackage{color}
\usepackage{float}

\graphicspath{{./figures_Kerr}{./figures_disk}}

\newcommand{\id}{\mathbb{I}}

\newcommand{\R}{\mathbb{R}}

\newcommand{\lp}{\left(}
\newcommand{\rp}{\right)}
\newcommand{\lc}{\left[}
\newcommand{\rc}{\right]}
\newcommand{\trans}{\top}

\newcommand{\Lag}{\mathcal{L}}

\newcommand{\pt}{p^t}
\newcommand{\prho}{p^{\rho}}
\newcommand{\pzeta}{p^{\zeta}}
\newcommand{\pphi}{p^{\phi}}

\newcommand{\ernst}{\mathcal{E}}

\newcommand{\vx}{\mathbf{x}}

\newcommand{\vp}{\mathbf{p}}

\newcommand*{\ii}{\mathrm{i}}
\newcommand*{\ee}{\mathrm{e}}
\newcommand*{\dd}{\mathrm{d}}

\begin{document}
\title{Visualisation of counter-rotating dust disks using ray tracing 
methods}

\author{Eddy B. de Leon}
\email{eddybrandon11@hotmail.com}
\address{Institut de Math\'ematiques de Bourgogne,
  Universit\'e de Bourgogne, 9 avenue Alain Savary, 21078 Dijon
  Cedex, France}

\author{J\"org Frauendiener}
\email{joerg.frauendiener@otago.ac.nz}
\address{Department of Mathematics and Statistics, 
  University of Otago,      
  PO Box 56, Dunedin 9054, New Zealand}

\author{Christian Klein}
\email{Christian.Klein@u-bourgogne.fr}
\address{Institut de Math\'ematiques de Bourgogne, Institut 
  Universitaire de France, 
  Universit\'e de Bourgogne, 9 avenue Alain Savary, 21078 Dijon
  Cedex, France}

\date{\today} 
\thanks{This work was partially supported by the EIPHI Graduate School (contract ANR-17-EURE-0002), the Bourgogne Franche-Comt\'e Region, the European fund FEDER, by the European Union Horizon 2020 research and innovation program under the Marie Sklodowska-Curie RISE 2017 grant agreement no. 778010 IPaDEGAN, and by the Marsden Fund Council from Government funding, managed by Royal Society Te Apārangi, New Zealand.}

\begin{abstract}
A detailed study of ray tracing in the space-time generated by a disk of counter-rotating dust is presented. The space-time is given in explicit form in terms of hyperelliptic theta functions. The numerical approach to ray tracing is set up for general stationary axisymmetric space-times and tested at the well-studied example of the Kerr solution. Similar features as in the case of a rotating black hole, are explored in the case of a dust disk.

The effect of the central redshift varying between a Newtonian disk and the ultrarelativistic disk, where the exterior of the disk can be interpreted as the extreme Kerr solution, and the transition from a single component disk to a static disk is explored. Frame dragging, as well as photon spheres, are discussed. 
\end{abstract}



\maketitle

\section{Introduction}\label{S.introduction}

A convenient way to study general relativistic effects compared to Newton's theory of gravitation is to trace light rays in relativistic space-times. While these are straight lines in Newtonian theory, light rays can be bent near strong masses in a relativistic setting, they can be trapped in certain regions or merely dragged along by rotating masses. Such studies are especially fruitful in stationary asymptotically flat space-times since a virtual camera can be placed at any point in the space-time to record what an observer would see at some distance from strongly gravitating objects. This approach allowed us to gain a deeper physical understanding of the Kerr solution describing a rotating black hole.

The first important exact solution to the Einstein field equations (EFE) in vacuum was the Schwarzschild solution in spherically symmetric vacuum. In 1963, Kerr found the exact stationary axisymmetric vacuum solution, which can be interpreted as a rotating black hole. In 1968, Ernst \cite{ernst} showed that the stationary, axisymmetric EFE in vacuum are equivalent to the Ernst equation
\[
  \Re(\ernst) \Delta \ernst = (\nabla \ernst)^2,
\]
where it is understood that $\Delta$ and $\nabla$ are the standard operators in $\mathbb{R}^3$ in cylindrical coordinates and where $\mathcal{E}$ is independent of the azimuthal coordinate.  Together with the relation $f=\Re(\ernst)$ and the quadratures of \eqref{axi} and \eqref{kxi}, the metric (\ref{eq:wlp}) can be reconstructed from a given Ernst potential. The Ernst equation was shown to be completely integrable in \cite{Mai,BZ,Neu} which allowed to construct large classes of exact solutions. Korotkin constructed solutions over a family of hyperelliptic curves of genus $g$ \cite{Kor} given by multi-dimensional theta functions. Kerr space-times can be constructed as the solitonic limit of the genus 2 solutions \cite{buch}.

A well-studied phenomenon in Kerr space-times (as in most curved space-times) is the bending of light, which can be observed by the difference in the apparent position of an object compared to the corresponding Newtonian situation. Ray tracing techniques have been widely used to simulate visualisations around black holes from the first picture simulation of a black hole \cite{luminet,Muller:2012} to the professional renderings used in the movie ``Interstellar''\cite{interstellar,interstellar2} to the first actual picture \cite{EHT}, where simulations were used in order to deduce the angular momentum of the black hole, as well as the inclination of its rotation axis with respect to the observer. To do this, each pixel of the 2-dimensional `film' or screen onto which the picture is projected is associated to a light ray, which is computed in the Kerr space-time via the geodesic equations. Since these equations are non-linear, they must in general be solved numerically, e.g., with a Runge-Kutta scheme. In the case of the Kerr solution, the computation of the Christoffel symbols which appear in the metric is straightforward, since they are given explicitly in terms of elementary functions. 

Our first goal is to set up a tool to solve geodesic equations in an efficient way, even if the metric functions are non-elementary as in the case of algebro-geometric solutions to the Ernst equation. To do this, we compute the metric functions numerically on a two-dimensional grid and use polynomial interpolation to obtain the functions for general position, a procedure which is both efficient and of high accuracy.

The Kerr solution in Weyl coordinates is used as a test example for this method, even though it is given in terms of elementary functions.  The two important phenomena we want to visualise are the shadow of the black hole due to photons being unable to escape from certain regions of the black hole space-time and the image of a non-gravitating accretion disk as seen by a distant observer.

Then, we proceed to study in detail a solution of the Ernst equation describing a family of counter-rotating dust disks given on hyperelliptic curves of genus 2 \cite{FK3}. Since this is a gravitating disk, it will affect the behavior of light in its vicinity. To explore this behavior, we study individual light rays with different initial conditions. This will indicate to what extent this space-time is similar to a Kerr space-time. Finally, we also want to generate the image of the disk seen by a distant observer. We assume that the disk is opaque in this setting. Due to a particular property of this space-time, analogous to the photon sphere in black hole space-times, some light rays take an infinite coordinate time to reach the disk. Thus, the observer will only see a portion of the disk, the rest being hidden by the shadow, as well as a portion of the background.

The paper is organised as follows. In Section \ref{sec:sav-spacetimes}, we present preliminary material on stationary axisymmetric vacuum space-times and its relation to the Ernst equation. In Section \ref{sec:sol-ernst}, we discuss a class of solutions to the Ernst equation on hyperelliptic Riemann surfaces of genus $g$ and the numerical computation of the necessary integrals.  In Section \ref{sec:cr-disk}, we discuss a special class of solutions on genus 2 hyperelliptic curves, which is physically interpreted as a family of counter-rotating dust disks. In Section \ref{sec:geodesics}, we detail the method to compute the metric functions and their derivatives via polynomial interpolation, which is more efficient than computing the functions explicitly. In Section \ref{sec:RT}, we show how to simulate images in stationary axisymmetric space-times by associating initial conditions for null geodesics to each pixel of the image. In Section \ref{sec:example-kerr}, we reproduce some known results for the Kerr solution by using polynomial interpolation whilst studying the solution in Weyl coordinates. In Section \ref{sec:example-cr-disk}, we show individual null geodesics in order to visualise the effect of the disk on light coming from different directions and to finish, we simulate the image of the counter-rotating disk seen by a distant observer. We add some concluding remarks in Section \ref{sec.out}.


\section{Stationary axisymmetric vacuum space-times}
\label{sec:sav-spacetimes}

In this section, we collect some of the known facts on stationary, axisymmetric vacuum space-times and the Kerr solution.

\subsection{Ernst equation}

Stationary axisymmetric space-times are characterised by two commuting Killing vectors, one asymptotically timelike $\partial_{t}$ and one spacelike $\partial_{\phi}$. In order to construct exact solutions to the vacuum Einstein equations in this case, it is convenient to introduce Weyl's cylindrical coordinates, see \cite{exac} and references therein. Since the coordinates $t$ and $\phi$ are adapted to the Killing vectors, the metric functions do not depend on them. The distance to the axis where the Killing vector $\partial_{\phi}$ vanishes is measured by $\rho$ ($\rho>0$), whereas $\zeta$ parametrises this axis. The metric in vacuum can then be written in Weyl-Lewis-Papapetrou form, see~\cite{exac}, \begin{equation}
    \dd s^{2}=-f(\dd t+a \dd\phi)^{2}+(\ee^{2k}(\dd\rho^{2} + \dd\zeta^{2})
    +\rho^{2}\dd\phi^{2})/f
    \label{eq:wlp},
\end{equation}
where the metric functions $f$, $a$ and $k$ depend on $\rho$, $\zeta$ only. 

It is convenient to introduce complex notation: put $\xi:=\zeta-\ii\rho$ and define the \emph{twist potential} $b$ via
\begin{equation}
  b_{\xi}=-\frac{\ii}{\rho}a_{\xi}f^{2}
  \label{bxi}.
\end{equation}
The potential $b$ is determined up to a constant for given $a$, $f$ which is in asymptotically flat space-times chosen such that $b$ vanishes at infinity. The Ernst potential $\mathcal{E}=f+\ii b$ was introduced in \cite{ernst}. In vacuum, it satisfies the Ernst equation,
\begin{equation}
    \mathcal{E}_{,\xi\bar{\xi}}-\frac{1}{2(\bar{\xi}-\xi)}(
    \mathcal{E}_{,\bar{\xi}}-\mathcal{E}_{,\xi}) = 
    \frac{2}{\mathcal{E}+
    \bar{\mathcal{E}}}\mathcal{E}_{,\xi}\mathcal{E}_{,\bar{\xi}}\;
    \label{eq:ernstcom}.
\end{equation}

Equation (\ref{eq:ernstcom}) is equivalent to the stationary axisymmetric Einstein equations in the sense that the metric functions (\ref{eq:wlp}) can be obtained from a given Ernst potential $\mathcal{E}$ via quadratures,
\begin{equation}
  a_{\xi}=2\rho\frac{(\mathcal{E}-\bar{\mathcal{E}})_{\xi}}{
    (\mathcal{E}+\bar{\mathcal{E}})^{2}}
  \label{axi},
\end{equation}
and
\begin{equation}
  k_{\xi}=(\xi-\bar{\xi})
  \frac{\mathcal{E}_{\xi}\bar{\mathcal{E}}_{\xi}}{
    (\mathcal{E}+\bar{\mathcal{E}})^{2}}\;.
  \label{kxi}
\end{equation}

It was shown in \cite{Mai,BZ,Neu} that the Ernst equation is completely integrable. This means that it can be treated as the integrability condition of an overdetermined linear differential system with an additional variable called \emph{spectral parameter}.  This structure makes it possible to construct large classes of solutions.

\subsection{Kerr solution}

The most prominent non-trivial solution to the Ernst equation\footnote{Ernst introduced his equation actually to provide a simple way to verify that the Kerr metric is a solution to the vacuum Einstein equations.} is the Kerr black hole given with $r_{\pm}= \sqrt{(\zeta\pm m\cos\varphi)^{2}+\rho^{2}}$ by
\begin{equation}
    \mathcal{E}=\frac{\ee^{-\ii\varphi}r_{+} + \ee^{\ii\varphi}r_{-}
    -2m\cos\varphi}{\ee^{-\ii\varphi}r_{+} + \ee^{\ii\varphi}r_{-}
    +2m\cos\varphi};
  \label{kerr5}
\end{equation}
with $X=(r_{+}+r_{-})/(2m\cos \varphi)$ and $Y=(r_{+}-r_{-})/(2m\cos \varphi)$ the function $a$ takes the form
\begin{equation}
    a=\frac{2m \sin \varphi(1-Y^{2})(1+X\cos \varphi)}{
    \cos^{2}\varphi X^{2}+\sin^{2}\varphi Y^{2}-1}
  \label{kerr6}.
\end{equation}
The coordinates $\rho$, $\zeta$ can be expressed in terms of the ``parabolic coordinates'' $X$ and $Y$,
\begin{equation}
    \zeta/(m\cos\varphi)=XY,\quad \rho/(m\cos\varphi)=\sqrt{(X^{2}-1)(1-Y^{2})}
    \label{kerr6a}.
\end{equation}
With the above setting, we can write the Ernst potential in the form
\begin{equation}
    \mathcal{E}=\frac{\cos \varphi X - \ii\sin \varphi Y-1}{
    \cos \varphi X - \ii\sin \varphi Y+1}
    \label{kerr5a},
\end{equation}
which implies
\begin{equation}
    f=\frac{\cos^{2}\varphi X^{2}-1+\sin^{2}\varphi Y^{2}}{
    (\cos\varphi X+1)^{2}+\sin^{2}\varphi Y^{2}}
  \label{kerr5b}.
\end{equation}

\begin{figure}[!htb]
  \includegraphics[width=0.49\hsize]{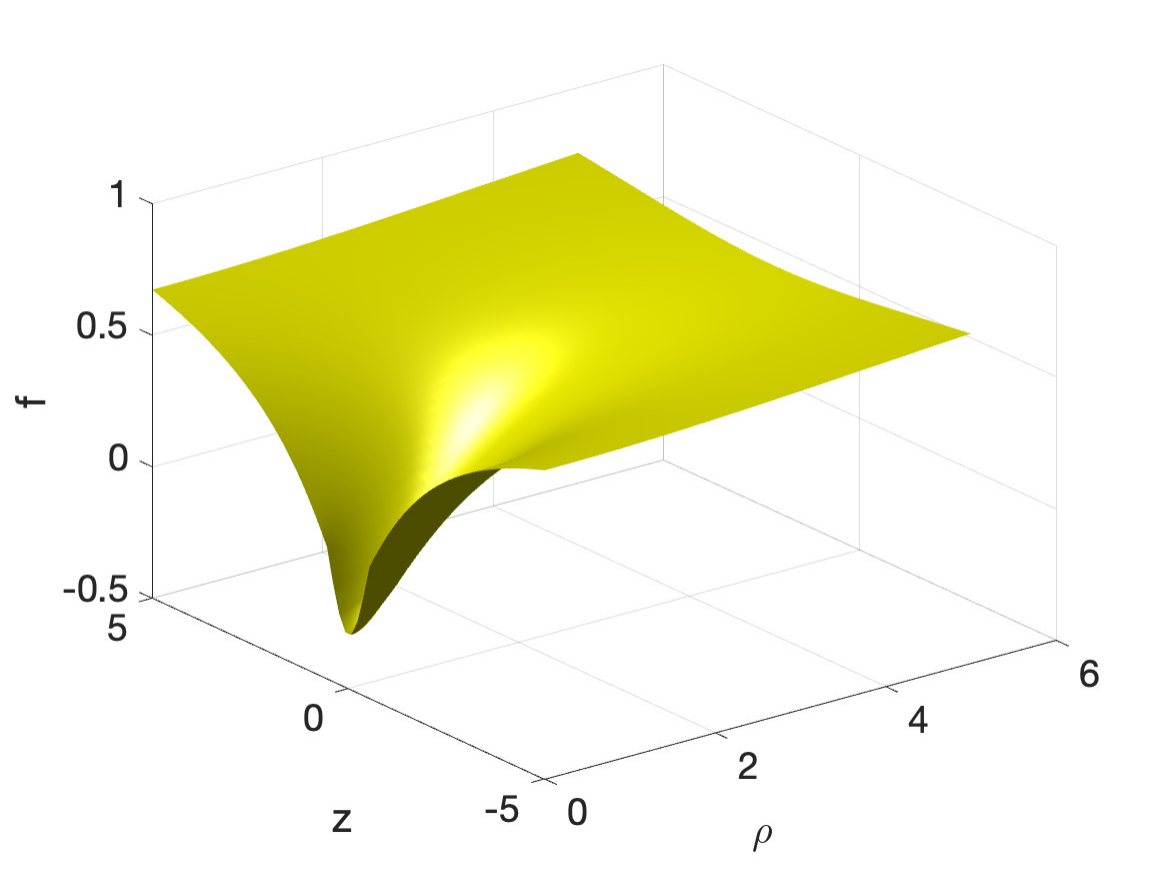}
  \includegraphics[width=0.49\hsize]{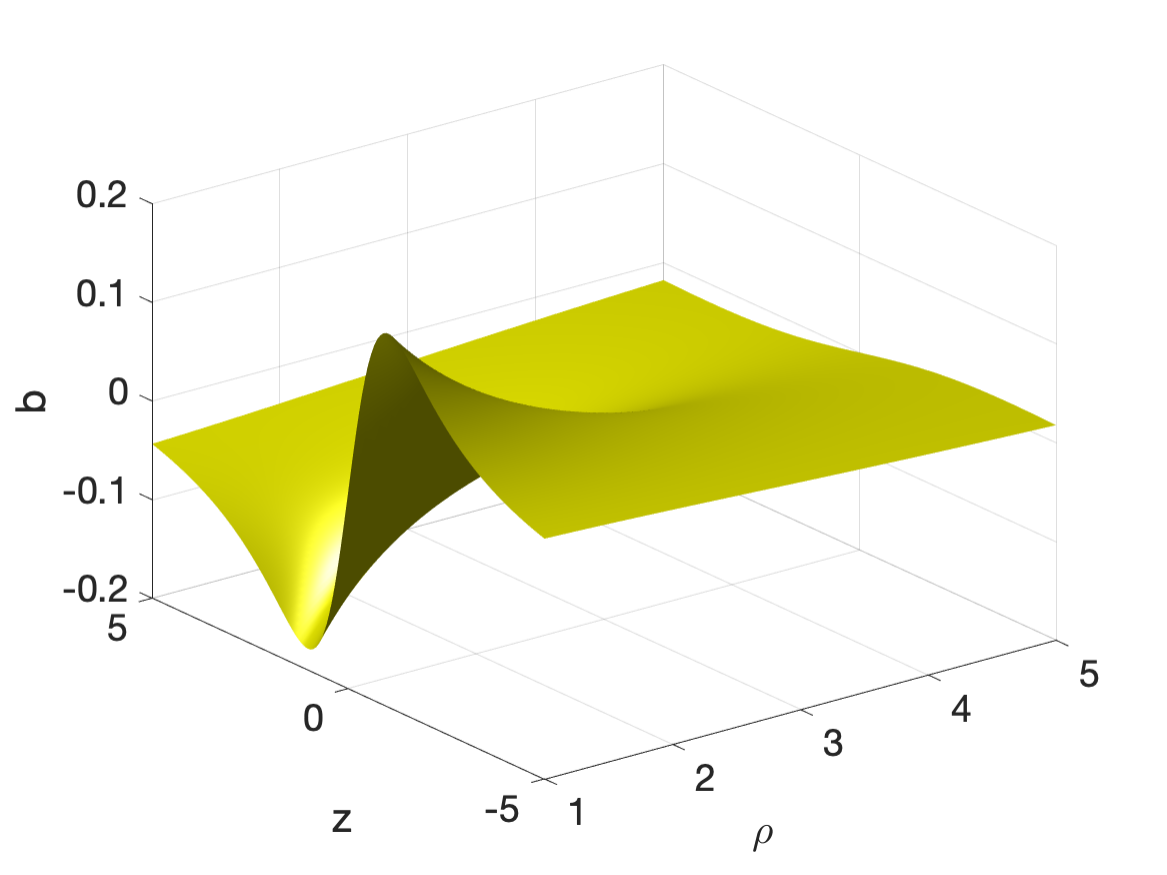}
  \caption{Ernst potential for the Kerr solution for $m=1$ and $\varphi=1$, on the left the real part, on the right the imaginary part.}
  \label{kerrernst}
\end{figure}

We show the Ernst potential for $m=1$ and $\varphi=1$ in Fig.\ref{kerrernst}. It can be seen that the real part is an even function in $\zeta$, whereas the imaginary part is odd, $\mathcal{E}(\rho,\zeta)=\bar{\mathcal{E}}(\rho,-\zeta)$. The space-time thus shows an \emph{equatorial symmetry}. Note, that $f$ is negative in the \emph{ergoregion}. It is delimited by the \emph{ergosphere} where an observer stationary at infinity will note an infinite redshift.

\begin{figure}[!htb] 
\includegraphics[width=0.49\hsize]{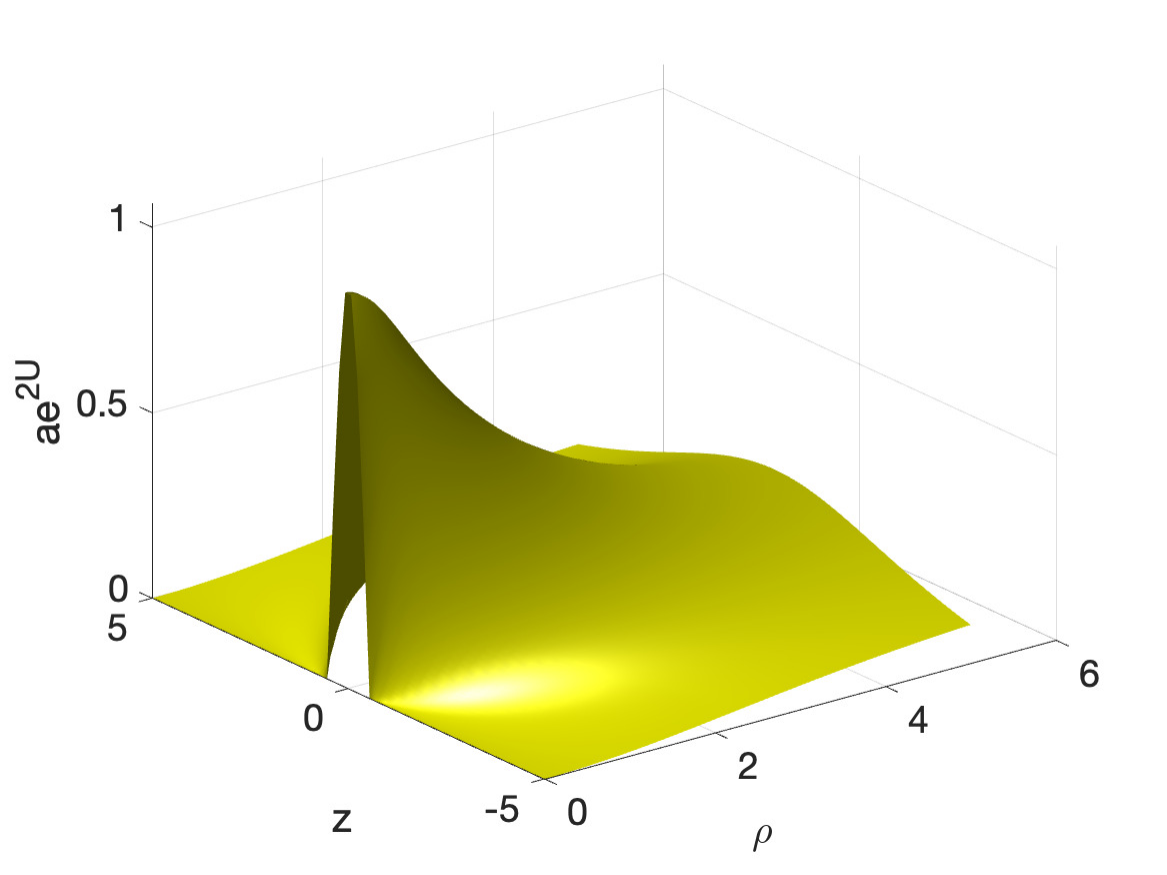} 
\includegraphics[width=0.49\hsize]{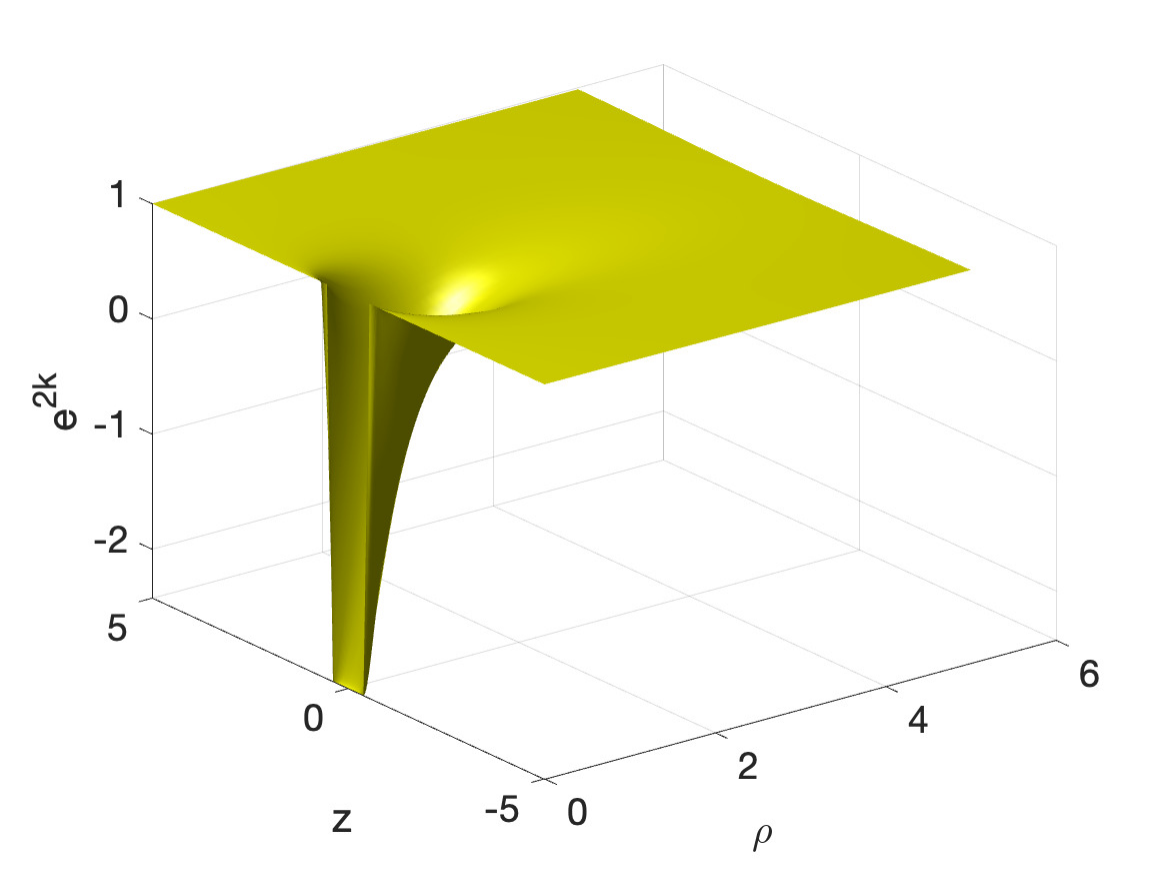} \caption{Kerr solution for $m=1$ and $\varphi=1$; the metric function $a$ on the left, and on the right $\ee^{2k}$.  } \label{kerrernstak} \end{figure}

The  metric function $k$ reads
\begin{equation}
    \ee^{2k}=\frac{m^{2}}{r_{+}r_{-}}  \left(\sin^{2}\varphi (Y^{2}-1)
    + \cos^{2}\varphi(X^{2}-1)\right)
    \label{kerr7}.
\end{equation}
In Weyl coordinates, the horizon is the part of the axis between $-m\cos\varphi$ and 
$m\cos\varphi$, where 
\begin{equation}
  \ee^{2k}=-\tan^{2}\varphi,\quad a=-1/\Omega_{BH}
  \label{kerr8},
\end{equation}
where $\Omega_{BH}$ is the angular velocity of the horizon, which is 
here $1/\Omega_{BH}=2m\cot \frac{\varphi}{2}$.  This implies that $a$ 
and $k$ are finite constants on the horizon.  In 
Fig.~\ref{kerrernstak}, we show the metric functions $ae^{2U}$ and 
$\ee^{2k}$ corresponding to the Ernst potential displayed in 
Fig.~\ref{kerrernst}. It can be seen that $a$ and $k$ vanish on the 
regular part of the axis.  

Normally, the Kerr solution is discussed in Boyer-Lindquist (BL) coordinates. The relation of the Weyl coordinates to the BL coordinates is given by
\begin{equation}
    \rho=\sqrt{r^{2}-2mr+m^{2}\sin^{2}\varphi}\sin \vartheta, \quad 
    \zeta=(r-m)\cos \vartheta,
    \label{boyer1}
\end{equation}
or for the parabolic coordinates
\begin{equation}
    m\cos \varphi X=r-m,\quad Y=\cos \vartheta
    \label{boyer2}.
\end{equation}
The horizon is located in the BL coordinates at $R=2m\cos^{2}\frac{\varphi}{2}$. Thus, it is spherical in BL coordinates, whereas it corresponds to a rod on the axis in Weyl coordinates.


\section{Algebro-geometric solutions to the Ernst equation}
\label{sec:sol-ernst}

In this section, we briefly discuss Korotkin's solutions to the Ernst equation in terms of multi-dimensional theta functions on a family of hyperelliptic Riemann surfaces and how to efficiently compute them numerically.

\subsection{Hyperelliptic solutions to the Ernst equation}

In \cite{Kor}, Korotkin showed that solutions to the Ernst equation can be given on a family $\mathcal{L}_{\xi}$ of hyperelliptic Riemann surfaces defined by
\begin{equation}
	\mu^{2}=(K-\xi)(K-\bar{\xi})\prod_{i=1}^{g}(K-E_{i})(K-F_{i})
	\label{hyper}
\end{equation}
where $(\mu,K)\in \mathbb{C}^{2}$, where the $2g$ constant branch points are pairwise distinct and either pairwise complex conjugate, $E_{i}=\bar{F}_{i}$, or real $E_{i},F_{i}$, $i=1,\ldots,g$. The genus of the hyperelliptic surface is $g$.

\begin{figure}[!htb]
\includegraphics[width=0.7\hsize]{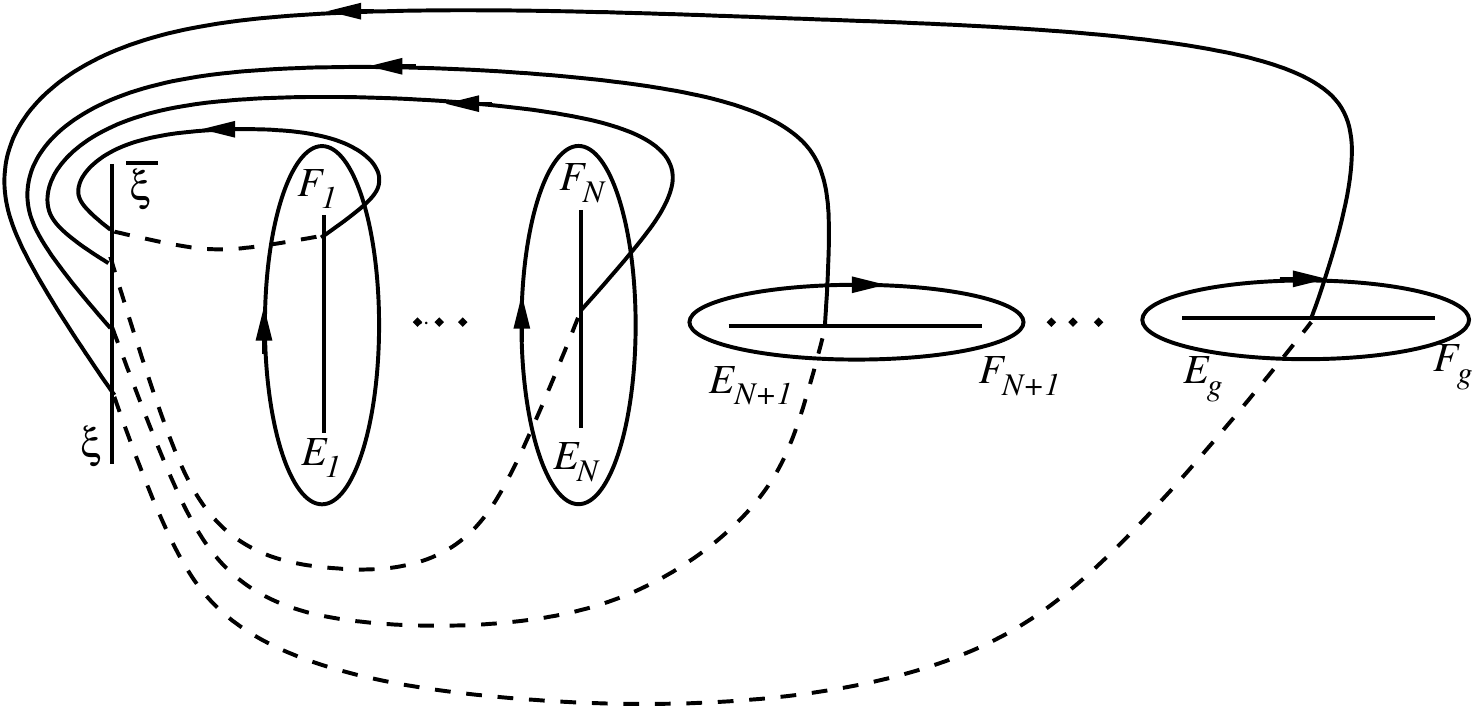}
\caption{Homology basis of the hyperelliptic surface (\ref{hyper}).  }
\label{homology}
\end{figure}

On $\mathcal{L}_\xi$ we introduce a canonical homology basis of cycles 
$a_{i}$, $b_{i}$, $i=1,\ldots,g$, see Fig.~\ref{homology}.
The 1-forms 
\begin{equation}
  \nu_{k} = \frac{K^{k-1}dK}{\mu(K)}
  \label{oneforms}
\end{equation}
form a basis of the holomorphic 1-forms on the surfaces $\mathcal{L}_{\xi}$. They are normalised by the conditions $\int_{a_{i}}^{}\omega_{k}=2\pi \ii\delta_{ik}$. Then the matrix $\mathbb{B}$ of $b$-periods,
\begin{equation}
	\mathbb{B}_{ik} = \int_{b_{i}}^{}\omega_{k},\quad i,k = 1,\ldots,g
	\label{rie}
\end{equation}
is a Riemann matrix, i.e., a $g\times g$ symmetric matrix with negative definite real part. 

This implies that the theta function with characteristics
\begin{equation}\label{theta}
  \Theta_{\mathrm{pq}}(\mathrm{z},\mathbb{B})=
  \sum\limits_{\mathrm{N}\in\mathbb{Z}^g}\exp\left\{\frac{1}{2}
    \left\langle\mathbb{B}\left(\mathrm{N+p}\right),
      \mathrm{N+p}
    \right\rangle+
    \left\langle \mathrm{z+q},\mathrm{N+p}
    \right\rangle\right\}
  \;,
\end{equation}
depending on $\mathrm{z}\in\mathbb{C}^g$ and with characteristics $\mathrm{p}$, $\mathrm{q}\in{\mathbb{R}}^g$ is an entire function of $\mathrm{z}$. 
Any theta function with characteristics is related to the Riemann theta function $\Theta:= \Theta_{00}$, the theta function with zero characteristics, via
\begin{equation}
  \Theta_{\mathrm{pq}}(\mathrm{z},\mathbb{B})=\Theta(\mathrm{z}
  +\mathbb{B}\mathrm{p} + \mathrm{q})\exp\left\{\frac{1}{2}
    \left\langle\mathbb{B}\mathrm{p,p}\right\rangle+
    \left\langle \mathrm{p,z} + \mathrm{q}\right\rangle
  \right\}\;.
  \label{thchar}
\end{equation}
The theta function has the periodicity properties 
\begin{equation}
  \Theta_{\mathrm{pq}}(\mathrm{z}+2\pi \mathrm{i} e_{j}) = 
  e^{2\pi \ii p_{j}}
  \Theta_{\mathrm{pq}}(\mathrm{z})\;,
  \quad 
  \Theta_{\mathrm{pq}}(\mathrm{z}+\mathbb{B}
  e_{j})=
  e^{- (z_{j}+q_{j}) - \frac{1}{2} B_{jj}}
  \Theta_{\mathrm{pq}}(\mathrm{z})\;
  \label{eq:periodicity},
\end{equation}
where $e_{j}$ is a vector in $\mathbb{R}^{g}$ consisting of
zeros except for a 1 in $j$-th position.  

The \emph{Abel map}, defined by the integral between two points $P$, $P_{0}$ on $\mathcal{L}_{\xi}$,
\[
  P \mapsto \int_{P_{0}}^{P}\nu_k
\]
is a bijective map from the surface into the \emph{Jacobian} $Jac(\mathcal{R}) = \mathbb{C}^{g}/\Lambda$, where $\Lambda$ is the lattice formed by the periods of the holomorphic one-forms,
\[
  \Lambda = \left\{2\pi \mathrm{i}\,\mathrm{m} +\mathbb{B}
    \,\mathrm{n}: \mathrm{m}, \mathrm{n}\in \mathbb{Z}^{g}\right\}.
\]

Korotkin \cite{Kor} showed that the following function is a solution
to the Ernst equation,
\begin{equation}
  \mathcal{E} = \frac{\Theta(\int_{\xi}^{\infty^{+}} + \mathrm{u})}{\Theta(\int_{\xi}^{\infty^{-}}+\mathrm{u})}e^{I}
  \label{korpot};
\end{equation}
here $\infty^{\pm}$ are the points on the two sheets of the 
hyperelliptic surface covering the infinite point in the complex 
plane (the labelling of the sheets has to be imposed at some base 
point that is not a branch point), and
\begin{equation}
  \mathrm{u}_{k}= \frac{1}{2\pi}\int_{\Gamma}^{}\ln G 
  \omega_{k},\quad I = \frac{1}{2\pi}\int_{\Gamma}^{}\ln G 
  d\omega_{\infty^{+}\infty^{-}},
  \label{path}
\end{equation}
where $\Gamma$ is some piecewise smooth contour on $\mathcal{L}_{\xi}$, where $\ln G$ is a H\"older continuous function on $\Gamma$, and where $d\omega_{\infty^{+}\infty^{-}}$ is the differential of the third kind with poles only at $\infty^{+}$ and $\infty^{-}$ with residues $+1$, respectively $-1$, normalised by the condition that all $a$-periods vanish. A different form of these solutions was discussed in \cite{KKS}.

Note that the dependence of the solutions (\ref{korpot}) on the physical coordinates $\rho$, $\zeta$ is exclusively via the moving branch points $\xi$, $\bar{\xi}$. Thus, in contrast to solutions to integrable evolution equations such as the Korteweg-de Vries equation, see \cite{algebro} and references therein, the physical coordinates do not enter the argument of the theta functions explicitly. Thus, they do not inherit the periodicity properties of the theta functions (\ref{eq:periodicity}). In the limit of coinciding branch cuts, $E_{i}\to F_{i}$, where the related period tends to infinity, \emph{solitons} emerge in the case of the nonlinear evolution equations, i.e., stable particle-like waves. In the case of the Ernst equation, the Kerr solution and multi-black holes can be obtained in this limit.

\subsection{Numerical computation on hyperelliptic Riemann surfaces}

To numerically compute the needed hyperelliptic quantities entering (\ref{korpot}), we follow the approach of \cite{FK15}, see also \cite{FK04,FK17}. The basic idea is that the periods of the hyperelliptic surface can be written in terms of integrals of the 1-forms (\ref{oneforms}) between branch points, e.g.,
\begin{equation}
  \int_{\xi}^{\bar{\xi}}\frac{K^{k}dK}{\mu(K)}=
  \int_{\xi}^{\bar{\xi}}\frac{K^{k}dK}{\sqrt{(K-\xi)(K-\bar{\xi})}\tilde{\mu}(K)},
  \quad k = 0,\ldots,g-1,
  \label{int1}
\end{equation}
where $\tilde{\mu}(K)=\prod_{i=1}^{g}\sqrt{(K-E_{i})(K-F_{i})}$. After the substitution $K-\zeta=\ii\rho\sin \frac{\pi}{2}s$, this is mapped to 
\begin{equation}
  \int_{-1}^{1}\frac{ds}{\tilde{\mu}(\zeta + \ii\rho \sin (s\pi/2))}
  \label{int2}
\end{equation}
with an analytic integrand. This integral can be conveniently computed with the Clenshaw-Curtis algorithm \cite{CC}, a spectral method, i.e., the numerical error decreases exponentially with the number of collocation points. There are some subtleties related to the fact that the Matlab root is branched along the negative real axis. This can introduce unwanted sign changes in (\ref{int2}) which are avoided by an analytic continuation of the root along the path of integration. To improve the numerical efficiency, it is convenient to address cases analytically where two or more branch points almost coincide. The reader is referred to \cite{FK15} for details. The line integrals in (\ref{korpot}) along $\Gamma$ are also computed with the Clenshaw-Curtis algorithm.

In the computation of the theta function, the series~(\ref{theta}) is approximated as a sum. The argument $\mathrm{z}$ is always reduced to the fundamental domain of the Jacobian, $\mathrm{z}=\mathrm{z}_{0}+2\pi \mathrm{i}\mathrm{m}+\mathbb{B}\mathrm{n}$, where $\mathrm{z}_{0}$ is in this fundamental domain defined by $\mathrm{z}_{0}=\mathbb{B}p+2\pi \ii q$ with $p_{i},q_{i}\in]-1/2,1/2]$, $i=1,\ldots,g$. The theta function is computed for the argument $\mathrm{z}_{0}$, and the periodicity relations (\ref{eq:periodicity}) then give the theta function for the argument $\mathrm{z}$. For details of the computation, the reader is again referred to \cite{FJK} and \cite{FKbuch}.

In \cite{FK04}, the metric functions were computed on a numerical 
grid in $\rho$, $\zeta$. In the equatorially symmetric case, we 
concentrate always on $\zeta\geq 0$.  We divided the $(\rho,\zeta)$ plane 
into quadrants, $\rho\leq1$, $\rho>1$ and $\zeta\leq 1$, $\zeta>1$. 
If $\rho>1$ or $\zeta>1$, we used $1/\rho$ or $1/\zeta$ respectively 
as a variable, thus compactifying the infinite intervals. Each of the 
intervals in both $\rho$ and $\zeta$ was then mapped to $[-1,1]$ 
where \emph{Chebyshev collocation points } are introduced, 
$s_{j}=\cos(j\pi/N)$, $j=0,\ldots,N$ for some $N\in\mathbb{N}$. To 
reach higher accuracies near the disk, we use here \emph{disk 
coordinates} as detailed in section \ref{subsec:cusps}. This choice of a numerical grid is motivated by two reasons: first, derivatives of the metric functions appear in the geodesic equations to be solved in the ray tracing algorithms. Though explicit formulae for these derivatives were given in \cite{KKS,buch}, it is numerically more efficient to use numerical differentiation on the grid introduced above. On the Chebyshev collocation points, the Chebyshev differentiation matrices given, for instance, in \cite{Tre} can be applied to approximate the derivatives. This leads again to a spectral method for which the numerical error decreases exponentially with $N$.  Secondly, the ray tracing algorithm will need these functions also on points in the space-time not being on the numerical grid. In such a case, we will use numerically efficient and stable \emph{barycentric interpolation}, see \cite{barycentric} for a review and a Matlab code for interpolation on Chebyshev collocation points.

Thus, the space-time functions are computed in the beginning on the numerical grid, derivatives are obtained numerically, and the values needed for the geodesic equation are then obtained via interpolation.

\section{Counter-rotating dust disk}
\label{sec:cr-disk}

In \cite{KR2}, a class of solutions to the Ernst equation has been given, which can be interpreted as a family of counter-rotating dust disks. It includes the disk \cite{NM} as a limiting case. In this section, we briefly summarise the physical aspects of this family of solutions. For details, the reader is referred to \cite{FK3,buch}.

In \cite{KR2}, an infinitesimally thin disk was considered, which means there is a discontinuity in the metric functions for $\rho\leq 1$ in the equatorial plane $\zeta=0$. This leads to a $\delta$-type energy-momentum-tensor. The difference in the extrinsic curvatures $K_{\alpha\beta}$ of the hypersurface $\zeta=0$ with respect to its embeddings into $V^{\pm}=\{\pm\zeta>0\}$ can be interpreted as the energy-momentum tensor of the disk taking the form
\begin{equation}
  S^{\mu\nu}=\sigma_{+} u^{\mu}_+ u^{\nu}_+ +\sigma_{-} u^{\mu}_- u^{\nu}_-\;
  \label{vac16.11},
\end{equation}
where the vectors $u^{\alpha}_{\pm}$ are a linear combination of the Killing vectors, $(u^\alpha_{\pm})=(1, 0, 0, \pm\Omega(\rho))$. Thus, the matter in the disk can be interpreted as two counter-rotating components of dust. We define
\begin{equation}
  \gamma(\rho)=\frac{\sigma_{+}(\rho)-\sigma_{-}(\rho)}{\sigma_{+}(\rho)
    +\sigma_{-}(\rho)}\;
  \label{vac20a},
\end{equation}
the weighted difference of the two components of counter-rotating dust.

In \cite{KR2}, an exact solution to the Ernst equation of the form (\ref{korpot}) for a counter-rotating disk with constant $\Omega$ and constant $\gamma$ in (\ref{vac16.11}) was given on a surface of genus 2. The constant (with respect to the physical coordinates) branch points are given by $E_{2}=\alpha - \ii\beta=\bar{F}_{2}=-F_{1}=-\bar{E}_{1}$ with
\begin{equation}
  \alpha=-1+\frac{\delta}{2}\;, \quad 
  \beta=\sqrt{\frac{1}{\lambda^2}+\delta-\frac{\delta^2}{4}}\;
  \label{eq37a}.
\end{equation}
The parameter $\delta$ varies between $\delta=0$ (only one component, i.e., $\gamma=1$) and $\delta=\delta_{s}$, 
\begin{equation}
  \delta_{s}=2\left(1+\sqrt{1+\frac{1}{\lambda^{2}}}\right)\;
  \label{eq37b},
\end{equation}
in the static limit, i.e., when $\gamma=0$. The function $G$ (see~\eqref{path}) is given by 
\begin{equation}
  G(\tau)=\frac{\sqrt{(\tau^{2}-\alpha)^2+\beta^2}+\tau^{2}+1}{
    \sqrt{(\tau^{2}-\alpha)^2+\beta^2}-(\tau^{2}+1)}\;
  \label{eq38}.
\end{equation}

Interesting limiting cases of the solutions are:
\begin{itemize}
\item  Newtonian limit $\lambda\to 0$: the redshift at the disk as measured at infinity tends to zero;
\item static limit $\delta\to\delta_{s}$ ($\gamma\to0$): the Ernst potential becomes real and solves the Euler-Darboux equation, i.e., the axisymmetric Laplace equation;
\item ultrarelativistic limit: for $\gamma=1$ (only one dust component), the solution has a limit where the redshift at the disk as measured at infinity diverges; this can be interpreted as the disk vanishing behind the horizon of the extreme Kerr solution ($\varphi=\pi/2$), see the discussion in \cite{buch} and references therein.
\end{itemize}


\section{Geodesics in curved space-times}
\label{sec:geodesics}

The goal of this section is to set up a numerical tool which allows us to efficiently solve the geodesic equations for any metric given in Weyl-Lewis-Papapetrou form \eqref{eq:wlp}, which describes a stationary axisymmetric vacuum space-time. We are particularly interested in solving geodesics corresponding to space-times constructed via the algebro-geometric solutions to the Ernst equation. Thus, such functions are non-elementary in general, and they are only given in terms of series. However, these functions are smooth, and they can be computed efficiently via barycentric interpolation, a polynomial interpolation technique based on spectral methods and thus, with exponential convergence, see \cite{barycentric}.

Let us recall that the equation for an affinely parametrised geodesic in a general space-time is given by the Euler-Lagrange equations
\begin{equation*}
    \frac{d}{ds} \frac{\partial \mathcal{L} }{ \partial \dot{x}^\mu} = \frac{\partial \mathcal{L}}{\partial x^\mu},
\end{equation*}
corresponding to the Lagrangian
\begin{equation} \label{eq:Lag}
    \mathcal{L}(x,\dot{x}) = \frac{1}{2} g_{\mu\nu}(x) \dot{x}^\mu \dot{x}^\nu.
\end{equation}
This yields a system of four second-order ordinary differential equations (ODEs), which can be converted to a system of eight first-order ODEs by introducing the variable $p^\mu=d\dot{x}^\mu/ds$. However, considering that the presence of Killing vectors leads to conserved quantities along geodesics, we can use those associated to the vector fields $\eta = \partial_t$ and $\tilde{\eta} = \partial_\phi$ to reduce the system to six first-order ODEs. In analogy with Kerr's solution, we call such conserved quantities the energy $E:=-g_{\mu\nu}\eta^\mu p^\nu$ and the angular momentum $L:=g_{\mu\nu}\tilde{\eta}^\mu p^\nu$ of the geodesic, i.e., 
\begin{equation} \label{eq:conserved_EL}
    E = f p^t + (fa) p^\phi, \quad L = -(fa) p^t+\Phi p^\phi,
\end{equation}
where $\Phi:= g_{\phi\phi} = (\rho^2 - (fa)^2)/f$. Furthermore, the value of the Lagrangian is also constant along each solution. Its value is the squared length of the tangent vector to the curve, which in the present case vanishes since we are determining light rays for which the tangent vector is null\footnote{In the case of the Kerr space-time there is an additional constant, the so called Carter constant~\cite{Carter:1968a}. Its use would make the geodesic equations analytically solvable by quadratures. However, to our knowledge this constant does not exist in the more general space-times considered here, so that we refrain from using it here.}. When following particles, the value of the Lagrangian would, of course, be $-1$.

Therefore, the equations of motion of photons or test particles in  Weyl coordinates are given by the system of ODEs
\begin{equation} \label{ode_geodesics}
  \begin{split}
    \frac{dt}{ds} &= \frac{1}{\rho^2} \left(E \Phi - (fa) L \right),\\
    \frac{d \rho}{ds} &= p^\rho, \\
    \frac{d \prho}{ds} & = \frac{1}{2h} \lc -f_\rho (p^t)^2 - h_\rho (p^\rho)^2 + h_\rho (p^\zeta)^2 +  \Phi_\rho (p^\phi)^2 - 2 (fa)_\rho p^t p^\phi - 2 h_\zeta p^\rho p^\zeta \rc, \\
    \frac{d \zeta}{ds} &= \pzeta, \\
    \frac{d \pzeta}{ds} & = \frac{1}{2h} \lc -f_\zeta (p^t)^2 + h_\zeta (p^\rho)^2 - h_\zeta (p^\zeta)^2 + \Phi_\zeta (p^\phi)^2 - 2(fa)_\zeta p^t p^\phi - 2 h_\rho p^\rho p^\zeta \rc, \\
    \frac{d\phi}{ds} &= \frac{1}{\rho^2} ((fa)E+fL),
  \end{split}
\end{equation}
with $h:=g_{\rho\rho}=g_{\zeta\zeta}=e^{2k}/f$.

Thus, given initial conditions $(t_0,\pt_0,\rho_0,\prho_0,\zeta_0,\pzeta_0,\phi_0,\pphi_0)$ we obtain the conserved quantities \eqref{eq:conserved_EL} first, then determine the initial tangent vector to be null and past-pointing and then solve the initial value problem (IVP)
\begin{equation}
  \label{eq:IVP}
  \left\{
    \begin{array}{cc}
      \frac{d \vx}{ds} = F(\vx),\\
      \vx(0) = \vx_0, 
    \end{array} \right.
\end{equation}
where $\vx=(t,\rho,\prho,\zeta,\pzeta,\phi)$ and $F:\R^6\to\R^6$ is the function whose components are given by \eqref{ode_geodesics}. We express $F$ in terms of the functions $fa$, $e^{2k}$ and their respective derivatives, since the quadratures of \eqref{axi} and \eqref{kxi} are more easily expressed in terms of these functions rather than $a$ and $k$ themselves. A standard way to solve non-linear IVPs is by using numerical methods, namely, given an initial value $\vx^{(0)}\in\R^6$, we approximate the solutions $\vx^{(1)}$, $\vx^{(2)}$, ..., $\vx^{(N)}$ via evaluations of $F$ at the preceding stages. For the examples used in this paper, we used Matlab's built-in function \texttt{ode45} combining 4th and 5th order explicit Runge-Kutta methods with an automatic control of the accuracy via an adaptive time step.

\subsection{Stopping criteria} \label{sub_sec:stopping}

For all the problems of interest in this paper, the time integration is stopped if one of the following criteria is met:

\begin{itemize}

\item $f(\vx^{(n)})<\epsilon$: the coordinate time to reach the ergosphere tends to infinity. Thus, we stop the time integration when the photon gets sufficiently close to the ergosphere, i.e., the $g_{tt}$ component of the metric is approximately zero.

\item $r^{(n)} > R_\infty$, with $r^2=\rho^2+\zeta^2$: the photon escapes to infinity if it crosses a sphere with a sufficiently big radius $R_{\infty}$ in the outwards direction. The reason is that the dragging effect of the rotating space-time is too small to attract the photon towards the origin at big distances. A necessary condition on the choice of $R_\infty$ is that the ``infinity sphere'' must contain all the photon spheres (see Subsection \ref{sec:photon_sphere}). 

\item $\Lag(\mathbf{x}^{n})>\varepsilon$: the value of the Lagrangian 
\eqref{eq:Lag} surpasses a predefined threshold $\varepsilon$. The 
Lagrangian is a conserved quantity along the geodesics. Thus, if the 
numerical solution yields a different value than the initial 
$\Lag(\mathbf{x}^{0})=0$, the solution is no longer physical. In 
practice, this condition is used to choose the tolerance of the 
numerical method applied to solve the IVP \eqref{eq:IVP}, since the computation of $\Lag(\mathbf{x}^{n})$ at every step would be computationally expensive. It is observed that the tolerance values for the \texttt{ode45} must be smaller than $\varepsilon$ in order for the condition $\Lag(\mathbf{x}^{N})<\varepsilon$ to be satisfied by the final point $\mathbf{x}^{N}$.
\end{itemize}
Besides these stopping criteria, we might consider others depending on
the problem under study, for instance, for the visualisation of
non-gravitating disks around rotating black holes or visualisation in
counter-rotating disk space-times. These additional criteria will be
mentioned in their respective sections.

\subsubsection{Error propagation}

The local truncation error of the Runge-Kutta approximation is of the order $O(h^5)$, where $h$ is the step size $h=s_{n+1}-s_n$. The adaptive step size of the Matlab routine \texttt{ode45} makes sure that the error satisfies $\epsilon_n \leq \texttt{RelTol}*|\vx^{(n)}| + \texttt{AbsTol} $, where \texttt{RelTol} and \texttt{AbsTol} are the relative and absolute tolerances of \texttt{ode45}. This means that the approximation of the solution $\vx^{(n)}$ will be precise for small values of $s$. However, this being a local truncation error, the total error accumulates as the integration time $s$ increases. This loss of accuracy affects mainly the photons that orbit the black hole one or several times.

Finally, checking the criterion $\Lag(\vx^{(N)})\leq \varepsilon$ at every time step is computationally expensive for the image simulation of extended objects. Instead, we use such a criterion to choose the tolerances for the \texttt{ode45} routine. As we observed in Subsection \ref{subsec:total_errors}, the necessary tolerances needed to assure that $\Lag(\vx^{(N)})\leq \varepsilon$ for all the null geodesics coming from a light source must be one order of magnitude smaller. E.g., the necessary tolerances in order for the Lagrangians to stay under $\varepsilon = 10^{-7}$ would be $\texttt{RelTol} = \texttt{AbsTol} = 10^{-8}$.

\subsection{Computation of the function $F$}

The Runge-Kutta methods require evaluations of the function $F$ at 
times\footnote{In this case, what is usually known as the time 
parameter corresponds to the affine parameter $s$.} $s_{n}$ and at several ``stages'' between $s_{n}$ and $s_{n+1}$ (depending on the order of the method) in order to find a solution $\mathbf{x}^{(n+1)}$ for the new time $s=s_{n+1}$.  Thus, in order to compute a numerical approximation to the solution, one needs to know how to evaluate this function, and the efficiency of this evaluation is a large contributor to the overall efficiency of the method. The computation is straightforward for certain metric functions, such as the Kerr solution since the function $F$ is given in terms of elementary functions. If this is the case, the geodesics can be computed immediately. However, our goal is to consider space-times whose metric functions are non-elementary functions such as the algebro-geometric solutions to the Ernst equation. These metric functions are given in terms of theta functions that correspond to Riemann surfaces parametrised by the physical coordinates $(\rho,\zeta)$ and therefore, the computational costs of calculating such functions increase exponentially with the genus of the underlying Riemann surfaces.

We take advantage of the fact that the metric functions are piecewise smooth and, therefore, approximate them via polynomial interpolation in order to perform the computations more efficiently. Due to the symmetries of the space-times of interest, the metric functions only depend on $\rho$ and $\zeta$. Moreover, these functions are symmetric with respect to $\zeta$ in the following sense: either, $g_{\mu\nu}(\rho,\zeta) = g_{\mu\nu}(\rho,-\zeta) $ or  $g_{\mu\nu}(\rho,\zeta)=0$.  Thus, it suffices to do the computations on the domain $U=[0,\rho_{\max}]\times [0,\zeta_{\max}]$ and then use the symmetry property for $\zeta<0$. The idea is to compute the metric functions $g_{\mu\nu}$ on a 2-dimensional grid and obtain the derivatives on this grid via spectral methods. Then use barycentric interpolation to compute the values of these functions at any point $(\rho,\zeta)$ in the domain covered by the grid. Finally, we substitute these values in \eqref{ode_geodesics} in order to compute the function $F$ and use this numerical value for the Runge-Kutta method.

\subsubsection{Spectral derivatives}

In spectral methods, we approximate a function by a basis of smooth 
functions, here polynomials, and then obtain its derivatives on a grid by differentiating these polynomials. Let us consider a smooth function $v:[-1,1]\to\R$ and a Chebyshev grid on the interval $[-1,1]$, namely, the grid with points $x_j=\cos(\pi j/N)$ with $j=0,\ldots,N$. The function $v$ is approximated by Lagrange polynomial interpolation with values $V_j$ at the grid points $x_j$. The column vector with components $V_j$ is known as the grid function. Thus, we have $v(x)\approx \sum_{j=0}^NV_j\ell_j(x)$, where $\ell_j$ is the unique $N$-th degree polynomial with $\ell_j(x_k)=\delta_{ik}$. Then, the derivative of $v(x)$ is the approximation given by the derivative of its polynomial interpolation, and it can be expressed as a matrix multiplication on the grid, i.e.,
\begin{equation}
    V^{(x)} = D_N * V,
\end{equation}
where $D_N$ is known as the Chebyshev differentiation matrix, and its components depend on the grid points, for details see~\cite{Tre}. If we consider the interval $[a,b]$ instead, we use the linear transformation $x=\frac{1}{2}(b-a)\hat{x}+\frac{1}{2}(b+a)$, where $\hat{x}\in [-1,1]$, and using the chain rule, the differentiation matrix for this interval is just $2(b-a)^{-1} D_N$.

\begin{figure} [htb]
  \centering
 \includegraphics[width=8cm]{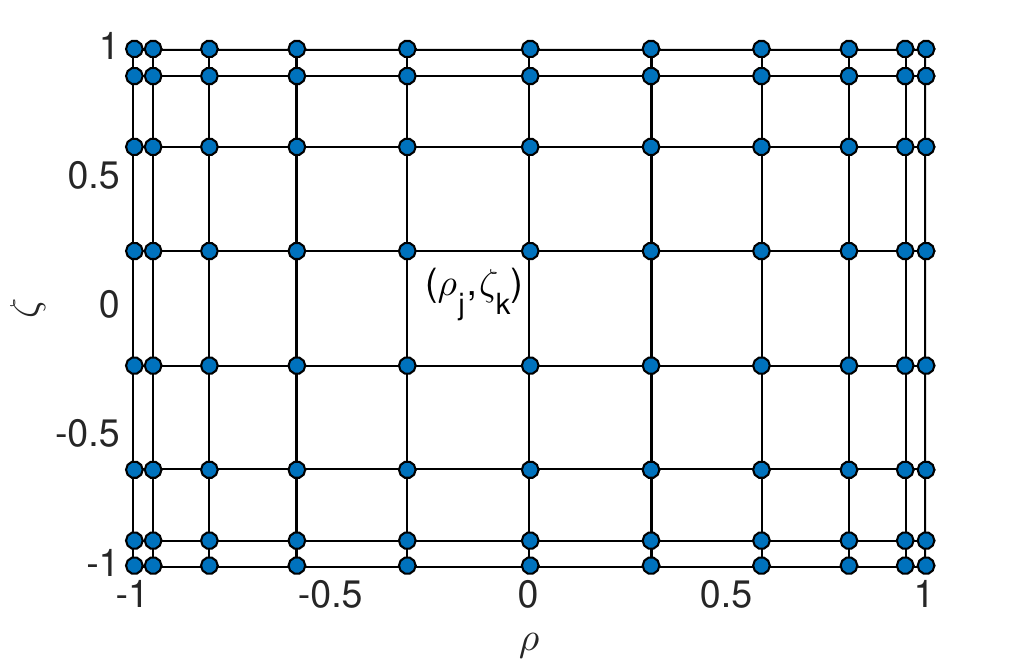}
  \caption{2-dimensional Chebyshev grid.} \label{fig:Cheb_grid_2D}
\end{figure}

In this work we consider the 2-dimensional extension of this formula, see \cite{Tre}. Let $v:U\to\R$ be a smooth function on the domain $U=[\rho_a,\rho_b]\times [\zeta_a,\zeta_b]\subset \R^2$ and $V_{k}^j=v(\rho_j,\zeta_k)$ its associated grid function, i.e., the evaluation of $v$ on the grid points shown in Fig.~\ref{fig:Cheb_grid_2D} with $\rho_j=\cos(\pi j/N_\rho)$ for $j=0,\ldots,N_\rho$ and $\zeta_k=\cos(\pi k/N_\zeta)$ for $k=0,\ldots,N_\zeta$.

Then, the derivatives of $v$ are approximated by
\begin{align*}
  \partial_\rho v(\rho,\zeta) & \approx \sum_{k=0}^{N_\zeta}\sum_{j=0}^{N_\rho} V_k^j \ell_j'(\rho) \ell_k(\zeta),\\
  \partial_\zeta v(\rho,\zeta) & \approx \sum_{k=0}^{N_\zeta}\sum_{j=0}^{N_\rho}  V_k^j \ell_j(\rho) \ell_k'(\zeta).
\end{align*}
This is the idea behind obtaining the derivatives at any $(\rho,\zeta)\in U$ via barycentric interpolation. However, it is more efficient to obtain the 2-dimensional spectral derivatives first and then apply barycentric interpolation to such grid functions. The reason is that the same grid function can be used for any point in the whole domain $U$, and the barycentric formula can be represented in matrix form, and it has exponential convergence for analytic functions, making it unnecessary to compute the Lagrange basis polynomials.

In order to approximate the derivatives $\partial_\rho v(\rho_j,\zeta_k)$ and $\partial_\zeta v(\rho_j,\zeta_k)$ by simple matrix multiplications, we write the grid function as a single column vector $\tilde{V}$. For that, we use the following ordering for the grid points: we start counting from bottom to top and then from left to right. Then the spectral derivatives with respect to $\rho$ and $\zeta$ can be expressed as

\begin{equation} \label{eq:2D_spectral_der} \begin{split}
    \tilde{V}^{(\rho)} &= (D_{N_\rho}\otimes \id_{N_\zeta}) * \tilde{V}, \\
    \tilde{V}^{(\zeta)} &= (\id_{N_\rho}\otimes D_{N_\zeta}) * \tilde{V},
\end{split}
\end{equation}
where $\otimes$ is the Kronecker product and $\id_{N}$ is the identity matrix of length $N+1$. With this step, we compute the derivatives of the metric functions $g_{\mu\nu}:\R^2 \to \R$ on the grid. In order to use barycentric interpolation, we reorder these vectors as square matrices in the same order as the 2-dimensional grid of Fig.~\ref{fig:Cheb_grid_2D} and denote them $V^{(\rho)}$ and $V^{(\zeta)}$.

\subsubsection{Barycentric interpolation}

Let $u:U\subset\R^2\to\R$ be a smooth function on the domain $U=[\rho_a,\rho_b]\times [\zeta_a,\zeta_b]$. We cover this domain with a grid as shown in Fig.~\ref{fig:Cheb_grid_2D}. Let $M$ be the matrix whose components are the values of $u$ on the grid points, i.e., $M_k^j = u(\rho_j,\zeta_k)$.

Thus, the value of the function $u$ at an arbitrary point $(\rho,\zeta)\in [\rho_a,\rho_b]\times [\zeta_a,\zeta_b]$ can be interpolated via a 2-dimensional extension of the barycentric interpolation~\cite{barycentric}. The 2-dimensional barycentric formula is given by
\begin{equation} \label{form:barycentric_interp}
    u(\rho,\zeta) \approx \frac{1}{a(\rho)b(\zeta)} \sum_{j=0}^{N_\rho} \sum_{k=0}^{N_\zeta} \frac{w^\rho_j w^\zeta_k}{(\rho-\rho_j)(\zeta-\zeta_k)} M_k^j,
\end{equation}
where $a(\rho)=\sum_j w^\rho_j/(\rho-\rho_j)$, $b(\zeta)=\sum_k w^\zeta_k/(\zeta-\zeta_k)$ and $w^\rho_j$ (respectively $w^\zeta_k$) are the weights on the one-dimensional grid on the interval $[\rho_a,\rho_b]$ (respectively $[\zeta_a,\zeta_b]$).  The barycentric weights are given by the formula
\[
  w_j= \frac{1}{\prod_{k\neq j} (x_k-x_j)},
\]
for a general grid $\{x_0,x_1,\ldots,x_N\}$ on the interval $[a,b]$. However, if we consider a Chebyshev grid, the weights take the simple form 
\begin{equation} \label{form:weights}
  w_j=(-1)^j \varepsilon_j,\qquad \varepsilon_j=\left\{ \begin{array}{ll}
1/2, & j=0 \text{ or } j=N, \\
    1, & \text{otherwise}.
\end{array}\right.
\end{equation}
The barycentric formula does not depend on the limits $a,b$ of the interval, since the grid on $[a,b]$ can be expressed in terms of the grid on $[-1,1]$ via the linear transformation $x=\frac{1}{2}(b-a) \hat{x}+\frac{1}{2}(b+a)$, with $\hat{x}\in[-1,1]$. It can be seen that the weights are multiplied by the same constant factor $2^N(b-a)^{-N}$, which cancels out in the barycentric formula \eqref{form:barycentric_interp}. Thus, we can use the same formula \eqref{form:weights} for the weights on Chebyshev grids, regardless of the intervals.

For computational purposes, we express the barycentric formula in terms of matrix multiplications. Let $A(\rho)$ and $B(\zeta)$ be the column vectors with entries $A_j(\rho)=w^\rho_j/(\rho-\rho_j)$, $ B_k(\zeta)=w^\zeta_k/(\zeta-\zeta_k)$ with $\rho\ne\rho_j$ and $\zeta\ne\zeta_k$. Then,
\begin{equation} \label{form:barycentric_matrix}
    u(x,y) \approx \frac{1}{a(\rho)\cdot b(\zeta)} B^\top(\zeta) * M * A(\rho),
\end{equation}
where `$*$' is the matrix multiplication.

\subsubsection{Algorithm}

Summarising, in order to obtain the value of the metric functions $g_{\mu\nu}(\rho,\zeta)$ and their partial derivatives $\partial_\rho g_{\mu\nu}(\rho,\zeta)$, $\partial_\zeta g_{\mu\nu}(\rho,\zeta)$ at any arbitrary point $(\rho,\zeta)\in [\rho_{\min},\rho_{\max}]\times [\zeta_{\min},\zeta_{\max}]$, we perform the following steps:
\begin{itemize}
\item Compute the metric functions $g_{\mu\nu}$ on the 2-dimensional
  grid $[\rho_{\min},\rho_{\max}]\times [\zeta_{\min},\zeta_{\max}]$
  with resolution $(N_\zeta+1) \times (N_\rho+1)$.
\item Compute the spectral derivatives of such functions via the
  formula \eqref{eq:2D_spectral_der}, where
  $V_k^j=g_{\mu\nu} (\rho_j,\zeta_k)$.
    \item Compute $g_{\mu\nu}(\rho,\zeta)$, $\partial_\rho g_{\mu\nu}(\rho,\zeta)$ and $\partial_\zeta g_{\mu\nu}(\rho,\zeta)$ at any required point $(\rho,\zeta)$ using the barycentric interpolation formula \eqref{form:barycentric_matrix}, considering the grid function $V$ and its spectral derivatives $V^{(\rho)}$ and $V^{(\zeta)}$ as the matrix $M$.
\end{itemize}

The values of the metric functions for $\zeta<0$ can be obtained by considering the equatorial symmetry $g_{\mu\nu}(\rho,-\zeta)=g_{\mu\nu}(\rho,\zeta)$, which in turn implies 
\begin{align*}
  \partial_\rho g_{\mu\nu}(\rho,\zeta) &= \partial_\rho g_{\mu\nu}(\rho,|\zeta|),\\
  \partial_\zeta g_{\mu\nu}(\rho,\zeta) &= {\rm sign}(\zeta) \cdot \partial_\zeta g_{\mu\nu}(\rho,|\zeta|).
\end{align*}
Thus, it is sufficient to use a domain with $\zeta_{\min}\geq 0$ for the interpolation.
Moreover, since we are considering metrics in Weyl-Lewis-Papapetrou form, it is sufficient to perform these computations on the functions $f=-g_{tt}$, $h=g_{\rho\rho}=g_{\zeta\zeta}$ and $fa=-g_{t\phi}$. The remaining metric function is $g_{\phi\phi}=(\rho^2-(fa)^2)/f$. Then, $g_{\phi\phi}$, $\partial_\rho g_{\phi\phi}$ and $\partial_\zeta g_{\phi\phi}$ can be computed via the interpolated values of $f$, $fa$ and their derivatives.

\subsection{Choice of domains} \label{subsec:domains}
The chosen domain $U\in\R^2$ on which the barycentric interpolation 
algorithm is to be applied should contain all the regions in which we intend to study the photon trajectories. However, since we use the symmetry relations when $\zeta<0$, a domain of the form $U=[0,\rho_{\max}]\times [0,\zeta_{\max}]$ is sufficient, where the upper limits $\rho_{\max}$ and $\zeta_{\max}$ are to be determined. 

\begin{figure} [htb]
    \centering
    \includegraphics[width=7cm]{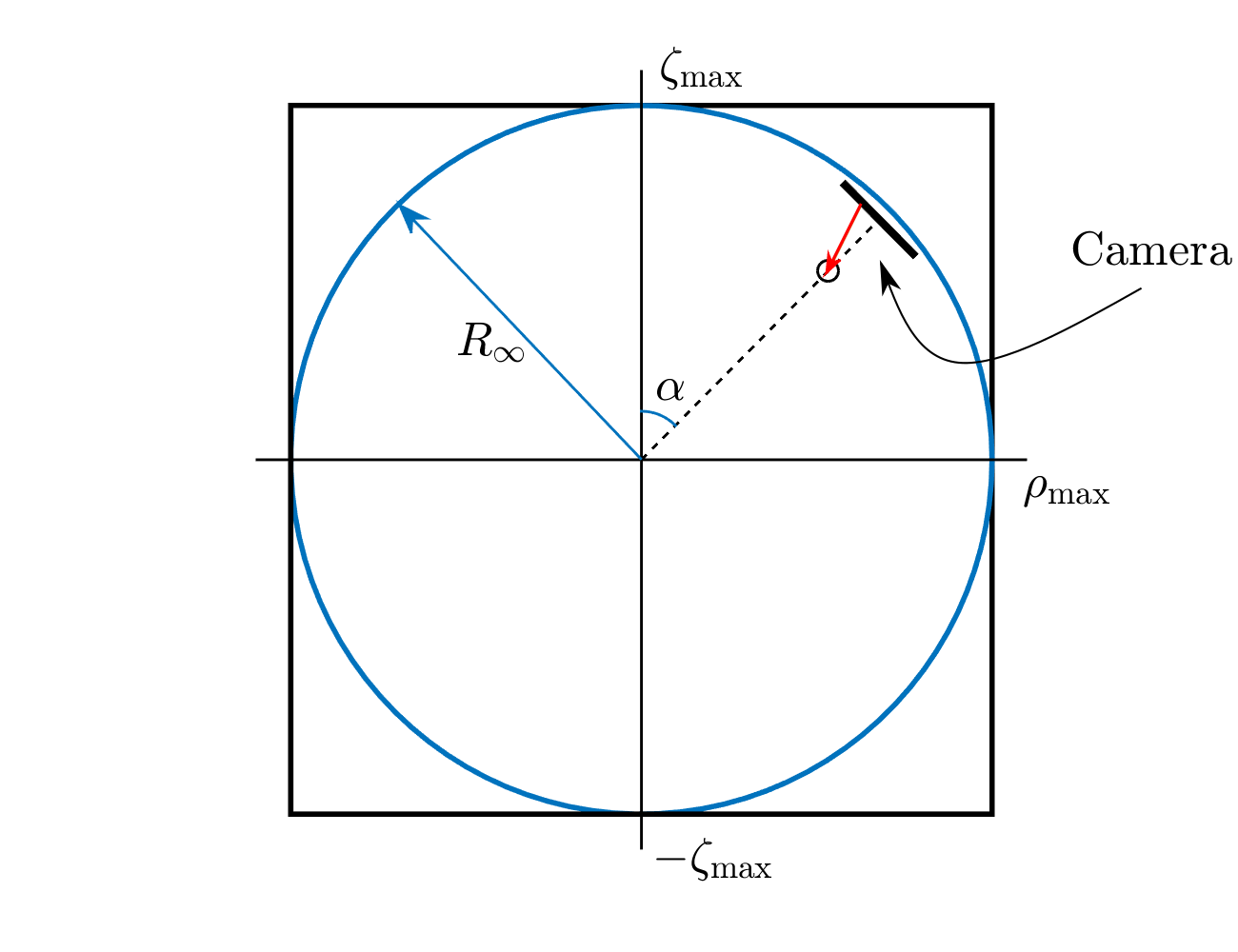}
    \caption{Projection of the virtual camera and the ball with radius $R_\infty$ on the $xz$-plane.} \label{fig:domain_Rinfty}
\end{figure}

In image simulation problems, we shoot the photons in the vicinities of a black hole or a counter-rotating disk from the virtual camera in Fig.~\ref{fig:camera} for various angles $\alpha\in [0,\pi/2]$. 
Then, the distance from the origin of the simulated photons will be less than the initial distance, unless they eventually escape to infinity. Therefore, an adequate condition on the radius $R_\infty$ used as a stopping criterion after which the photon is expected to escape to infinity is that the ball with radius $R_\infty$ should contain the virtual camera, as shown in Fig.~\ref{fig:domain_Rinfty}. A detailed description of the virtual camera is presented in Section \ref{sec:RT}, as well as the explicit value of $R_\infty$.

We recall that the numerical integration is terminated if $r=\sqrt{\rho^2+\zeta^2}>R_\infty$ (for a sufficiently big $R_\infty$) and the photon is traveling in outwards direction. The necessary condition on the choice of $R_\infty$ is that the ball with radius $R_\infty$ must contain all the photon spheres, which is the case for the choice we have made since we consider a distant observer. 

In other words, all the geodesics we are interested in are inside the 
ball with radius $R_\infty$, shown in Fig.~\ref{fig:domain_Rinfty}. 
Therefore, considering that we use the symmetry of the metric to 
evaluate the functions for negative $\zeta$, a necessary condition on 
the chosen domain $U=[0,\rho_{\max}]\times [0,\zeta_{\max}]$ for the 
interpolation algorithm is $B_\infty \subset D$, where $B_\infty$ is 
the portion of the ball on the first quadrant $B_{\infty}:=\{ 
(\rho,\zeta)\in\R_+^2 \mid \rho^2+\zeta^2\leq R_\infty^2 \}$. Since 
we are considering a rectangular domain, the upper limits must 
satisfy $\rho_{\max},\zeta_{\max}\geq R_\infty$ in order to meet the 
condition $B_{\infty} \subset D$.

\begin{figure} [htb]
  \centering
  \includegraphics[width=8cm]{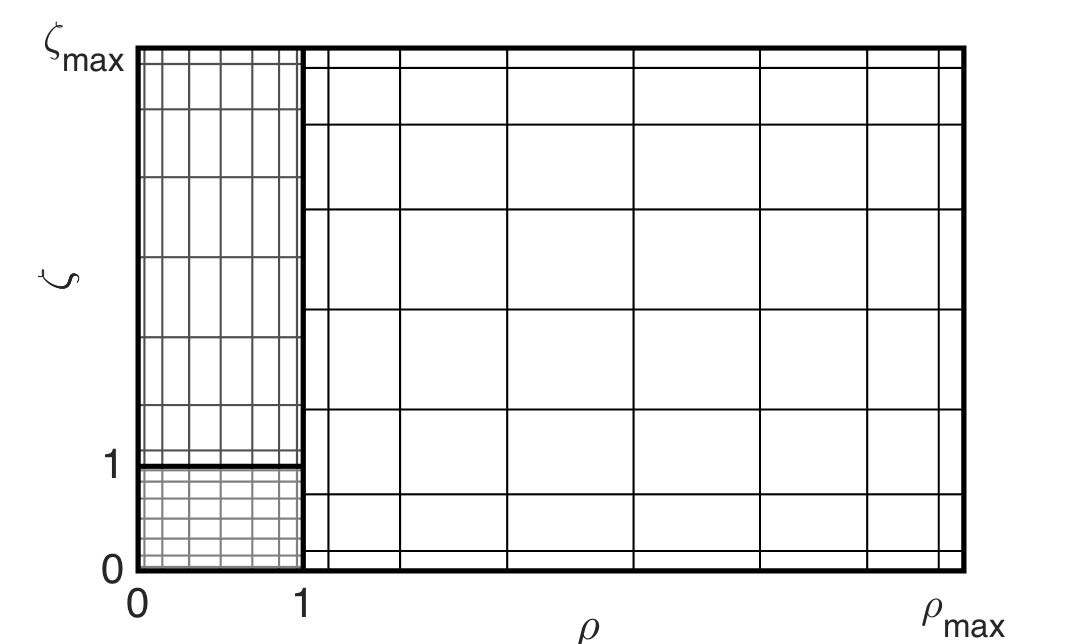}
  \caption{Chebyshev grids on the subdomains $U_1$, $U_2$ and $U_3$.}
  \label{fig:grid_multidomain}
\end{figure}

\textbf{Remark:} It is important to recall that the metric functions $g_{\mu\nu}: U\to\R$ must be smooth on the square domain $U$ on which we intend to do the interpolations. If this is the case, the numerical errors will decrease exponentially in dependence of the grid resolution and thus, convergence to machine precision is attained rapidly. Otherwise, this convergence cannot be assured. If the metric is piecewise smooth, then the domain $U$ can be divided into subdomains on which the metric is smooth, e.g., the division shown in Fig.~\ref{fig:grid_multidomain}.

However, there are cases in which the metric functions are not everywhere smooth, such as those corresponding to the two classes of solutions studied in this paper, namely, the Kerr metric in Section \ref{sec:sav-spacetimes} and the counter-rotating disk solutions in Section \ref{sec:cr-disk}. 
The first one, the Kerr metric, has cusps at the endpoints of the horizon $(\rho,\zeta)=(0,\pm m \cos\varphi)$ since it involves $\sqrt{(\zeta\pm m\cos\varphi)^2 + \rho^2}$. This means that the interpolated functions cannot achieve machine precision in Weyl coordinates if this point is in the interpolating domain. 
The second class of solutions we study is the counter-rotating disk, whose metric functions have a cusp at $(\rho,\zeta)=(1,0)$, the rim of the disk. Although the Ernst potential is discontinuous on the line $\zeta=0$ with $\rho<1$, the metric functions are continuous and the only issue is the cusp.  
In order to evaluate functions at any point $(\rho,\zeta)\in U$ for metrics with issues of this type, we perform the interpolation in adapted coordinates in terms of which the functions are smooth.  

\subsection{Cases where the metric is not everywhere smooth in Weyl 
coordinates} \label{subsec:cusps}

\subsubsection{Kerr metric} 
\begin{figure} [htb]
    \centering
    \includegraphics[width=6cm]{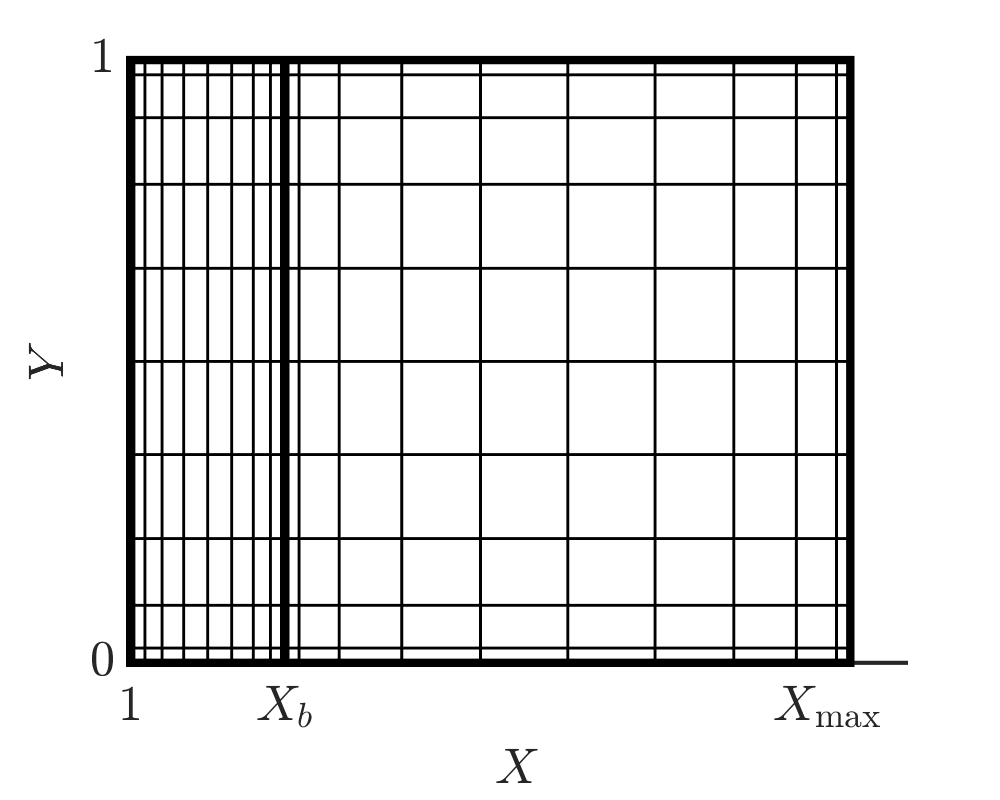}\includegraphics[width=6cm]{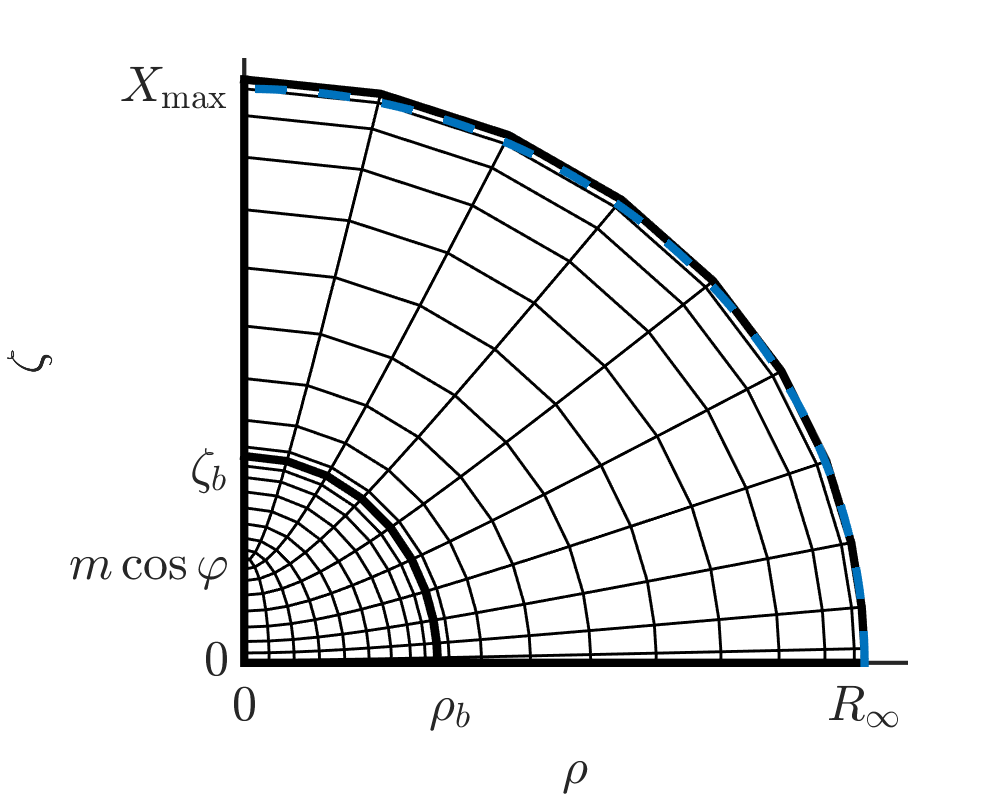}
    \caption{Mapping of the rectangular domain $U'$ onto $B_\infty$. The boundary of $B_\infty$ is indicated by the dashed line.} \label{fig:domain_kerr}
\end{figure}

For the Kerr metric functions, we can use parabolic coordinates whose relation with $(\rho,\zeta)$ is given by \eqref{kerr6a}. It can be seen that the metric functions \eqref{kerr6}, \eqref{kerr5b}, \eqref{kerr7} are rational functions in these coordinates. Thus, we interpolate the numerators and denominators on the domain $U'=[1,X_{\max}]\times [0,1]$. The condition on the upper limit $X_{\max}$ is that the image of the map $\psi:U'\to \R^2$ defined by $(X,Y)\mapsto (\rho,\zeta)$ must contain $B_{\infty}$. It can be seen from the relation \eqref{kerr6a} that the preimage of $B_{\infty}$ is given by 
$$B'_{\infty}=\left\{(X,Y)\in [1,+\infty)\times [0,1] \mid X^2+Y^2\leq 1+ \left(R_\infty / (m\cos\varphi) \right)^2 \right\}.$$
Then, considering that $Y\in[0,1]$, a domain satisfying $B'\subset U'$ and therefore $B_{\infty}\subset \psi(U')$ is given by $U'=[1,X_{\max}]\times [0,1]$, where the upper limit must satisfy $X_{\max}\geq \sqrt{1+(R_\infty/(m\cos\varphi))^2}$. Thus, for the Kerr examples considered in this paper, we evaluate $g_{\mu\nu}$, $\partial_X g_{\mu\nu}$ and $\partial_Y g_{\mu\nu}$ at any point $(X,Y)\in U'$ via barycentric interpolation. Therefore, in order to evaluate the metric functions and their derivatives at any $(\rho,\zeta)\in B_{\infty}$, we map this point to $(X,Y)\in U'$, compute the interpolations at $(X,Y)$ and then use the chain rule
\begin{align*}
    \partial_\rho g_{\mu\nu} & = \partial_X g_{\mu\nu} \cdot X_\rho + \partial_Y g_{\mu\nu} \cdot Y_\rho, \\
    \partial_\zeta g_{\mu\nu} & = \partial_X g_{\mu\nu} \cdot X_\zeta + \partial_Y g_{\mu\nu} \cdot Y_\zeta,
\end{align*}
as well as the symmetry relations if $\zeta<0$. Moreover, since the 
metric functions vary slowly away from the origin, it is more 
efficient to perform the interpolations on multiple subdomains of 
$U'$. E.g., we can choose $U'_1=[1,X_b] \times [0,1]$ and $U'_2 = 
[X_b,X_{\max}] \times [0,1]$, as shown in Fig.~\ref{fig:domain_kerr}, where $\rho_b$ and $\zeta_b$ are in dependence of $X_b$. Otherwise, in order to attain machine precision a higher grid resolution would be necessary. For the examples shown in this paper, we used the values $X_b=5$, $X_{\max} = \sqrt{1+(R_\infty/(m\cos\varphi))^2}$ and grids with $$N_{X}=N_{Y}=30.$$

\subsubsection{Counter-rotating disk} For the counter-rotating disk metric functions, we use the so-called disk coordinates\footnote{Notice that, depending on the parameters $\lambda$ and $\delta$, the rim of the disk might be inside the ergosphere and in such cases, it will not be necessary to cover that region with a Chebyshev grid and the interpolation can still be done in Weyl coordinates using multiple rectangular domains.} (see \cite{FK04,BT} for more details), related to $(\rho,\zeta)$ via the homomorphism 
\begin{align*}
    \psi: \R_+ \times [0,\pi/2] &\to \R_+\times \R, \\
    (\eta,\theta) &\mapsto ( \rho,\zeta ),
\end{align*}
defined by $\rho = \cosh(\eta)\cos(\theta)$, $ \zeta = \sinh(\eta)\sin(\theta)$. 
Using these relations, the preimage of $B_{\infty}$ is given by 
$$B'_{\infty}=\{ (\eta,\theta)\in \R_+ \times [0,\pi/2] \mid \sinh^2(\eta) + \cos^2(\theta) \leq R_\infty^2 \}.$$
Thus, considering that $\theta\in[0,\pi/2]$ and the condition $B'_{\infty}\subset U'$, the domain must be of the form $U'=[0,\eta_{\max}] \times [0,\pi/2]$ with the upper limit satisfying $\eta_{\max}\geq {\rm arcsinh}(R_\infty)$. 

\begin{figure} [htb]
    \centering
    \includegraphics[width=6cm]{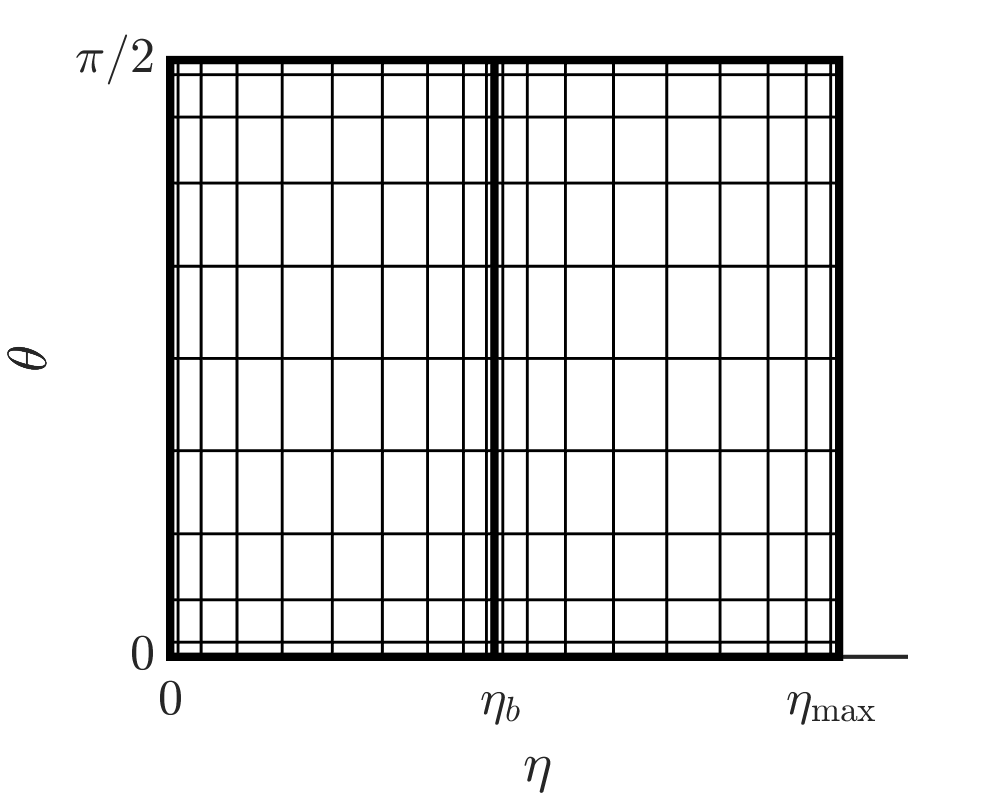} 
    \includegraphics[width=6cm]{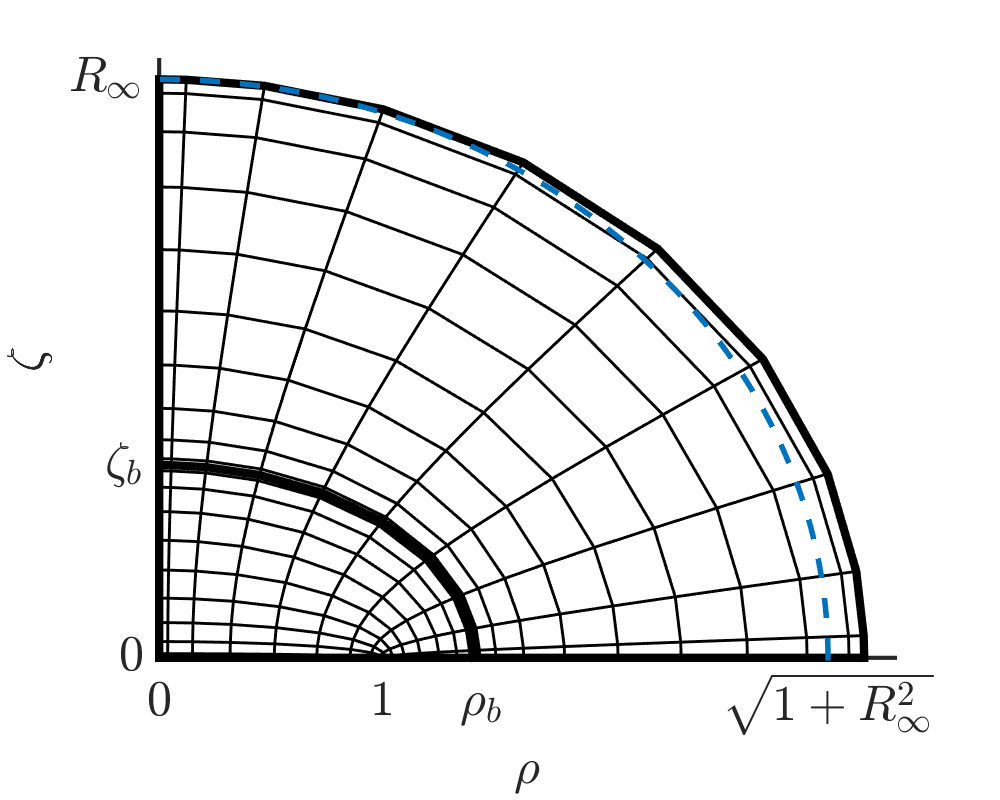}
    \caption{Mapping of the rectangular domain $U'$ onto $B_\infty$. The boundary of $B_\infty$ is indicated by the dashed line.} \label{fig:domain_disk}
\end{figure}

Analogous to the computations for Kerr, the evaluation of the metric functions and their derivatives at any $(\rho,\zeta)\in B_{\infty}$ is obtained by mapping this point to $(\eta,\theta)\in U'$, computing the interpolations of $g_{\mu\nu}$, $\partial_\eta g_{\mu\nu}$ and $\partial_\theta g_{\mu\nu}$ at $(\eta,\theta)$ and then using the relations
\begin{align*}
    \partial_\rho g_{\mu\nu} & = \partial_\eta g_{\mu\nu} \cdot \eta_\rho + \partial_\theta g_{\mu\nu} \cdot \theta_\rho, \\
    \partial_\zeta g_{\mu\nu} & = \partial_\eta g_{\mu\nu} \cdot \eta_\zeta + \partial_\theta g_{\mu\nu} \cdot \theta_\zeta.
\end{align*}
Moreover, since the metric functions vary slowly away from the origin, it is more efficient to perform the interpolation on multiple subdomains of $U'$. E.g., we can choose $U'_1=[1,\eta_b] \times [0,\pi/2]$ and $U'_2 = [\eta_b,{\rm arcsinh}(R_\infty)] \times [0,\pi/2]$, as shown in Fig.~\ref{fig:domain_disk}, where $\rho_b$ and $\zeta_b$ are in dependence of $\eta_b$. Otherwise, a higher grid resolution would be necessary to attain machine precision if we were to use a single domain $U'$. In this paper, we used the values $\eta_b={\rm arcsinh}(1)$, $\eta_{\max}={\rm arcsinh}(R_\infty)$ and grids with 
$$N_{\eta}=N_{\theta}=30.$$ 

Finally, we recall that unlike the Kerr metric, the counter-rotating 
disk metric functions (and in general, algebro-geometric metric 
functions) can only be computed numerically, this being the main reason to compute them via interpolation due to its high computational cost. Thus, it is essential to compute the grid functions to machine precision in order to minimise error propagation, since they are the input in the barycentric interpolation algorithm. Recall, that the algebro-geometric solutions only hold if the branch points of the underlying Riemann surfaces do not coincide; which means that in order to evaluate the metric functions for the counter-rotating disk on the axis and the equatorial plane (i.e., when $\rho=0$ or $\zeta=0$), we need to use formulas on degenerated Riemann surfaces, which are given explicitly in \cite{FK04}.

\begin{figure}[htb]
\centering
  \includegraphics[width=0.3\textwidth]{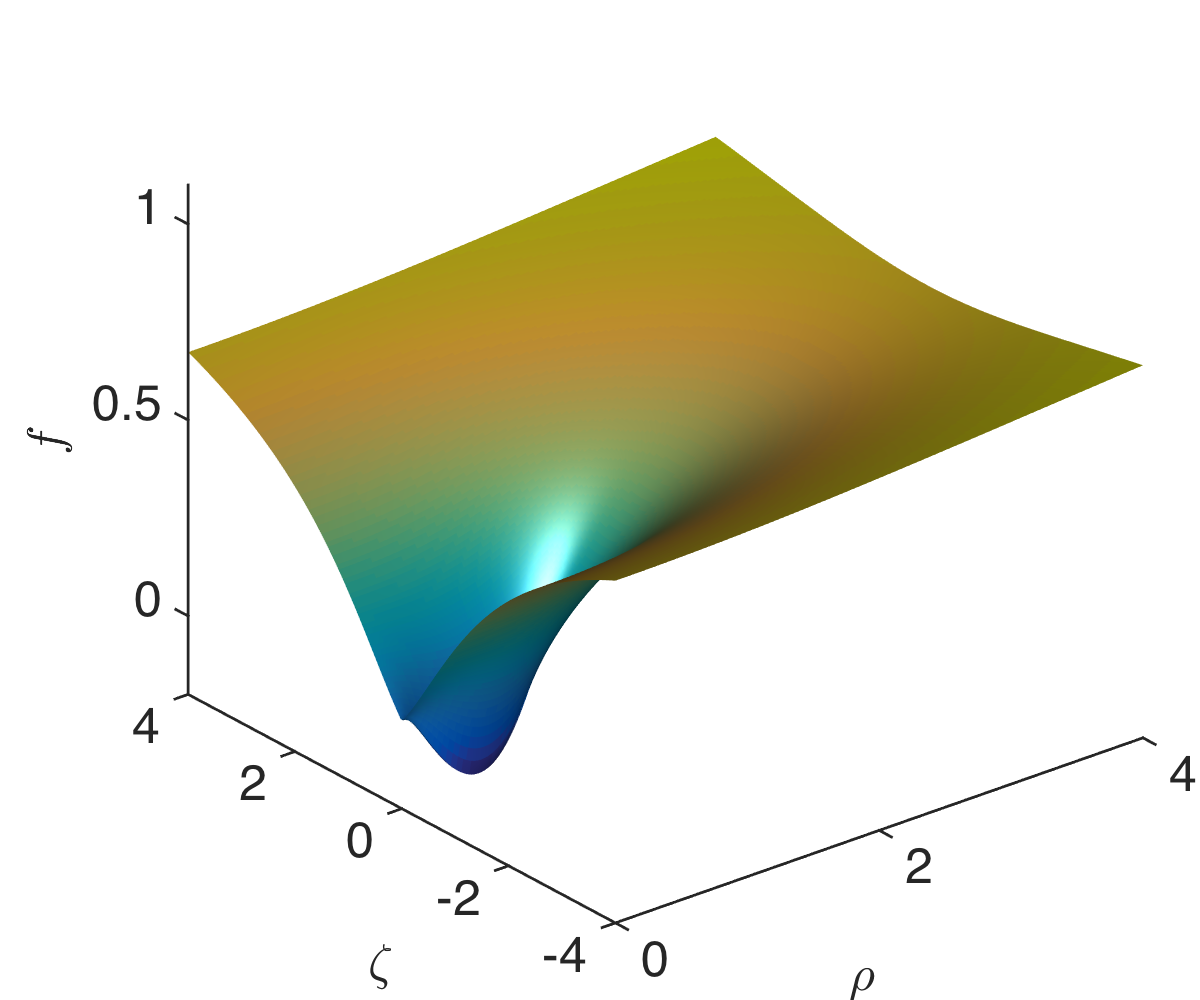} \includegraphics[width=0.3\textwidth]{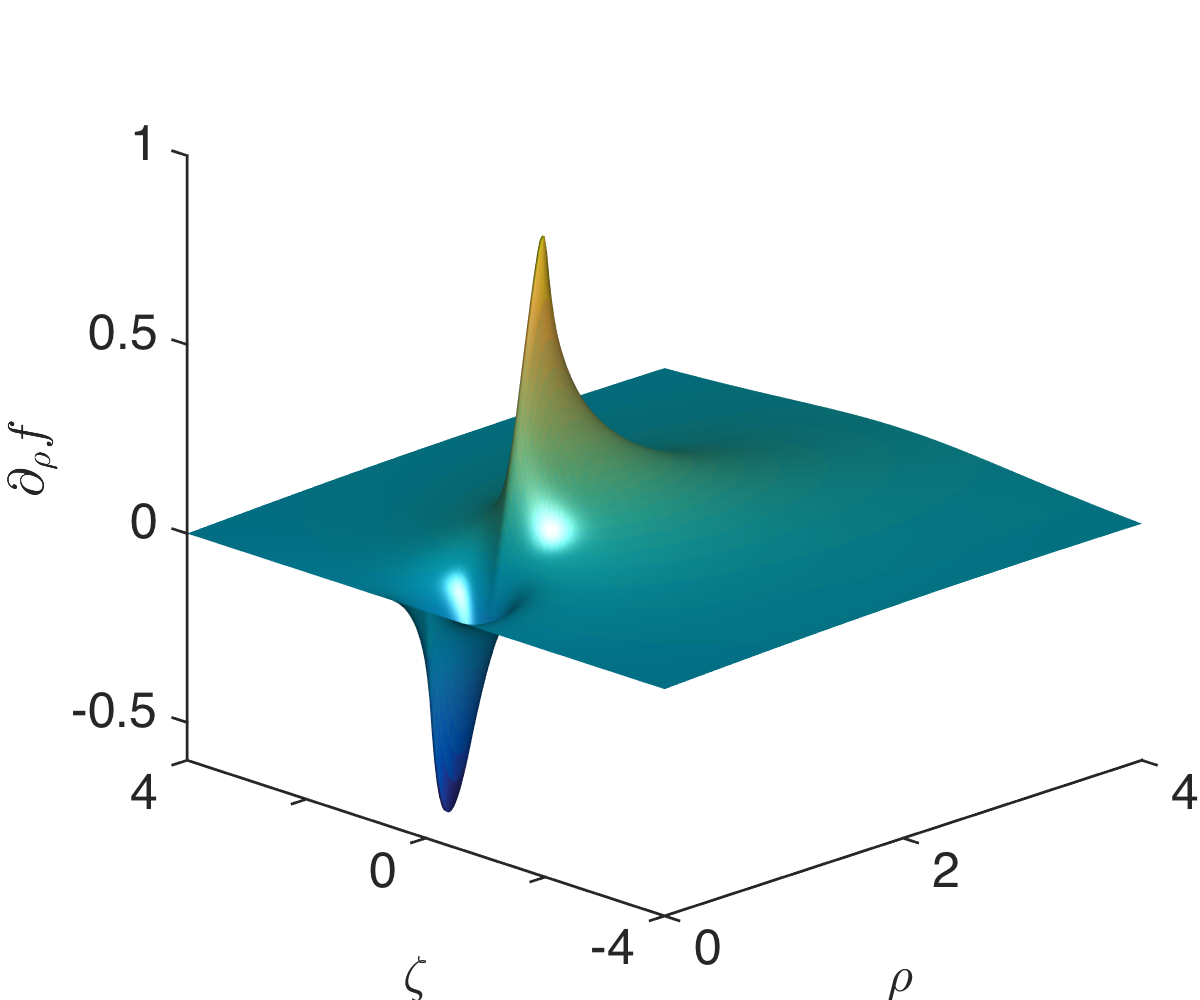} \includegraphics[width=0.3\textwidth]{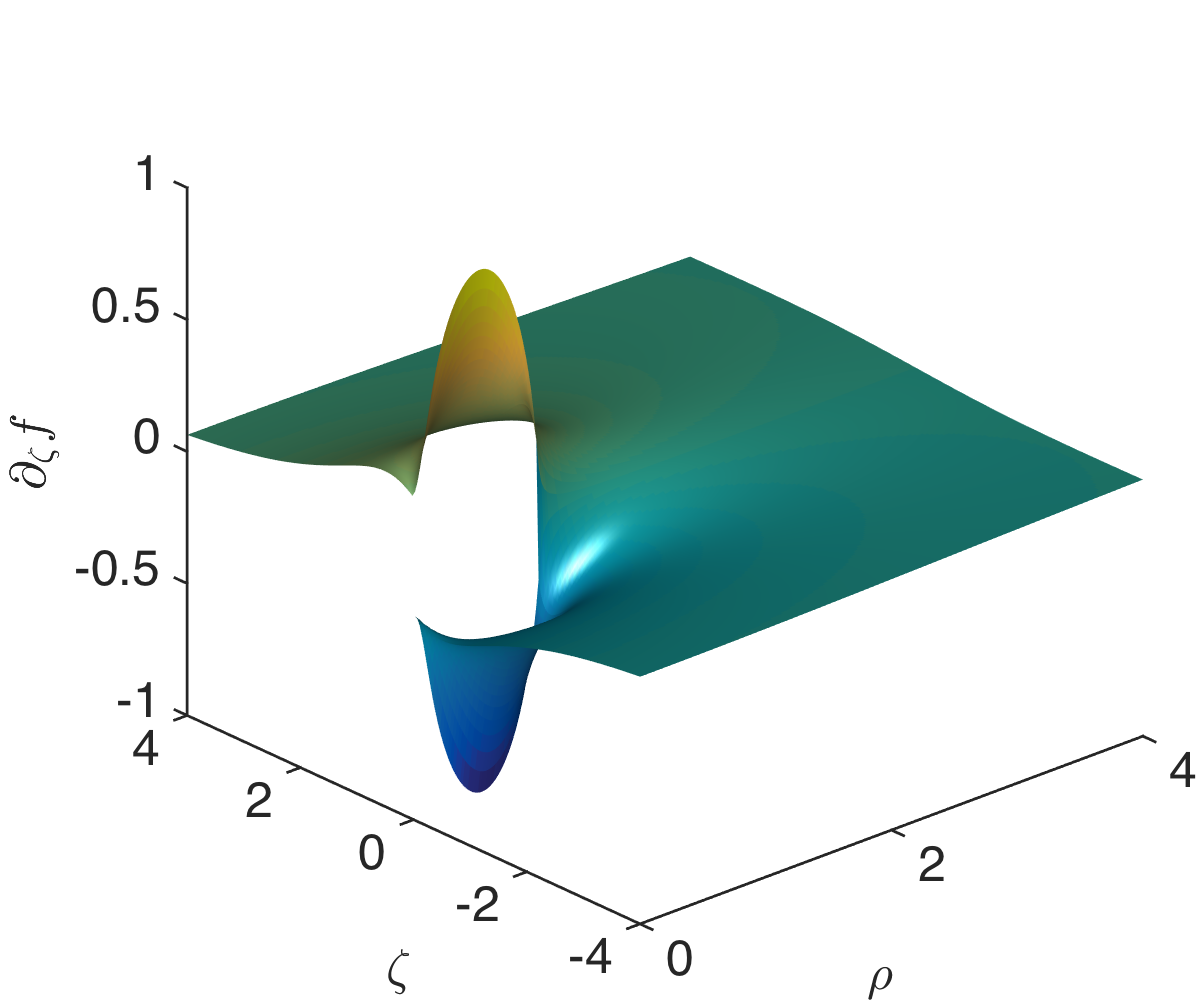}
  \caption{Barycentric interpolation of $f$ and its
    derivatives on a $200\times 200$ equispaced grid.} \label{fig:metric_f}
\end{figure}

For illustration purposes, we show the application of barycentric 
interpolation to the counter-rotating disk with parameters $\lambda=10.12$ and $\delta=0.856$. We compute the values of the metric function $f$, as well as its derivatives for points $(\rho,\zeta)\in \tilde{U}$, where $\tilde{U}=[0,4]\times [-4,4]$. These interpolated functions are shown in Fig.~\ref{fig:metric_f}, and they were computed on an equispaced grid of size $200\times 200$ on $\tilde{U}$.

\section{Relativistic ray tracing} \label{sec:RT}

In this section, we briefly describe the method we use to generate images of a particular situation or scene. This has been described in many places before; see, for instance, \cite{luminet, mueller, Muller:2012,interstellar}. In short, the idea is to select a pixel, i.e., a point in a predefined screen of a given resolution, and follow the light ray that hits this pixel backwards in time to find out where it originated from. The pixel is filled with a color that is determined by this point of origin, usually located on a body of interest, possibly modified by any transformation that could be affected by physical processes on the photon during its travel. A typical example would be the Doppler effect due to gravitational redshift.

Thus, the entire procedure is quite straightforward, and it can be separated in a clean way into different parts:
\begin{itemize}
\item the `camera',
\item the screen,
\item the light ray,
\item the bodies.
\end{itemize}
The camera, together with the screen, serves to select the initial conditions for the light rays which are sent into the past. The camera is placed in a fixed location, pointing in the direction of the scene. It contains the screen, the extent of which defines the solid angle that is captured. The screen has a certain resolution, i.e., a 2-dimensional array of pixels. Each of these pixels defines a direction into which a light ray is started. The light ray is obtained by solving the geodesic equation for a null geodesic with initial conditions determined by the camera and a pixel on the screen. In contrast to non-relativistic ray tracing, this is much more complicated because a system of ODEs must be solved, while in the non-relativistic case, the rays are obtained as -- possibly broken -- straight lines.

As the light rays propagate into the scene, they may hit any bodies that make up the scene. In general, these bodies must be defined by their world tube, i.e., the collection of world lines defined by its constituent points. Therefore, in situations when both the space-time and the bodies change in time, the detection of `hits' between light rays and bodies becomes very time consuming and difficult. For this reason, most relativistic visualisations have been done in stationary space-times with (almost) static bodies.

Each body carries information about the optical properties of its surface, such as opacity, textures, etc. These are used to determine the fate of the light ray. If the surface is opaque, then the light ray ends there, and the color of the corresponding pixel is determined by the texture. If the surface is partly reflexive and partly transparent, then new light rays must be started according to the refractive properties defined on the surface. These are followed back into the past until they hit other bodies.

\subsection{Initial conditions associated with individual pixels}
\begin{figure} [htb]
    \centering
    \includegraphics[width=8cm]{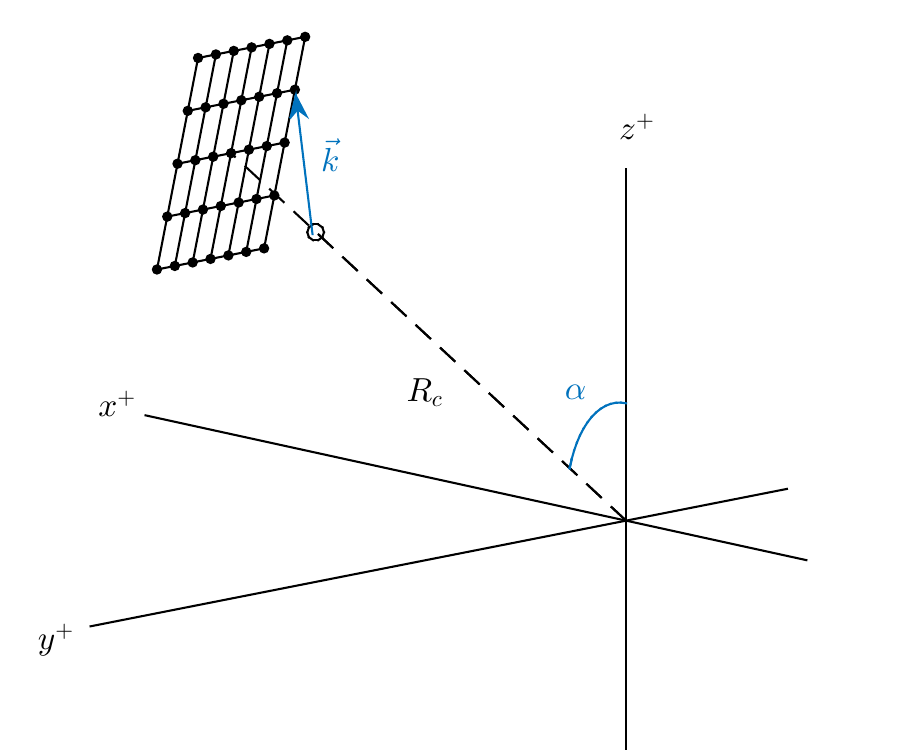}
    \caption{Diagram of the camera with an inclination $\alpha$ and one incoming light ray. }   \label{fig:camera}
\end{figure}

More precisely, we simulate the picture taken by a \textit{camera obscura}, whose diagram is shown in Fig.~\ref{fig:camera}. It is composed of an aperture (the place in which light enters the camera) and a film (the 2-dimensional screen on which the image is projected). It is placed pointing towards the origin of the coordinate system, and its normal vector (the principal axis) forms an angle $\alpha$ with respect to the positive $z$-axis. 

The technique consists in associating an event $\vx$ to each pixel $(i,j)$ on the screen as well as the four-momentum $\vp$ of the incoming light ray that hits that pixel. Then, we use this information as the initial conditions of the IVP \eqref{eq:IVP}, but with the opposite sign for $\vp$ since we are interested in computing the source of each light ray. Therefore, we need to solve an IVP for each pixel, i.e., we use the initial value to trace the photon back to the light source, record that event and associate this information to the corresponding pixel. 

Let us consider a screen with the following physical properties: width $d_H$, height $d_V$, vertical resolution $I_V$, horizontal resolution $I_H$ and focal length $f_L$. 
The camera is positioned at a distance $R_c$ from the origin (object distance) and with an inclination angle $\alpha$ with respect to the $z$-axis.
Therefore, the event associated with each pixel $(i,j)$ has components
\begin{equation} \label{eq:IC}
\begin{array}{ll}
   t_{ij} = 0,  & \rho_{ij} = \sqrt{x_{ij}^2+y_{ij}^2},\\    
    \zeta_{ij} = z_{ij}, &  \phi_{ij} = \arctan(y_{ij}/2), 
\end{array}
\end{equation}
where
\begin{align*}
    x_{ij} &= \cos \alpha \cdot \tilde{x}_{ij} + \sin\alpha \cdot (R_c + f_L),\\
    y_{ij} &= d_H \lc \frac{1}{2} - \frac{(j-1)}{(I_H-1)} \rc,\\
    z_{ij} &= -\sin\alpha \cdot \tilde{x}_{ij} + \cos\alpha \cdot (R_c + f_L),
\end{align*}
with
\begin{equation*} 
    \tilde{x}_{ij} = d_V \lc \frac{1}{2} - \frac{(i-1)}{(I_V-1)} \rc,
\end{equation*}
as shown in Appendix \ref{append_camera}.
The four-momentum associated to the pixel $(i,j)$ has spatial components
\begin{align*}
    p^\rho_{ij} &= \cos\phi_{ij} \cdot p^x_{ij} + \sin\phi_{ij} \cdot p^y_{ij}, \\
    p^\zeta_{ij} &= ( \cos\alpha \cdot R_c - z_{ij})/r_c = (\sin\alpha \cdot \tilde{x}_{ij} - \cos\alpha\cdot f_L)/r_c ,\\
    p^\phi_{ij} &= \frac{1}{\rho_{ij}} \left( \cos\phi_{ij} p^y_{ij} - \sin\phi_{ij} p^x_{ij} \right),
\end{align*}
with
\begin{align*}
    p^x_{ij} &= ( \sin\alpha \cdot R_c - x_{ij})/r_c = -( \cos\alpha \cdot  \tilde{x}_{ij} + \sin\alpha \cdot f_L)/r_c,\\
    p^y_{ij} &= - y_{ij}/r_c,\\
\end{align*}
where $r_c=\sqrt{f_L^2+\tilde{x}_{ij}^2+y_{ij}^2}$ is the normalising constant such that $|\vec{p}_{ij}|^2=1$. The remaining component $p_{ij}^t$ is chosen by solving the quadratic equation $g_{\mu\nu} p_{ij}^\mu p_{ij}^\nu=0$ for the variable $p_{ij}^t$ and choosing the solution with negative sign since we are interested in tracing the photon back in time.

\subsection{Explicit value of $R_\infty$} \label{subsec:value_R_infty}
The condition we imposed on $R_{\infty}$ in Subsection \ref{subsec:domains} for the choice of the domains for the barycentric interpolation algorithm is that the virtual camera must be fully contained in the sphere with radius $R_{\infty}$. In terms of the parameters of the camera, this condition is equivalent to
\begin{equation} \label{eq:R_infty}
    R_\infty \geq \sqrt{(R_c+f_L)^2 + (d_H/2)^2 + (d_V/2)^2}.
\end{equation}
The values of these parameters used throughout the whole paper are summarised in 
Table \ref{table:camera}. 

\begin{table}[htb]
	\centering
\begin{tabular}{|c|c|c|c|c|}
	\hline
	$R_c$ & $f_L$ & $d_H$ & $d_V$ & $R_\infty$ \\
	\hline
	20 & 0.1 & 0.1 & 0.1 & 21 \\
	\hline
\end{tabular}
		\caption{Values of the parameters of the virtual camera.}
	\label{table:camera}
\end{table}
The only parameter that will be movable is the focal length $f_L$, 
which will depend on the required angle of view of the observer. Analogous to a normal camera, the angle of view decreases as the focal length increases and vice versa.
However, this length will always be of the order of magnitude of $d_H$ and $d_V$. Thus, $R_\infty=21$ will always satisfy the condition \eqref{eq:R_infty} regardless of the chosen value for $f_L$.


\section{Example: Kerr solution} \label{sec:example-kerr}

We start by testing this method with the solution shown in Section 
\ref{sec:sol-ernst}, which is just the Kerr solution in Weyl coordinates. The aim is to use these simulations as testbeds 
for the numerical methods outlined above. Kerr space-times have been 
widely studied, so we know what to expect, i.e., there are known 
phenomena that we will be looking for, such as the shadow of a black 
hole. If the camera is pointed directly towards the black hole, then 
a shadow is projected onto the screen of the camera, due to the light rays that are trapped by the gravitational pull of the black hole. This shadow is centered if the black hole is not rotating; however, it will be shifted to one side if the black hole is rotating, due to the frame-dragging effect.
Other objects of interest are the apparent images of accretion disks 
around black holes. In order for the Kerr space-times shown in Section \ref{sec:sol-ernst} to still model the space-time around these objects, we assume the disk to be static and non-gravitating, namely, the disk does not change 
over time and its mass is negligible.

\subsection{Photon spheres}
\label{sec:photon_sphere}

Let us recall that in the Schwarzschild space-time, there exists a region with spherical symmetry, known as the photon sphere, in which photons whose initial four-momenta have a vanishing radial component will orbit on the surface of this sphere. In fact, the orbits will be circular due to the spherical symmetry of the Schwarzschild metric. This surface is also called the ``last photon orbit" since it is the smallest stable orbit a photon can have around this black hole. Thus, a photon crossing this surface from the outside will necessarily fall into the horizon. Therefore, an observer looking at the direction of the black hole with a luminous background will see a circular ``shadow'' corresponding to the photons that cross the photon sphere and, therefore, do not reach the observer. This shadow is also observed in Kerr space-times, but in this case, there are several photon spheres with their radii depending on the four-momenta of the orbiting photon \cite{bardeen}. The consequence of this on the shadow is that it will no longer look circular from the perspective of a distant observer.

The first test aims to reproduce these shadows using the metric in the form \eqref{eq:wlp} with the solution to the Ernst equation \eqref{kerr5} giving the Kerr solution. We do this test for several values of the rotation parameter $\varphi$. For this we consider a luminous background and determine the pixels of the screen that are not reached by any light ray. The projected shadow will depend on the inclination of the camera with respect to the rotation axis. 

\begin{figure} [htb]
    \centering
    \small $\varphi=0$ \hspace{1.9cm} $\varphi=0.5$ \hspace{1.9cm} $\varphi=1.0$ \hspace{1.9cm} $\varphi=1.5$ \par \medskip
    \fbox{\includegraphics[width=0.21\textwidth]{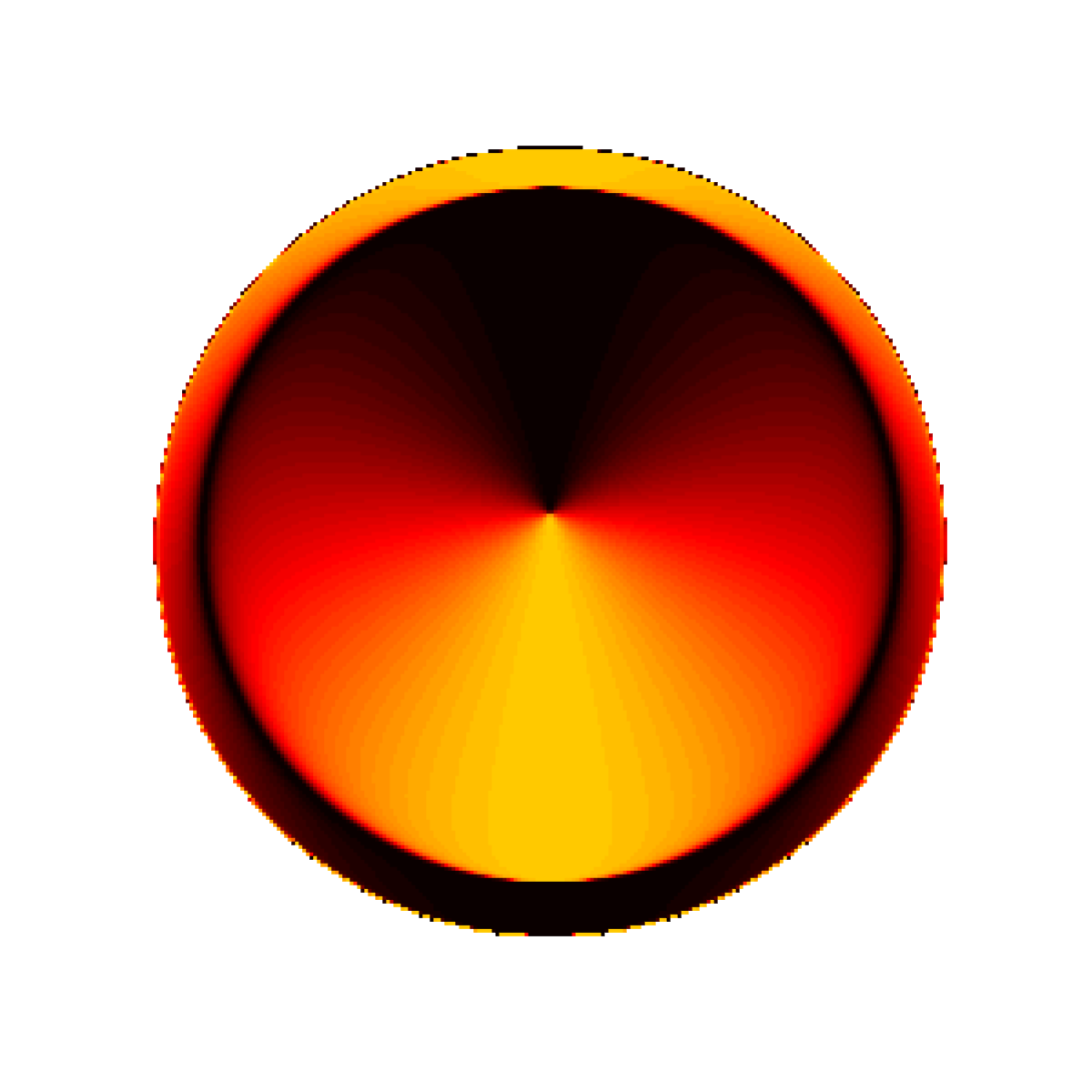} } 
    \fbox{\includegraphics[width=0.21\textwidth]{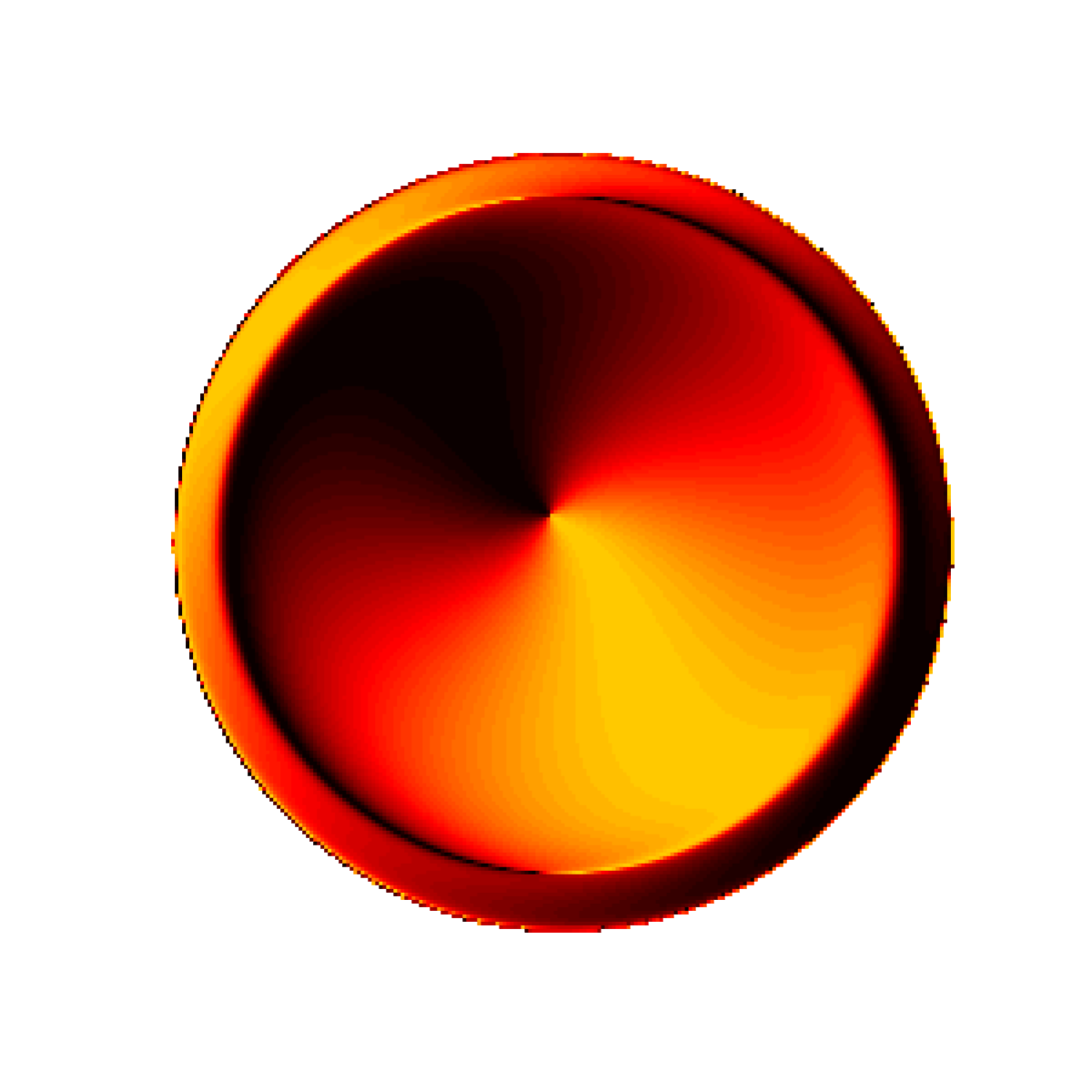} }
    \fbox{\includegraphics[width=0.21\textwidth]{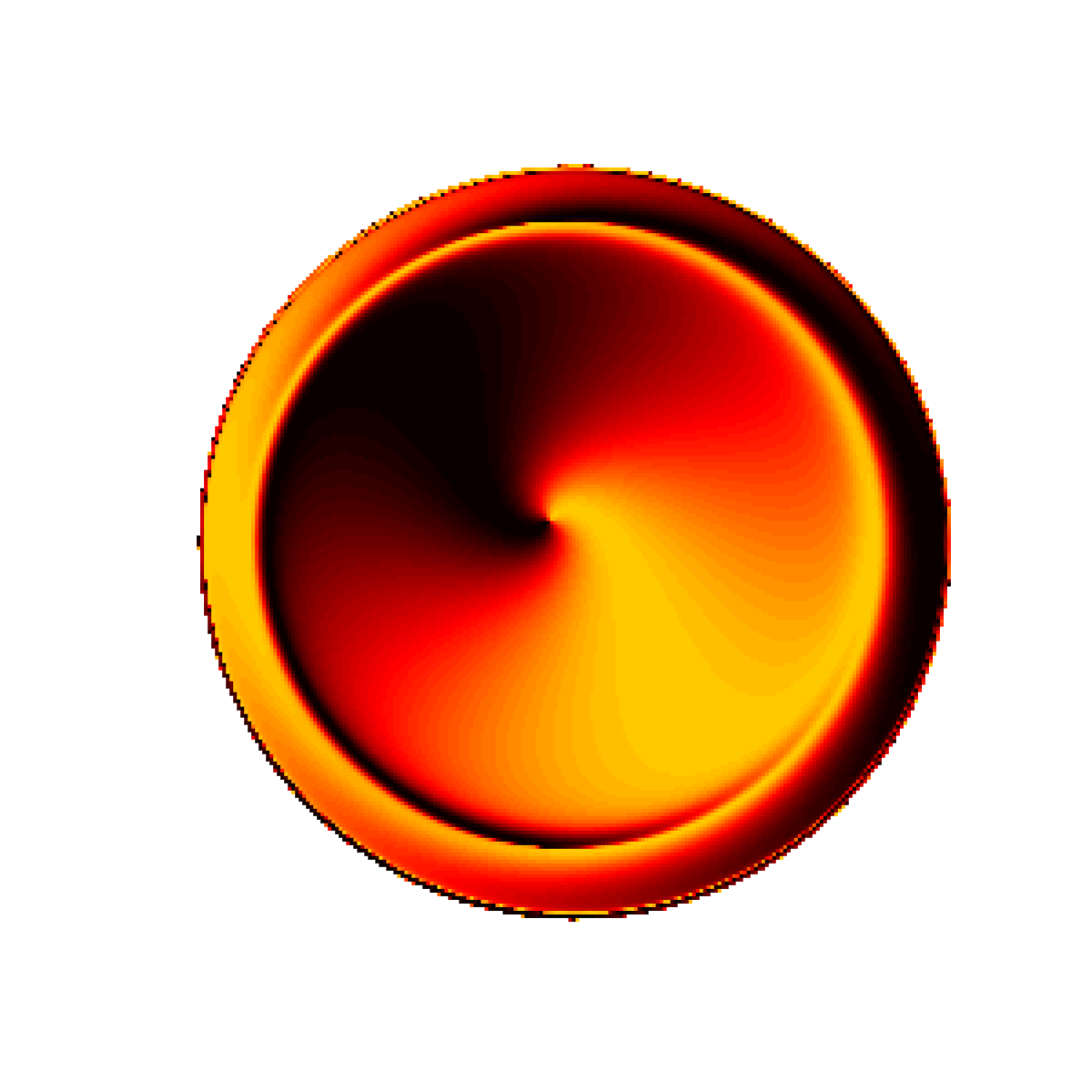} } 
    \fbox{\includegraphics[width=0.21\textwidth]{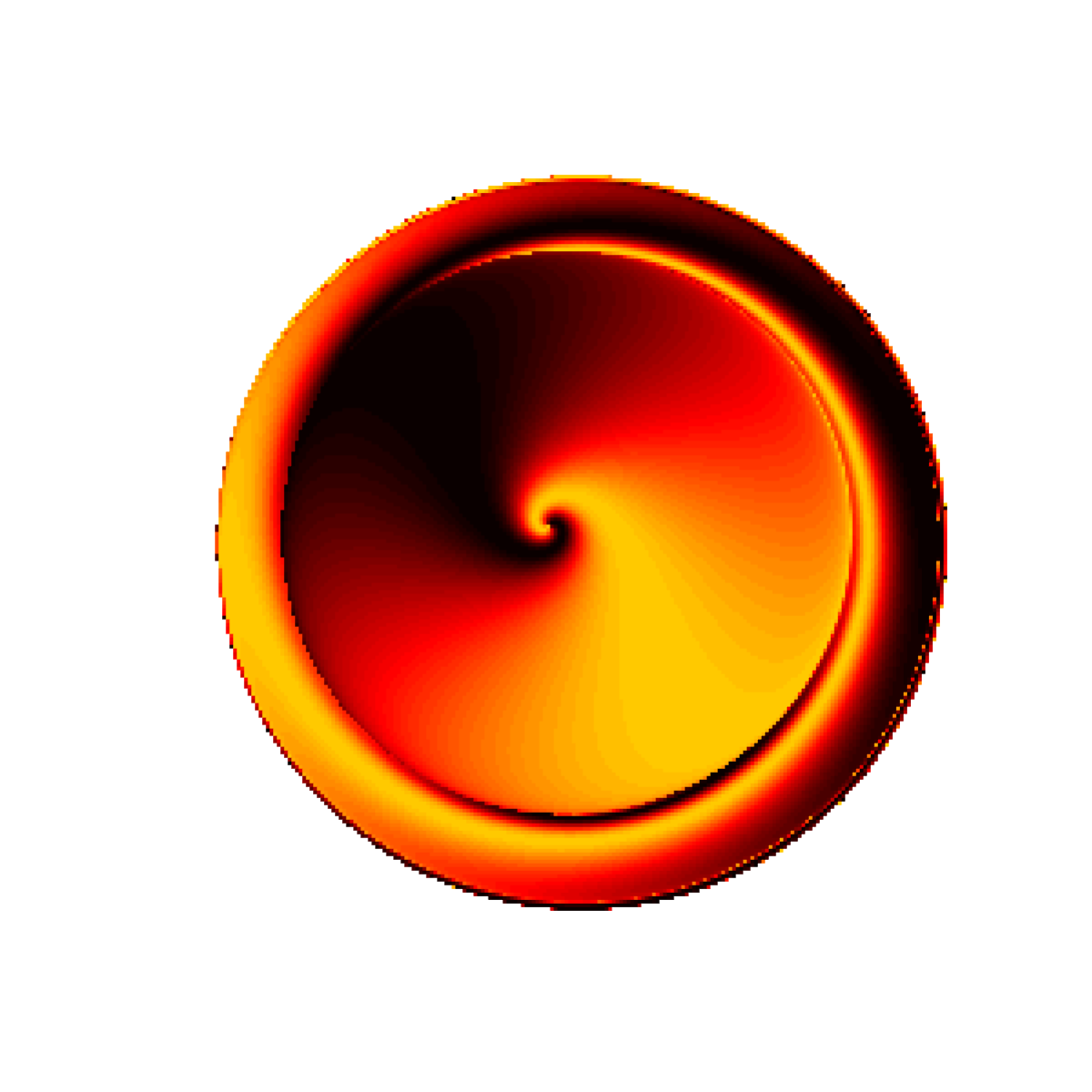} }
    \fbox{\includegraphics[width=0.21\textwidth]{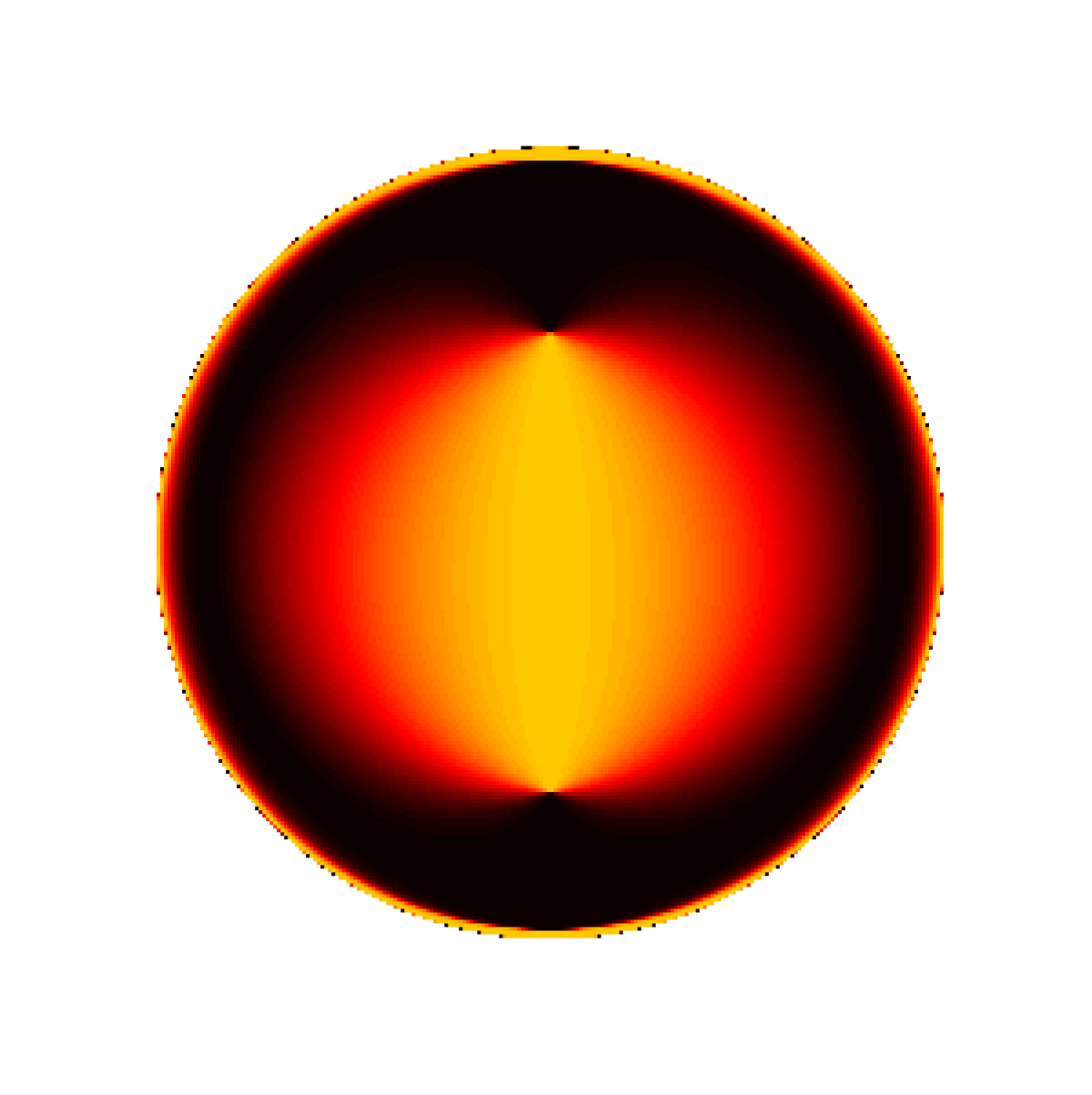} } 
    \fbox{\includegraphics[width=0.21\textwidth]{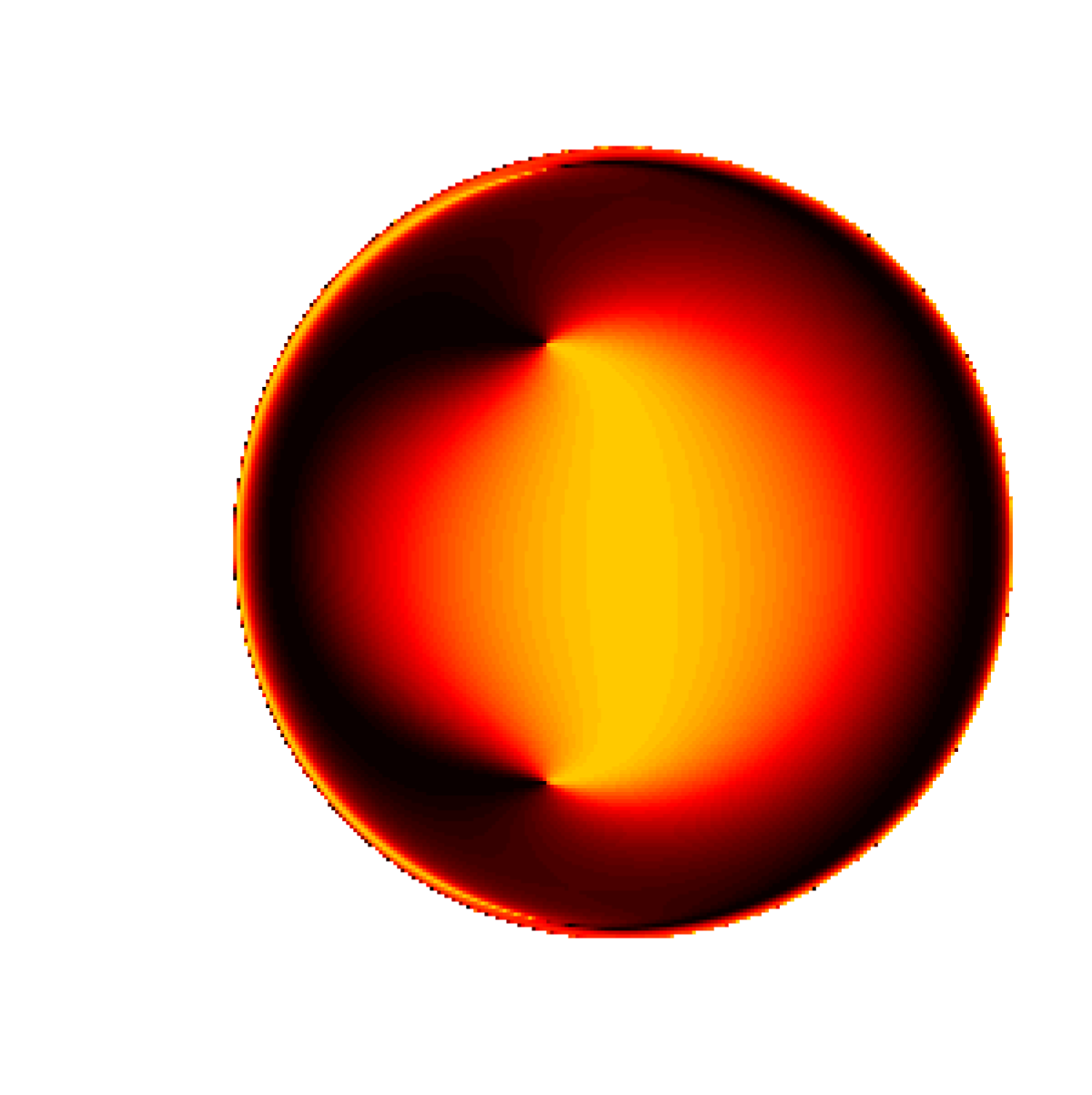} }
    \fbox{\includegraphics[width=0.21\textwidth]{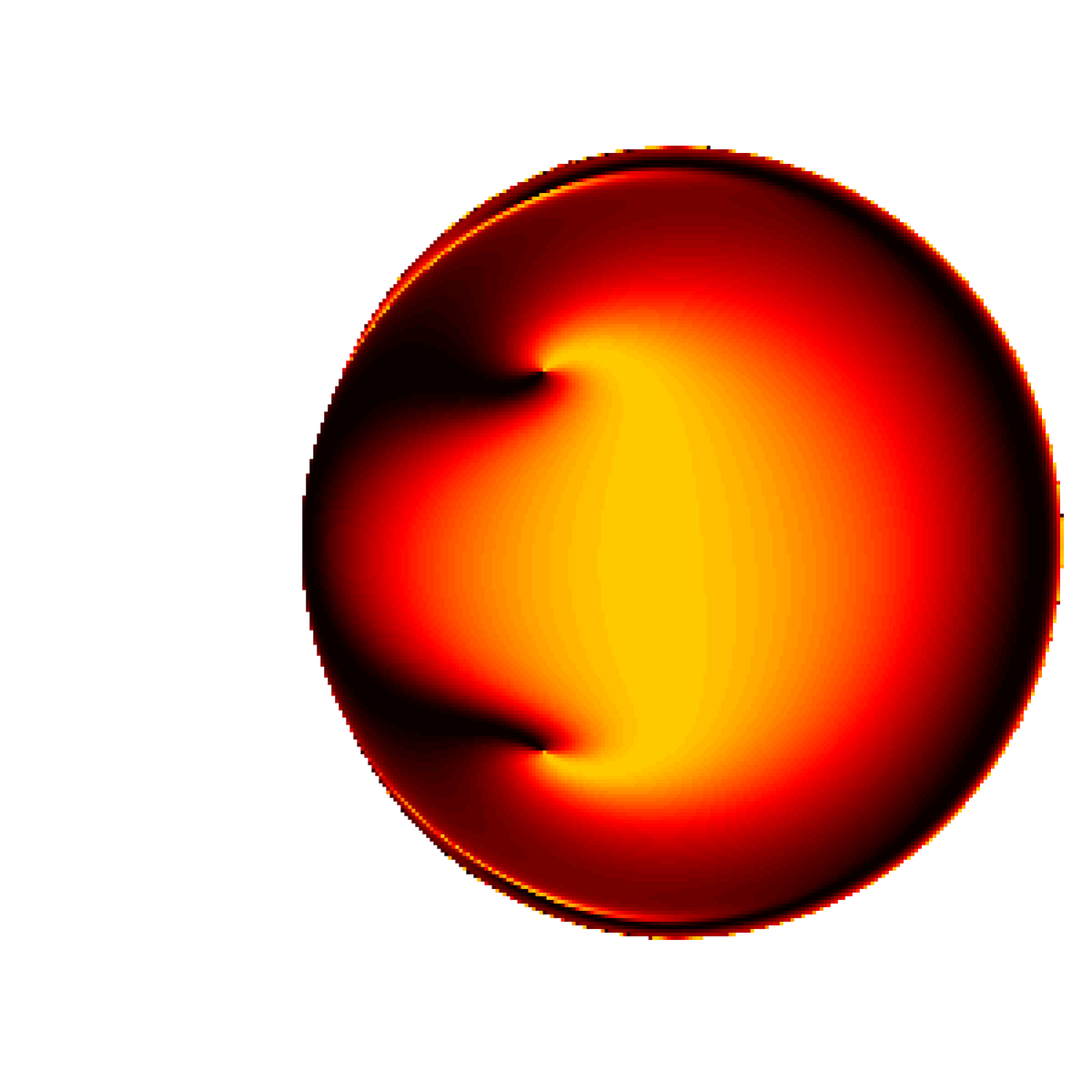} } 
    \fbox{\includegraphics[width=0.21\textwidth]{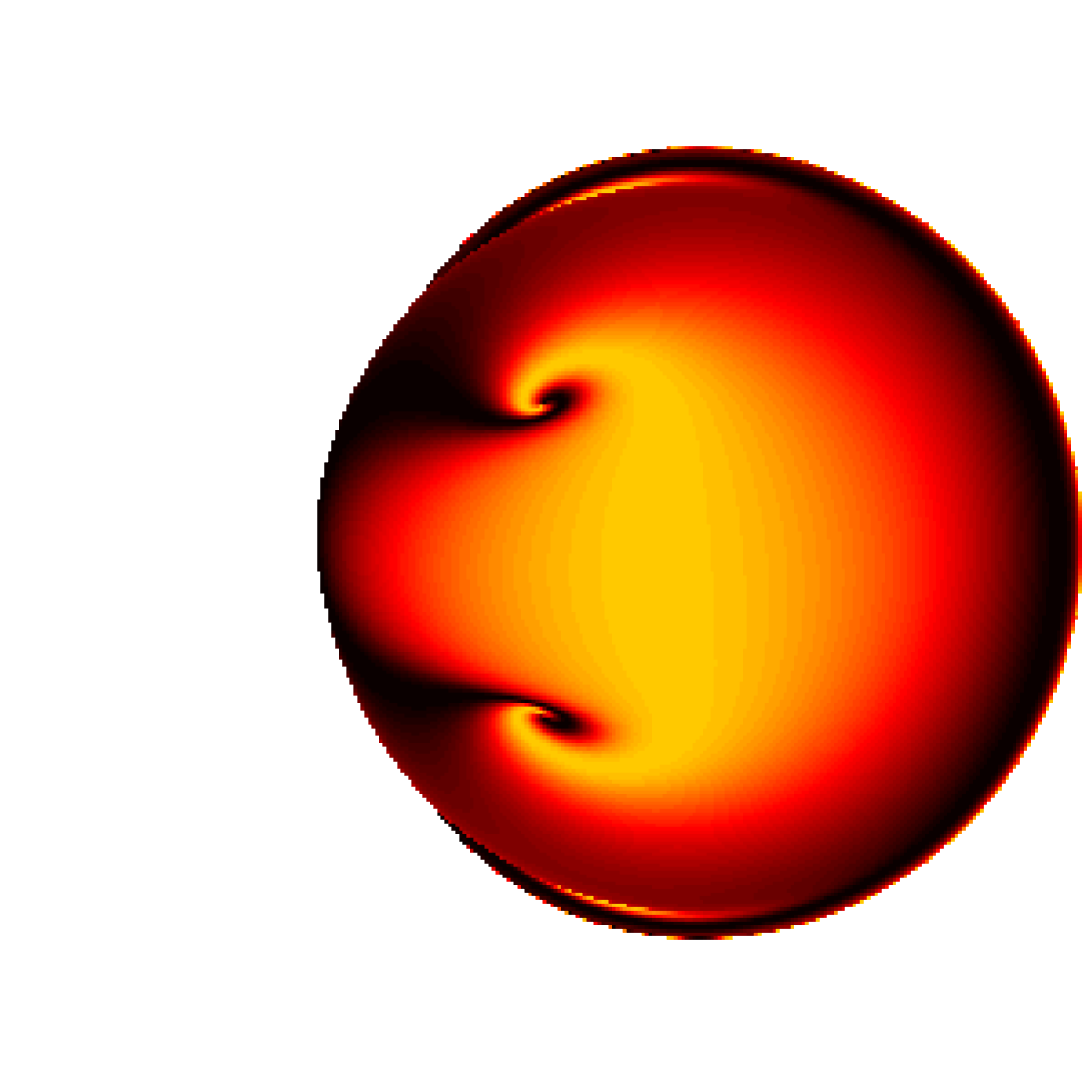} }
    
    \caption{Shadow of several Kerr black holes. From top to bottom, the angles of the camera with respect to the axis are $\alpha=10^\circ$ and $\alpha=80^\circ$.}  \label{fig:shadow_kerr}
\end{figure}

Fig.~\ref{fig:shadow_kerr} shows the shadow of these black holes with different rotation parameters $\varphi$ as perceived by a camera with the characteristics mentioned in Section \ref{sec:RT}, a focal length $f_L = 0.15$ and a resolution of $I_H\times I_V = 300\times 300$ pixels. In terms of the rotation parameter, $\varphi=0$ corresponds to the Schwarzschild black hole and $\varphi=\pi/2$ to the extreme Kerr black hole. The white part represents the luminous background, and the rest represents the shadow of the black hole. Additionally, in order to better visualise the effect of the rotation of the black hole in the vicinity, we add artificial coloring to the shadow in dependence of the coordinate angle $\phi$ in which the infalling photon approaches the ergosphere (recall that it takes an infinite coordinate time to reach the ergosphere; thus, we stop the computations when the photons get sufficiently close to this surface).

As observed in Fig.~\ref{fig:shadow_kerr}, the shadow of the black hole shifts to the left as the rotation parameter $\varphi$ increases, since the bending effect increases with the angular momentum of the black hole. This shift becomes more evident as the inclination $\alpha$ of the camera with respect to the $\zeta$-axis increases.

\subsection{Thin accretion disk}
One of the objects that has been widely studied in relativistic astrophysics is the accretion disk around a black hole. The first simulation of a thin accretion disk was done for a Schwarzschild black hole \cite{luminet,Muller:2012}, and then in 2019, the Event Horizon Collaboration (EHT) released the first actual picture of a black hole with an accretion disk \cite{EHT}. 

For these examples we consider an ideal thin disk around the black hole, namely, a non-gravitating disk. This means that the contribution of the disk to the stress-energy tensor is negligible; therefore, the solution \eqref{kerr5a} still holds.

\begin{figure} [htb]
    \centering
    \includegraphics[width=12cm,height=5cm]{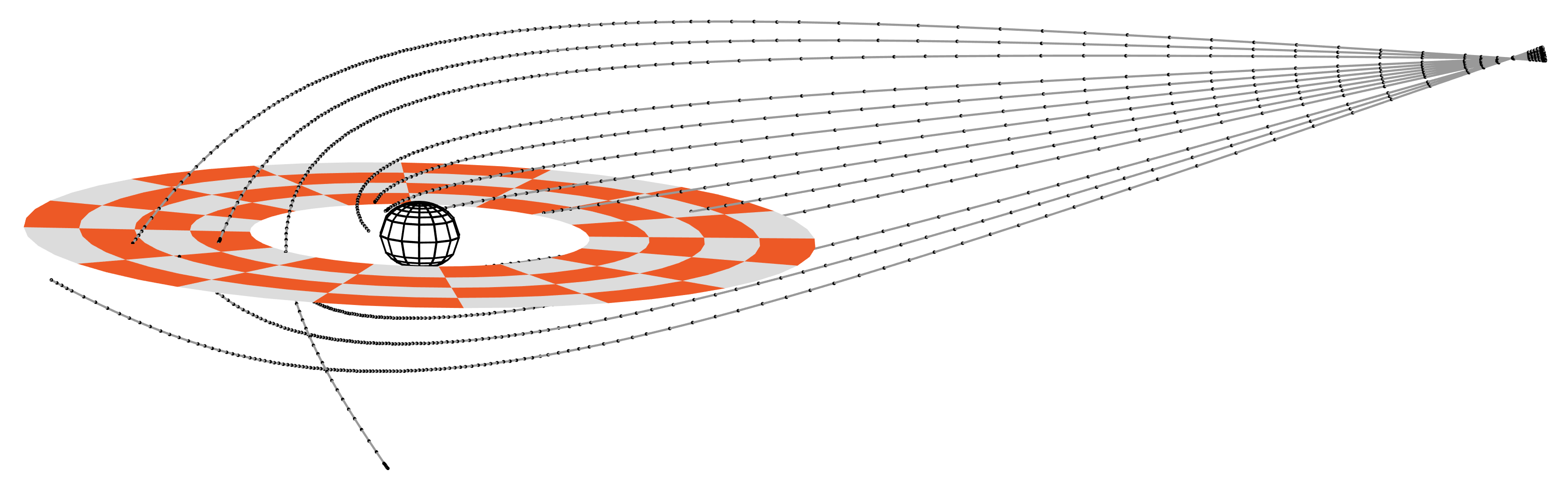}
    \caption{3D representation of a thin disk around a Kerr black hole with some light rays. The surface in the middle is the ergosphere.} \label{fig:Kerr-disk-3d}
\end{figure}

We simulate the apparent image of the thin disks in various Kerr space-times using a camera with the characteristics mentioned in Section \ref{sec:RT}, a focal length $f_L = 0.10$ and a resolution of $I_H\times I_V = 300\times 300$ pixels. Recall that a smaller focal length is necessary in order to have a greater angle of view. Fig.~\ref{fig:Kerr-disk-3d} shows the three-dimensional representation of the thin disk and some of the geodesics corresponding to the initial conditions \eqref{eq:IC}.

\begin{figure} [htb]
    \centering
   \small $\varphi=0.1$ \hspace{2.8cm} $\varphi=0.5$ \hspace{2.8cm} $\varphi=1.0$ \par 
    \includegraphics[width=0.3\textwidth]{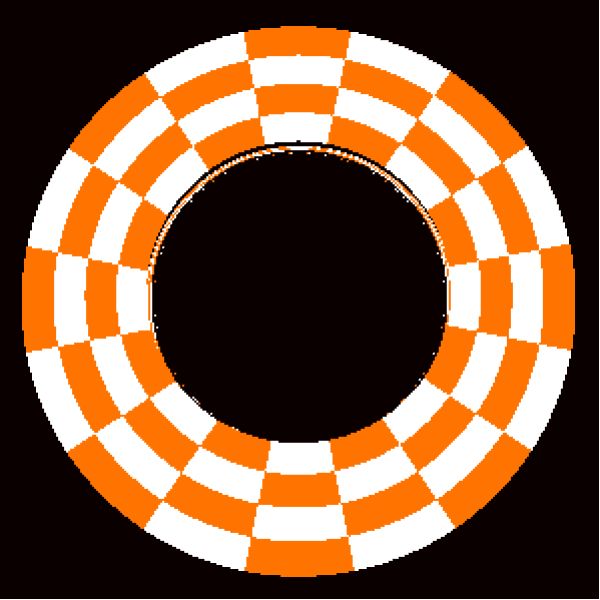} \includegraphics[width=0.3\textwidth]{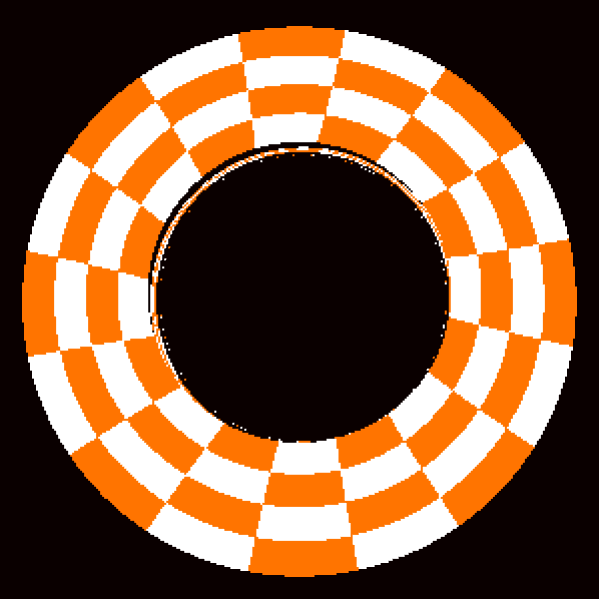} \includegraphics[width=0.3\textwidth]{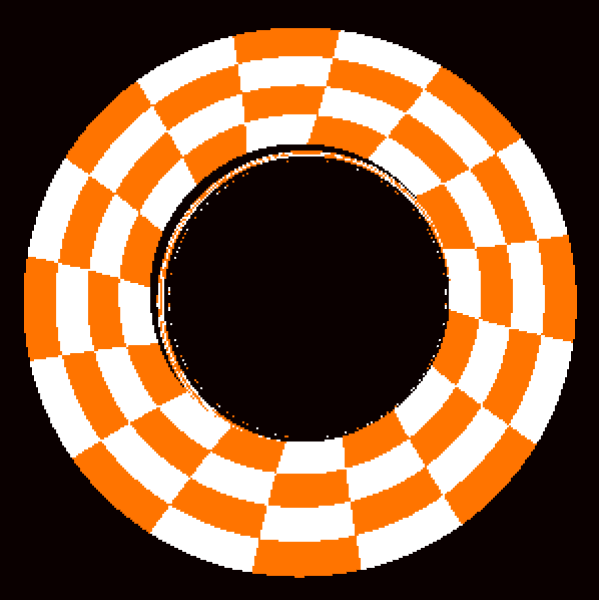} \par
    \includegraphics[width=0.3\textwidth]{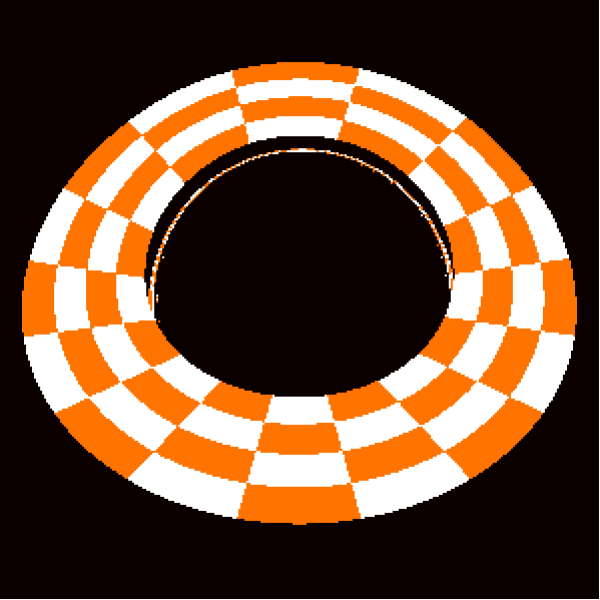} \includegraphics[width=0.3\textwidth]{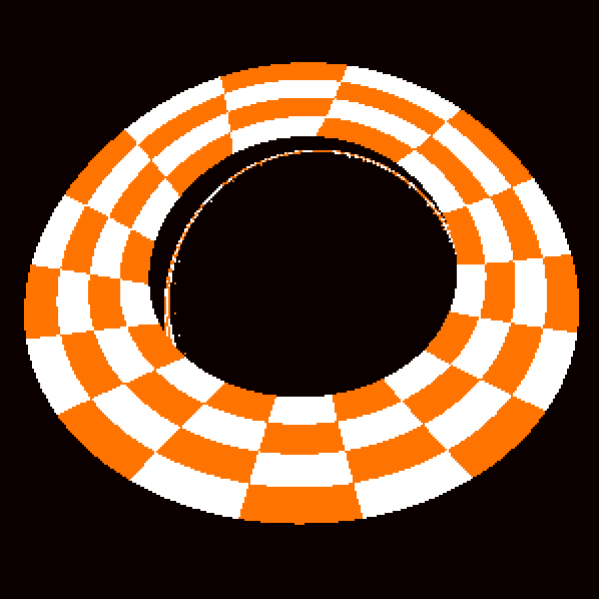} \includegraphics[width=0.3\textwidth]{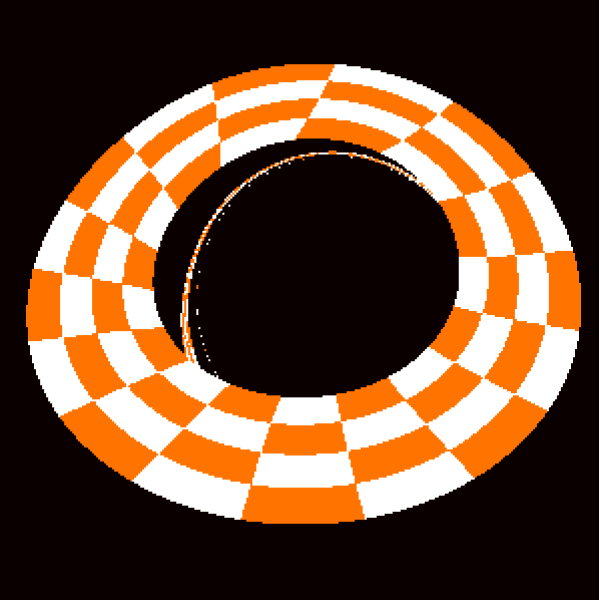} \par
    \includegraphics[width=0.3\textwidth]{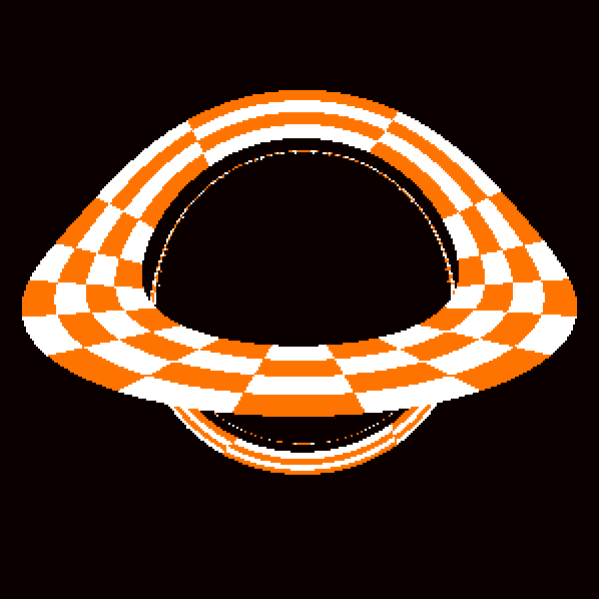} \includegraphics[width=0.3\textwidth]{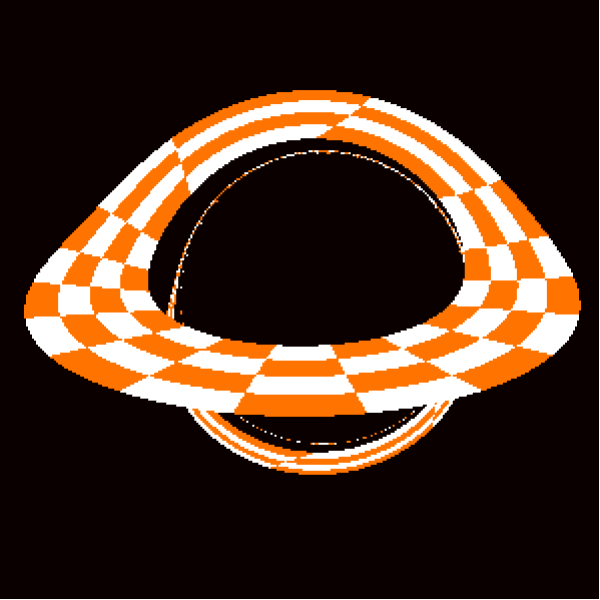} \includegraphics[width=0.3\textwidth]{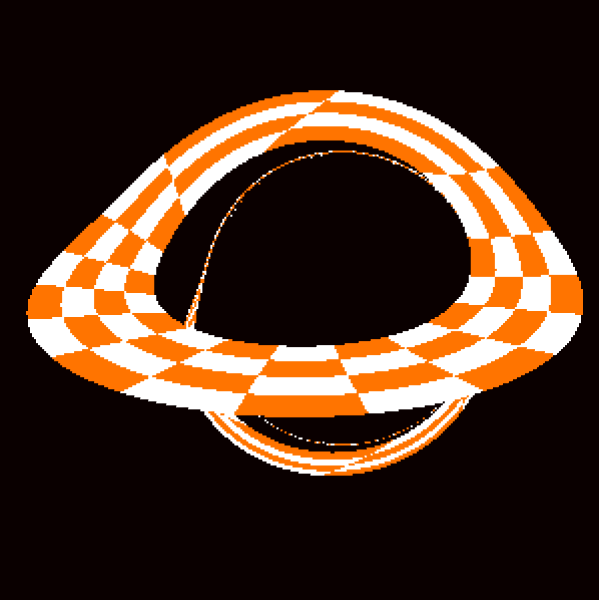} \par
    \includegraphics[width=0.3\textwidth]{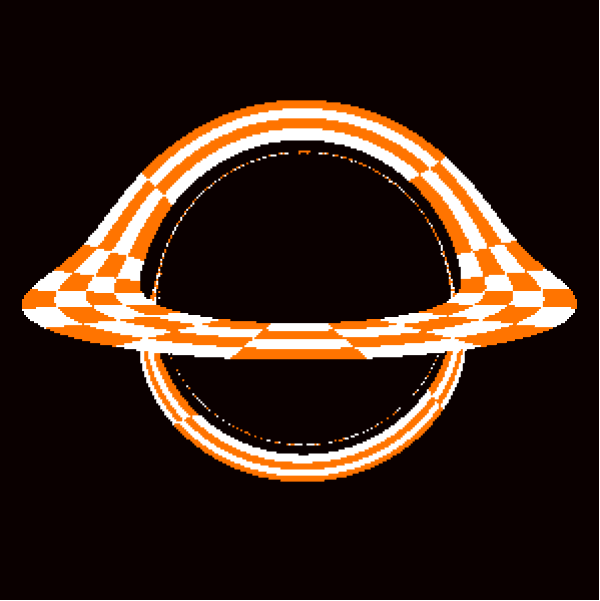} \includegraphics[width=0.3\textwidth]{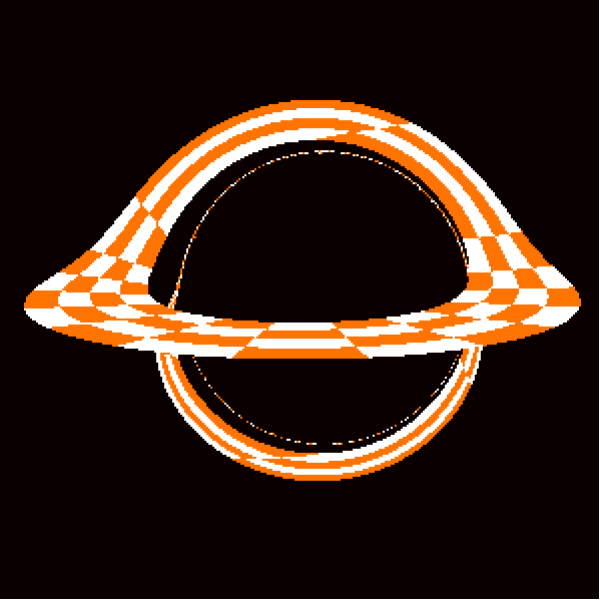} \includegraphics[width=0.3\textwidth]{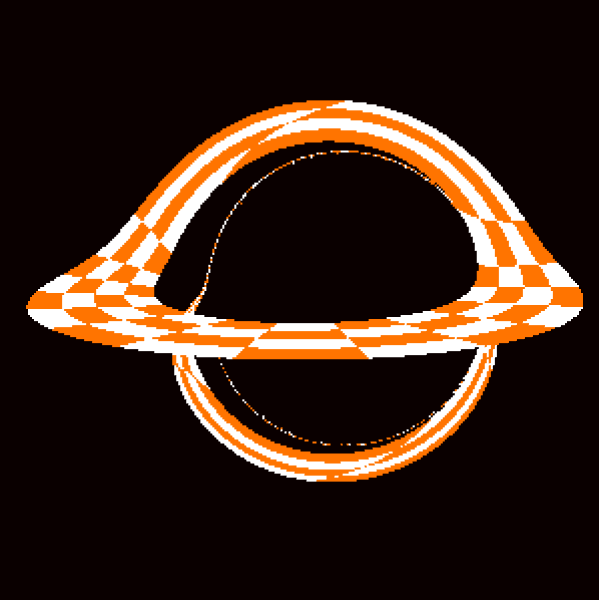}
 
    \caption{Ideal thin disk around Kerr black holes with various rotation parameters $\varphi$ and at different angles $\alpha$ with respect to the $z$-axis. From top to bottom, the inclinations of the camera are $\alpha=10^\circ$, $\alpha=45^\circ$, $\alpha=70^\circ$ and $\alpha=80^\circ$. }
    \label{fig:Kerr-disks}
\end{figure}

The distortion of the disk image is minimal when the camera is located with a small inclination $\alpha$ with respect to the axis. However, this distortion becomes more apparent as $\alpha$ increases. Fig.~\ref{fig:Kerr-disks} shows how the distortion of the image increases as the camera inclination gets close to $\alpha=90^\circ$. An interesting feature is that objects behind the shadow can be seen due to the bending of light in the vicinity of the black hole. In fact, some objects are visualised multiple times, as one can observe in Fig.~\ref{fig:Kerr-disks}. Due to the bending effect, the light emitted by the disk in the downward direction is also received by the camera, as seen in Fig.~\ref{fig:Kerr-disk-3d}.

The following phenomena are observed in the simulated images: the shadow of the black hole, the photon ring and the visibility of the lower face of the disk. The shadow is due to the photons crossing the photon spheres, which means that they fall into the black hole horizon and, therefore, do not reach the observer, as mentioned in the previous subsection. 

Photon rings are due to some photons orbiting the photon spheres one or several times until they reach the observer. A photon that passes close to a photon sphere with tangential velocity can orbit this surface an arbitrary number of times (depending on how close it is to the surface) and then escape to infinity. What we observe in the figure is the first photon ring, namely, the photons emitted by the disk that made one orbit around the photon spheres before they reached the camera. However, increasing the resolution near the ring would show the second and other subsequent rings.

The lower face of the disk is observed due to the bending of light, 
which is attracted towards the black hole. As shown in Fig.~\ref{fig:Kerr-disk-3d},  these effects can be seen in the study of individual photons for counter-rotating disk space-times.

This example serves both to show that the solution to the Ernst equation \eqref{kerr5a} reproduces the same properties of the Kerr solution in the Boyer-Lindquist coordinates and to show that the computation of the functions for the geodesic equations \eqref{ode_geodesics} can be done efficiently via barycentric interpolation even if the functions are elementary as in the case of Kerr space-times. 

In this particular example, the Kerr space-time, we can check the 
accuracy of these simulations in an explicit manner. This is done by 
comparing the outcomes of the simulations using barycentric 
interpolation and the simulations using the explicit values of the metric functions and their derivatives. These simulations coincided perfectly for the examples shown in Fig.~\ref{fig:Kerr-disks}. In both cases, we used the same inputs for the virtual camera. The only difference is in the use of barycentric interpolation. This comparison cannot be made for more general space-times, such as the algebro-geometric solutions to the Ernst equations. However, one can estimate the accuracy of these simulations by plotting the Lagrangian at the point where the iteration of the light ray corresponding to each pixel is terminated.


\subsubsection{Total errors in the simulation} \label{subsec:total_errors}
\begin{figure} [htb]
 \includegraphics[width=0.45\textwidth]{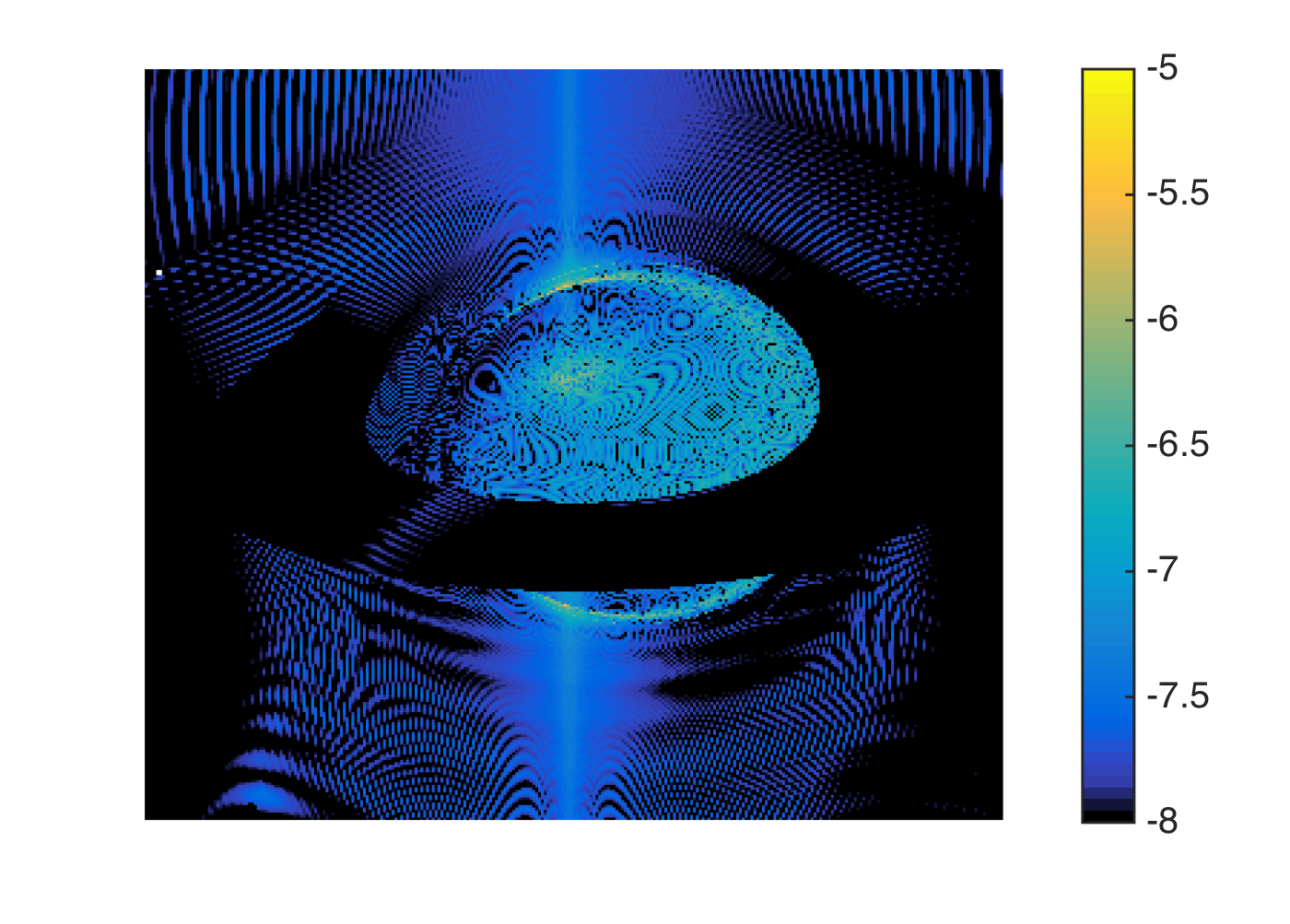}
  \includegraphics[width=0.45\textwidth]{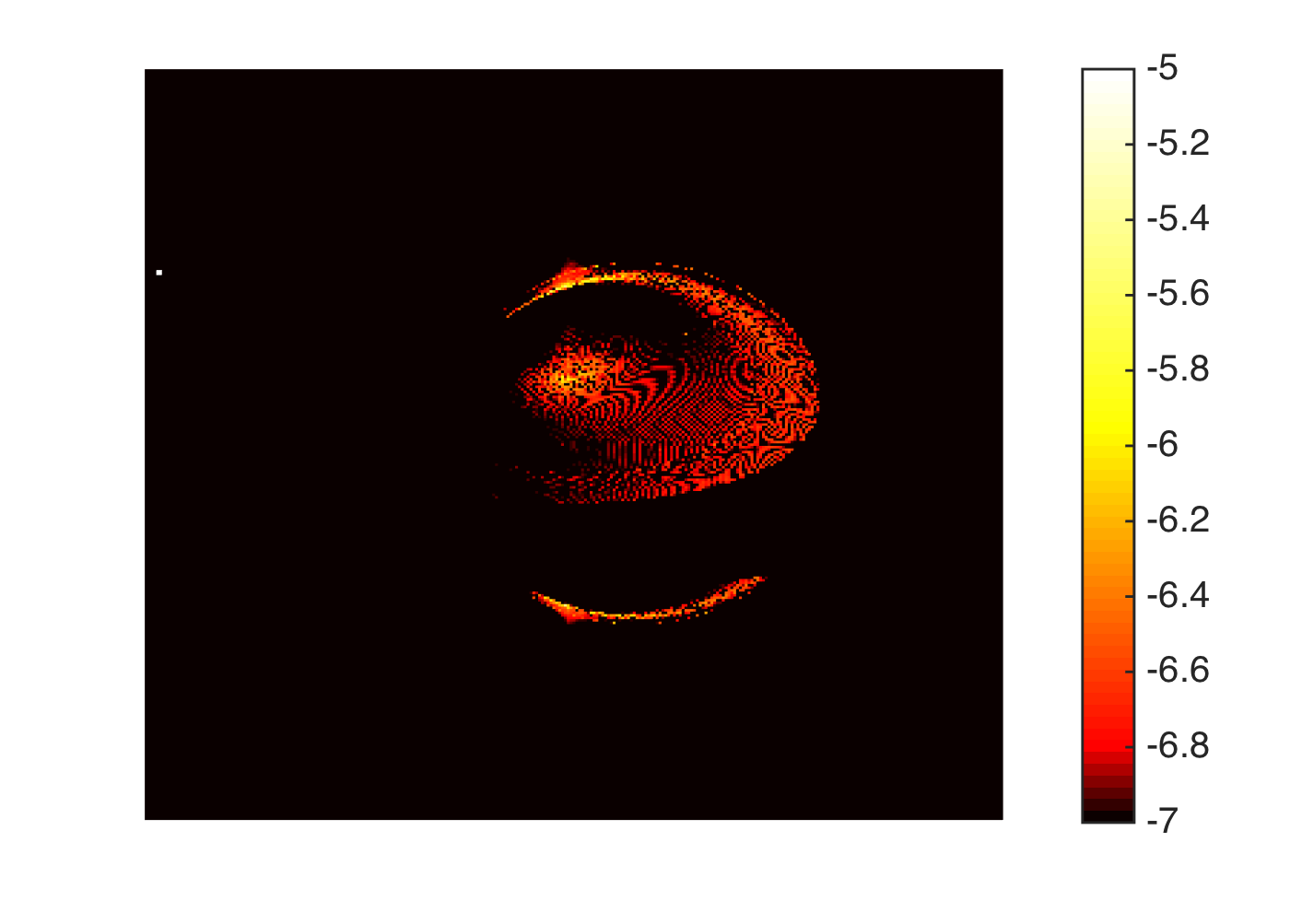}
\caption{Logarithmic plot of the Lagrangian $|\Lag(\vx^{(N)})|$ at the stopping point of each pixel. The left-hand side shows a color plot of the pixels for which $\log_{10} |\Lag(\vx^{(N)}) | \geq -8$ and the right-hand side shows the plot for which $\log_{10} |\Lag(\vx^{(N)}) | \geq -7$.}
    \label{fig:lagrangian_kerr}
\end{figure}

In general, the errors arising from the different numerical methods used in the image simulations can be tracked by computing the Lagrangian $\Lag(\vx^{(N)})$ of each pixel, where $\vx^{(N)}$ is the point where the iteration is terminated. The value of this Lagrangian should be zero, but since we are computing these null geodesics numerically, we expect it to fall under certain threshold $\varepsilon$, i.e., we impose the condition $\Lag(\vx^{(N)})\leq \varepsilon$. In such a case, we say that the geodesic has precision $\varepsilon$. Fig.~\ref{fig:lagrangian_kerr} shows the logarithmic color plot of the Lagrangian in dependence of the pixels of the simulated image in a Kerr space-time corresponding to a rotation parameter $\varphi=1$ and a virtual camera with inclination $\alpha=70^\circ$. For ease of reference, we only color the pixels whose Lagrangians are greater than $10^{-8}$ (respectively $10^{-7}$). Thus, the pixels in black are those whose geodesics have precision $10^{-8}$ (respectively $10^{-7}$). For the examples discussed here, we used the tolerances $\texttt{RelTol}=\texttt{AbsTol}=10^{-8}$ for the \texttt{ode45} routine. 
As observed in Fig.~\ref{fig:lagrangian_kerr}, the precision of the pixels is at most one order the magnitude greater than such the chosen tolerances for \texttt{ode45}, as long as the light ray does not approach the ergosphere. Otherwise, $ |\Lag(\vx^{(N)}) |$ could deviate from zero even more. However, since we are interested in visualising objects outside the ergosphere, the tolerances \texttt{RelTol} and \texttt{AbsTol} can be chosen to be one order of magnitude smaller than the prescribed precision $\varepsilon$ for $ |\Lag(\vx^{(N)}) |$.

\section{Ray tracing for the counter-rotating dust disk} \label{sec:example-cr-disk}

We now show the image simulation in the counter-rotating dust disk 
space-times presented in Section \ref{sec:cr-disk}. The metric functions are computed on Chebyshev grids with the method detailed in Section \ref{sec:cr-disk}, and then we use barycentric interpolation to compute the metric functions at any other point $(\rho,\zeta)$ in the domain covered by the grids. We perform the simulations to visualise images in the vicinity of the counter-rotating disks using the ray tracing technique in the same way we did for the Kerr solution. One fundamental difference with respect to the non-gravitating disk mentioned above is that a portion of the counter-rotating disk will be inside the ergoregion, and therefore, light emitted from this part of the disk would take an infinite coordinate time to cross the surface. The implication for the apparent image seen by a distant observer is that one portion of the disk will be hidden by a shadow. However, the visible part can appear multiple times in the image simulation.

Prior to any visualisation of any extended body in counter-rotating disk space-times, we first analyse the motion of individual photons for various parameters $\delta$ and $\lambda$, which parametrise the family of solutions to the Ernst equation. 

\subsection{Trajectories}
We simulate several individual trajectories for specific space-times 
with different parameters $\delta$ and $\lambda$. We are mainly 
interested in studying the differences between space-times whose 
ergospheres cover the disk almost entirely (except in the middle of 
the doughnut-like surface) and those that are smaller than the disk, 
as seen in Fig.~\ref{fig:3D_disk}. The size of the ergosphere and the 
value of the disk's angular momentum depend on $\lambda$ and 
$\delta$, see \cite{FK3}. We have listed the physical parameters for 
the examples below in Table \ref{table}, the central redshift 
$z_{R}$ of photons emitted from the center of the disk and detected 
at infinity, the relative density $\gamma$, the mass $M$ and the 
angular momentum $J$. 
\begin{table}[tbp]
	\centering
\begin{tabular}{|c|c|c|c|c|c|}
	\hline
	$\lambda$ & $\delta$ & $z_{R}$ & $\gamma$ & $M $ & $J$  \\
	\hline
	10.12 & 1.2 & 4.92 & 0.91 & 0.64 & 0.4 \\
	\hline
	10.12 & 0.865 & 5.64 & 0.95 & 0.81 & 0.66 \\
	\hline
	10.12 & 0.7 & 6.26 & 0.97 & 0.95 & 0.92 \\
	\hline
\end{tabular}
		\caption{Parameters for the examples studied below}
	\label{table}
\end{table}

We want to simulate initially-parallel light rays and then observe the bending of the trajectories. To observe this effect, we choose initial conditions of the form:
\begin{itemize}
\item[(i)] Photons in the $xy$-plane directed in the negative $x$-direction for the same $x_0$ and for several equispaced values for $y_0$. The initial conditions for these light rays have the form
  \begin{equation*}
    \left\{
      \begin{array}{llll}
        t_0 = 0, & \rho_0=\sqrt{x_0^2+y_0^2}, & \zeta_0 = 0, & \phi_0=\arctan{(y_0/x_0)},\\
        p^t_0 , & p^\rho_0 = - \frac{x_0}{\rho_0} , & p^\zeta_0 =0, &  p^\phi_0 = \frac{y_0}{\rho_0^2},
      \end{array}
    \right.
  \end{equation*}
\item[(ii)] Photons on the $xz$-plane directed in the negative $x$-direction for the same $x_0$ and several equispaced values for $z_0$, i.e.,
  \begin{equation*}
    \left\{
      \begin{array}{llll}
        t_0 = 0, & \rho_0=x_0, & \zeta_0 = z_0, & \phi_0=0,\\
        p^t_0 , & p^\rho_0 = -1, & p^\zeta_0 =0, &  p^\phi_0=0,
      \end{array}
    \right.
  \end{equation*}
\item[(iii)] Photons on the $xy$-plane in the negative $y$-direction for the same $y_0=0$ and several values for $x_0$, i.e.,
  \begin{equation*}
    \left\{
      \begin{array}{llll}
        t_0 = 0, & \rho_0=x_0, & \zeta_0 = 0, & \phi_0=0,\\
        p^t_0 , & p^\rho_0 = 0, & p^\zeta_0 =0, &  p^\phi_0 = -\frac{1}{\rho_0},
      \end{array}
    \right.
  \end{equation*}
\item[(iv)] Photons on the $xz$-plane in the positive $z$-direction for the same $z_0=0$ and several values for $x_0$, i.e.,
  \begin{equation*}
    \left\{
      \begin{array}{llll}
        t_0 = 0, & \rho_0=x_0, & \zeta_0 = 0, & \phi_0=0,\\
        p^t_0 , & p^\rho_0 = 0, & p^\zeta_0 = 1, &  p^\phi_0=0,
      \end{array}
    \right.
  \end{equation*}
\end{itemize}
In every case we compute the light rays in the future direction, thus $p_0^t$ is chosen as the positive solution of the quadratic equation $g_{\mu\nu}p_0^\mu p_0^\nu=0$. 
Recall that in ray tracing we choose the negative solution $p_0^t$ 
since we compute the light rays backwards in time, and this is 
equivalent to choosing the positive $p^t_0$ in a space-time rotating in the opposite sense. 

We expect analogous behaviors to those observed in Kerr space-times: (i) should indicate the dragging direction of the space-time, (ii) should attract the particles towards the centre of the disk (origin of the coordinate system).
On the other hand, (iii) and (iv) will indicate whether the photons will fall towards the disk (or its ergosphere) or whether they will escape to infinity, depending on the initial conditions. 
Due to the conservation of angular momentum, the photons starting on the $xy$-plane remain on the plane, implying that the photon in (iii) that orbits the photon sphere should trace a circular trajectory, as opposed to (iv) which might have a more complicated trajectory. However, we expect to find differences as well.

The following are the light rays corresponding to conditions (i) and (ii). We set the positions with $x_0=10$ for all the photons and different equally spaced values $y_0\in [-3,3]$ and $z_0\in [-3,3]$.

\begin{figure} [htb]
    \centering
    \includegraphics[width=0.45\textwidth]{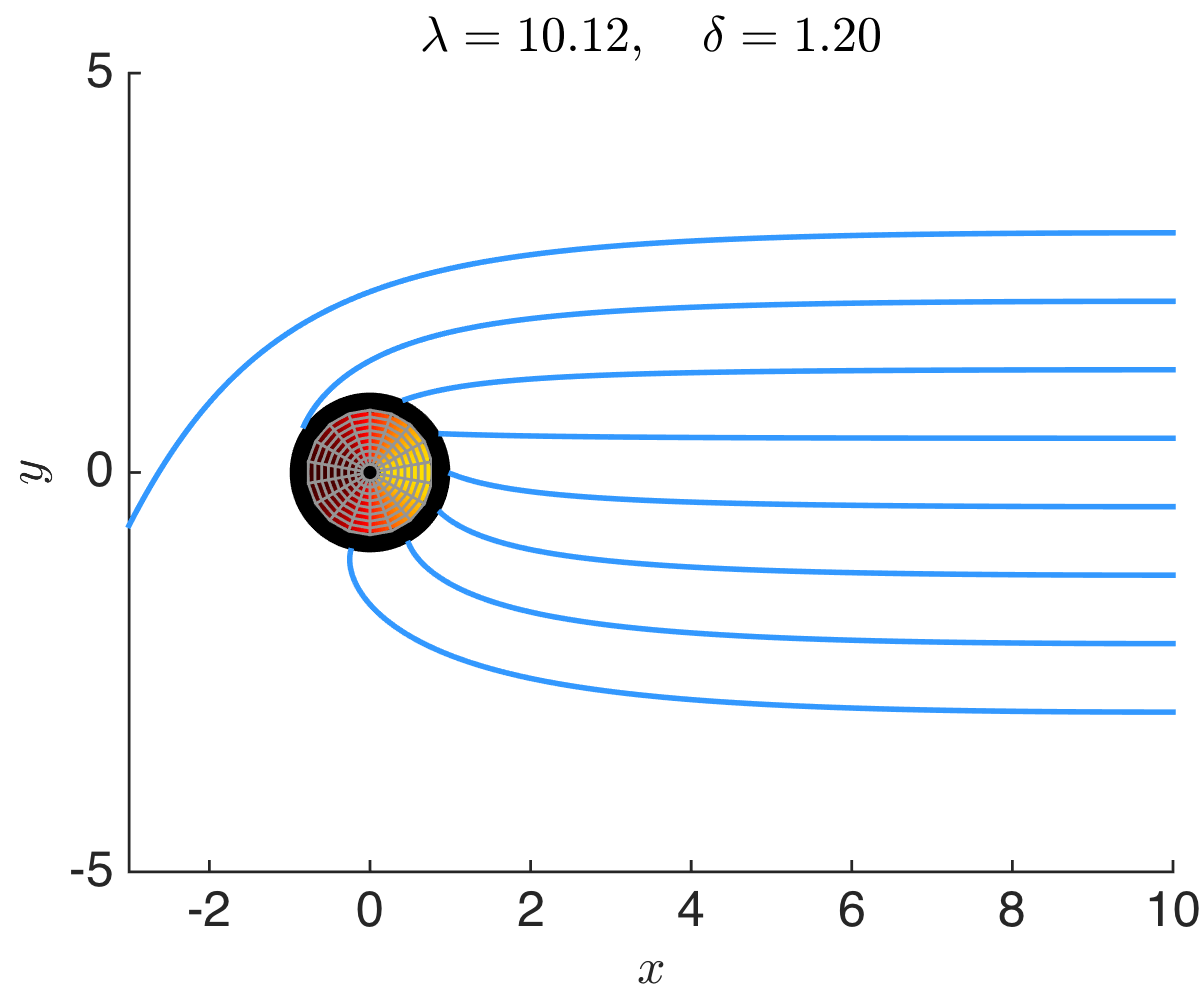} 
    \includegraphics[width=0.45\textwidth]{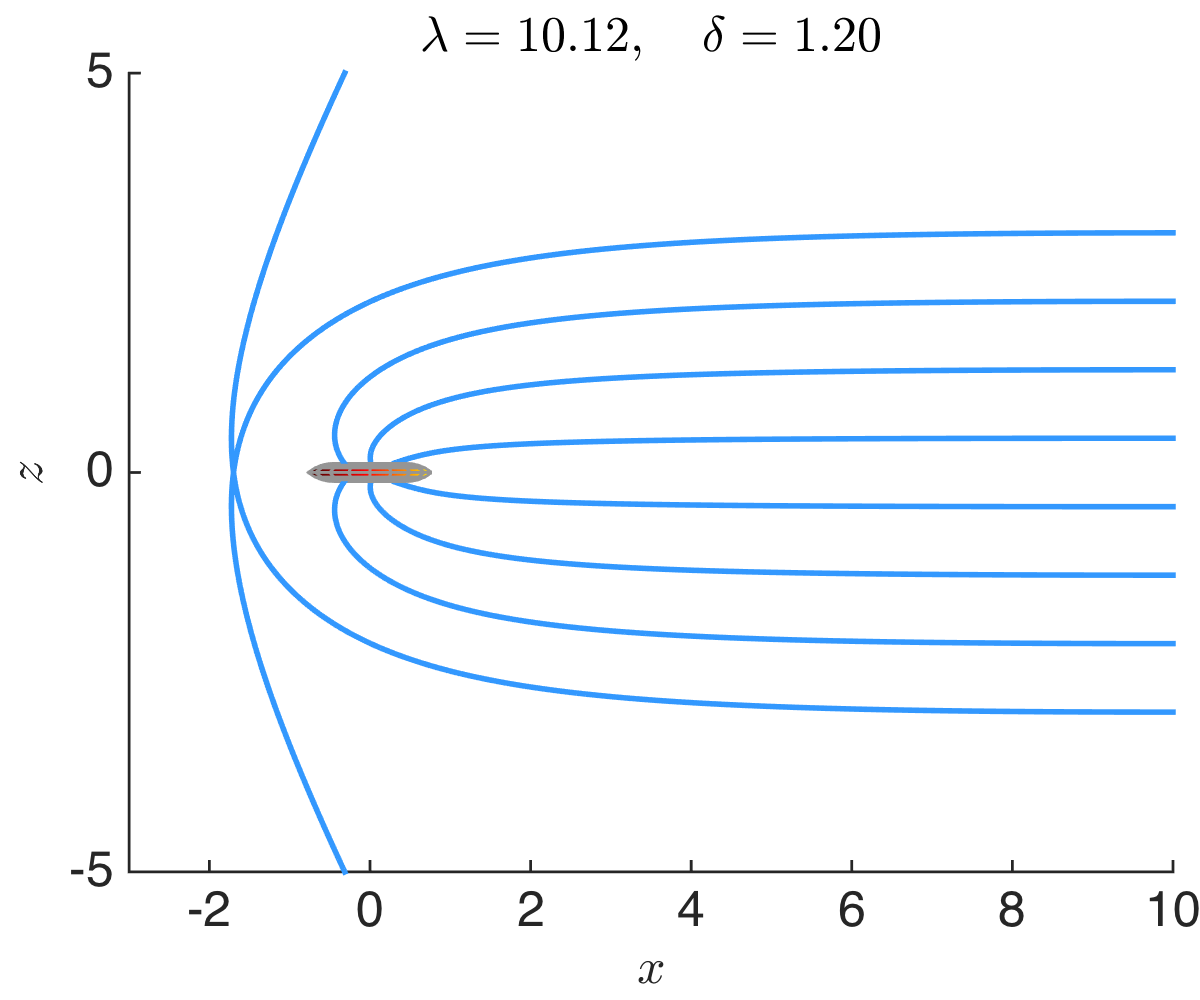} 
    \includegraphics[width=0.45\textwidth]{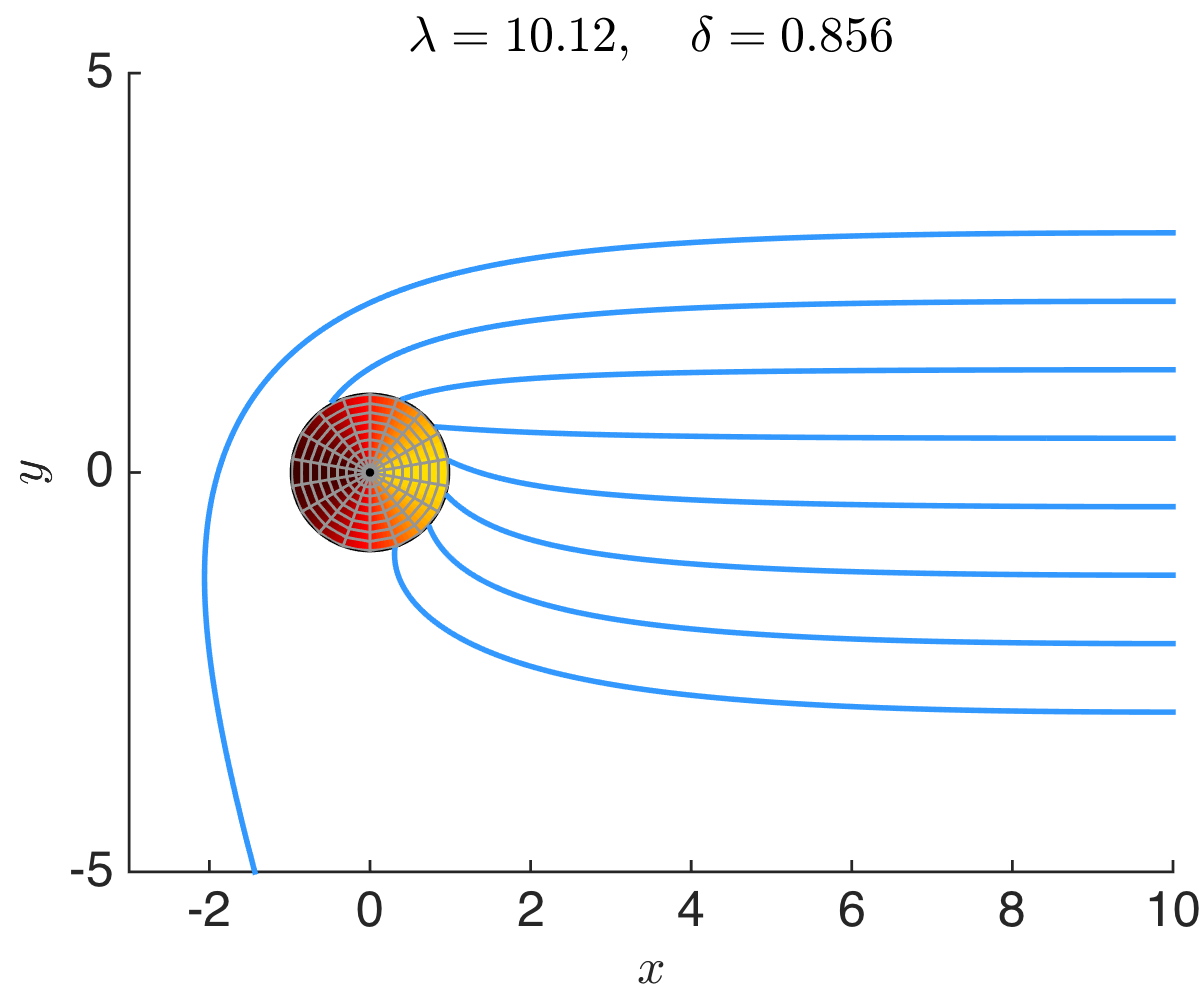} 
    \includegraphics[width=0.45\textwidth]{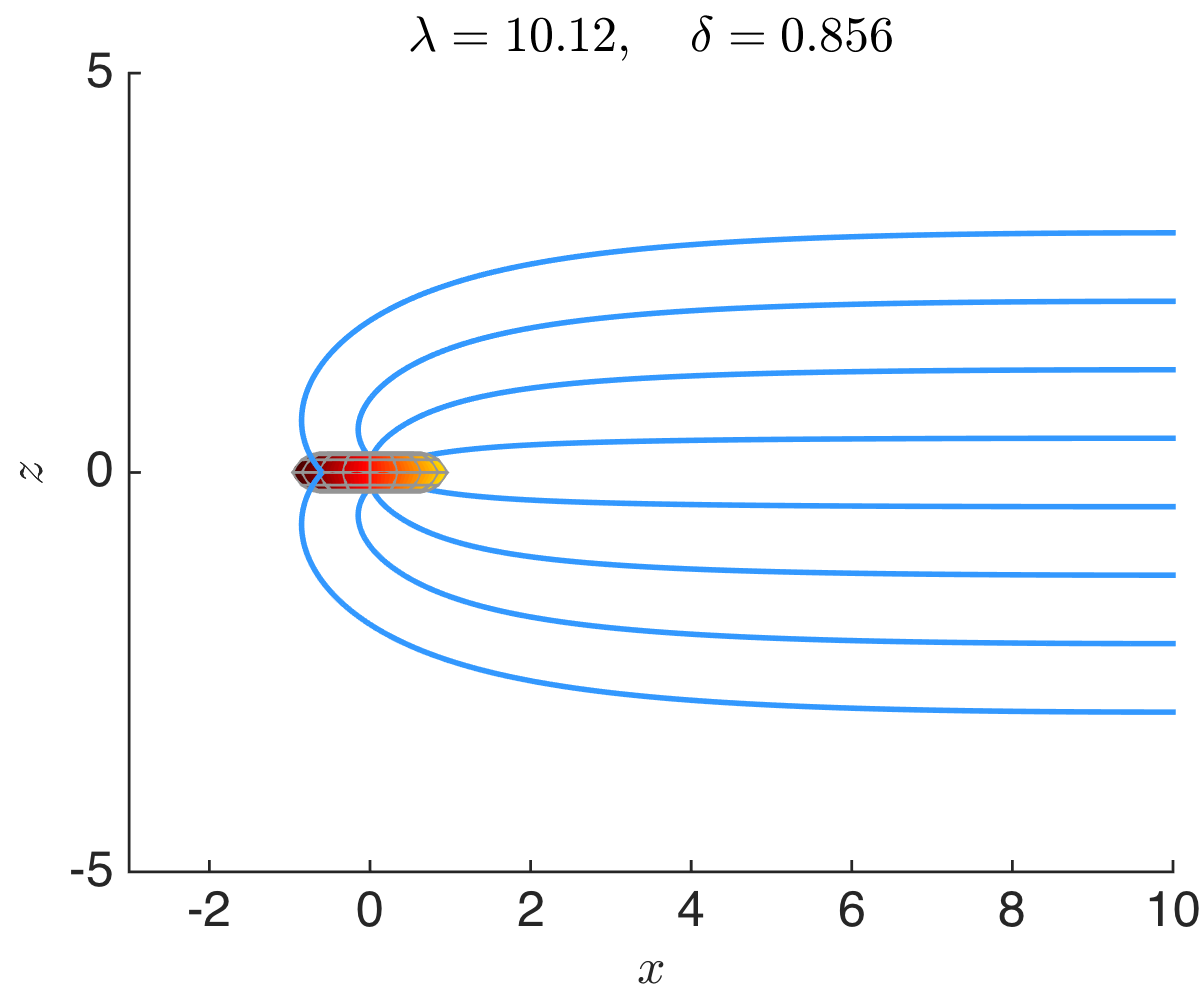} 
    \includegraphics[width=0.45\textwidth]{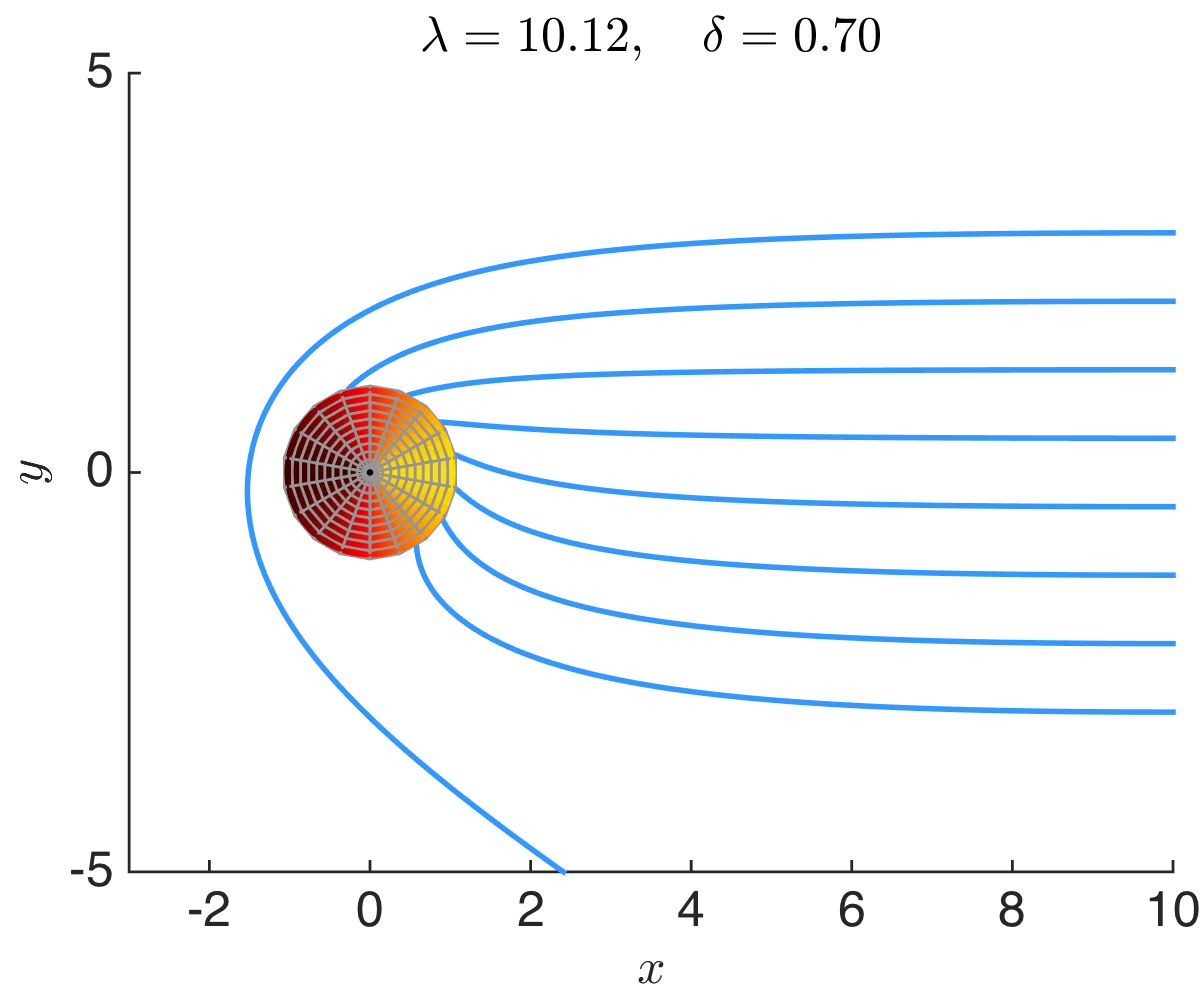} 
    \includegraphics[width=0.45\textwidth]{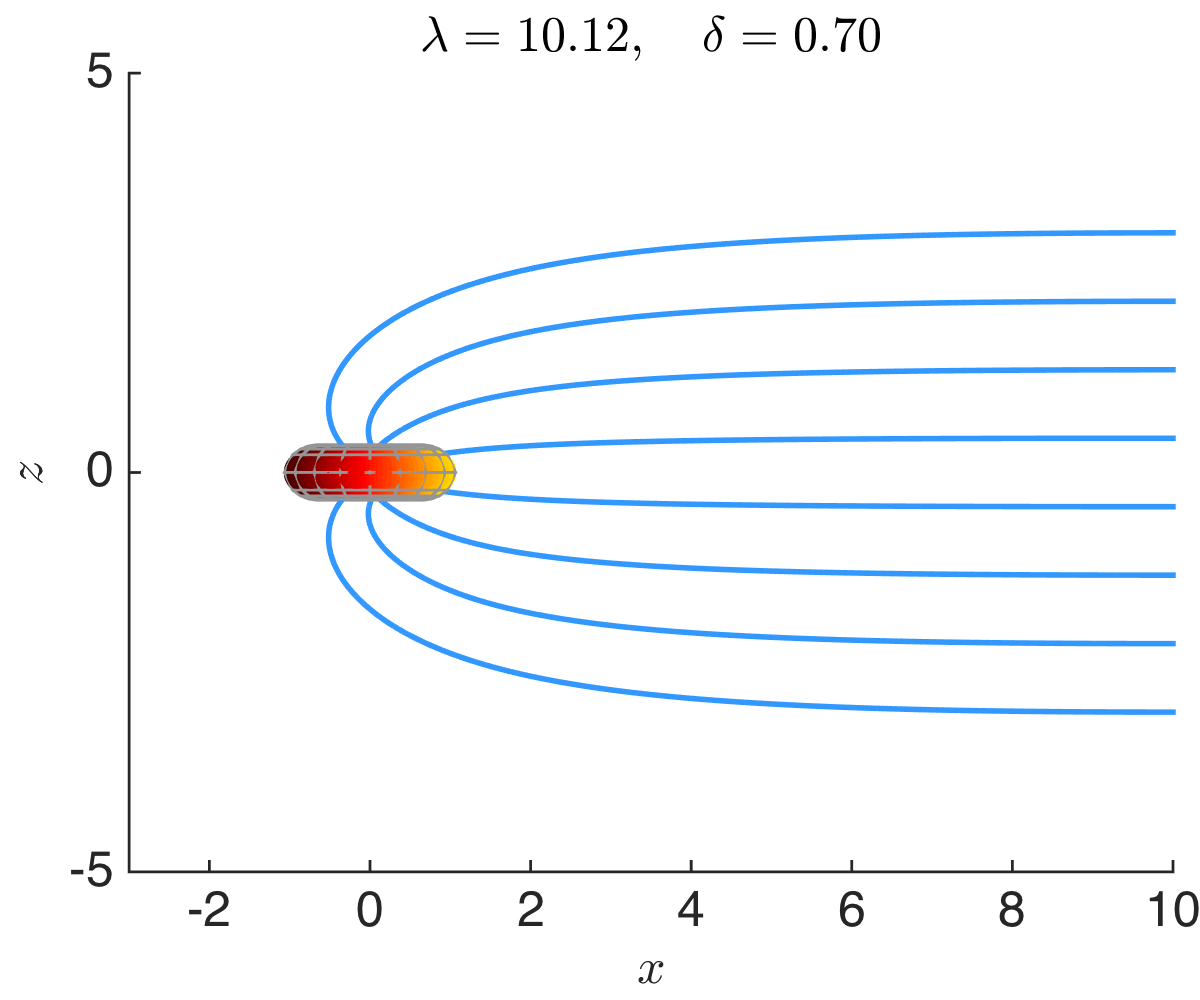}
    
    \caption{Initially-parallel photons in the disk space-times. } \label{fig:parallel_i_ii}
\end{figure}

It can be observed that the bending of light increases with the size of the ergosphere, e.g., all the photons in the example with $\lambda=10.12$ and $\delta=0.70$ in Fig.~\ref{fig:parallel_i_ii} with initial conditions of the type (ii) are trapped in the surroundings of the ergosphere. Moreover, the initial conditions of the type (i) are an indicator of dragging effect in the counter-rotating disk space-times. The size of the ergospheres is inversely proportional to the parameter $\delta$ in these examples, where $\lambda$ is fixed.

Now, we are going to study photons with initial conditions of the form (iii), namely, photons whose initial velocities are tangent to a circle of radius $\rho_0$ on the equatorial plane. We are particularly interested in photons that trace circular orbits. In Kerr space-times, there are two possible initial conditions for which the null geodesics project to circular orbits, known as the last photon orbits \cite{bardeen}. One of these conditions corresponds to a photon going in the direction of the black hole (prograde) and the other going in the opposite direction (retrograde). The prograde orbit is smaller since the dragging effect of the black hole increases the angular momentum of the photon, and the retrograde orbit is bigger since the angular momentum of the photon is decreased by the dragging effect.

\begin{figure} [htb]
    \centering
     \includegraphics[width=0.4\textwidth]{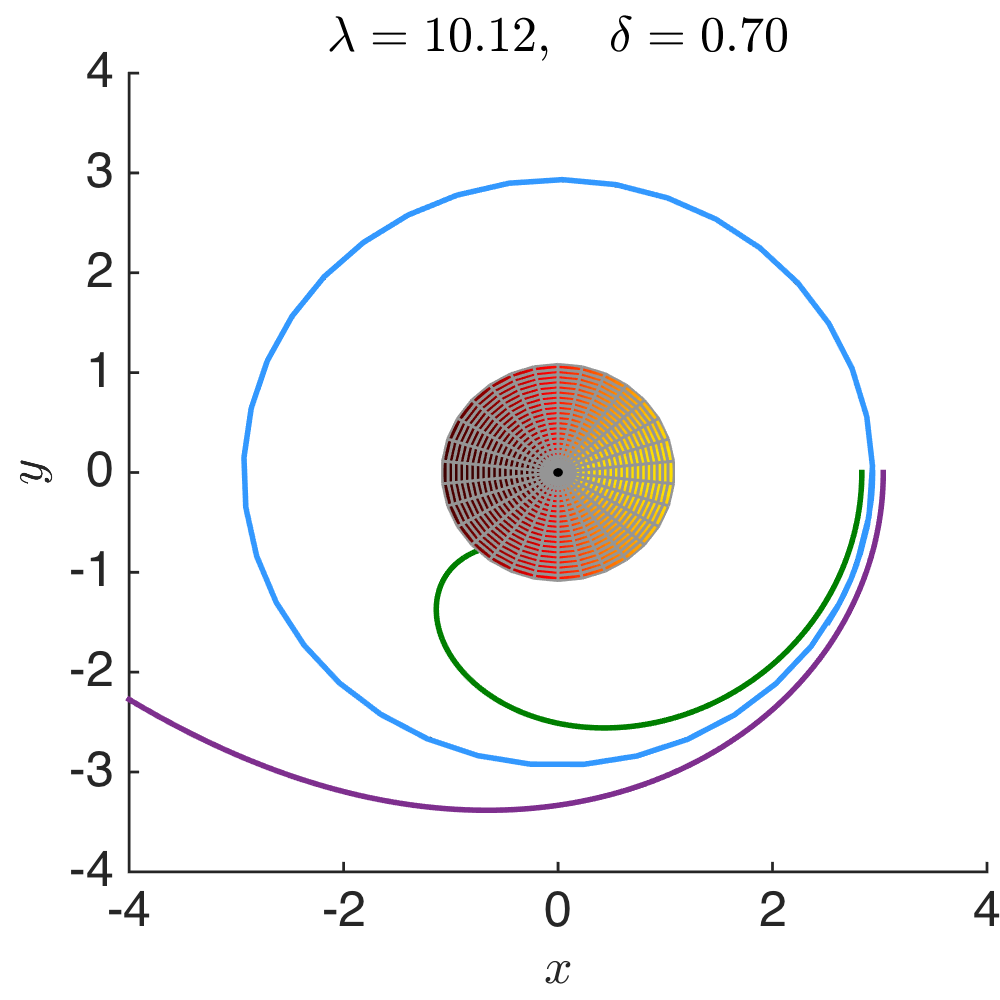} 
    \includegraphics[width=0.4\textwidth]{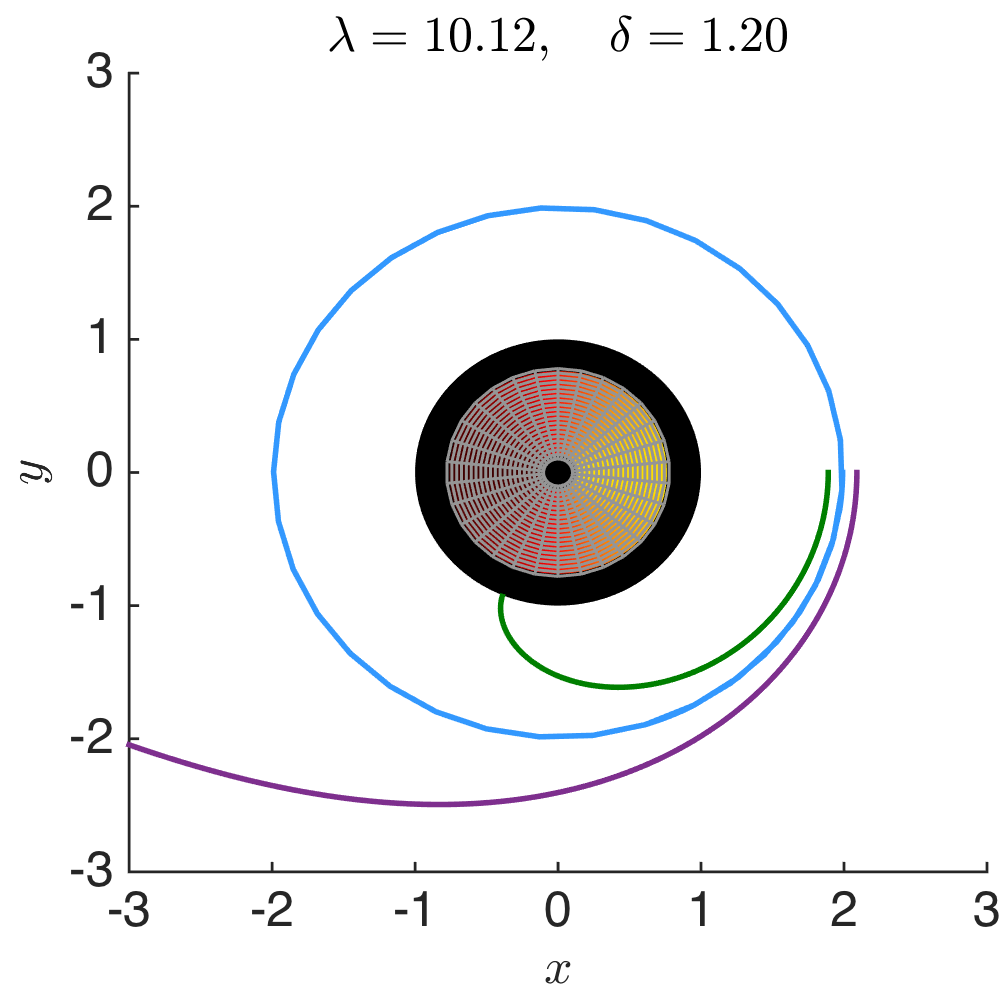} 
    \caption{Photons with circular orbits.} \label{Ph_Sph_xy}
\end{figure}

Choosing the following initial radii
\begin{align*}
    \rho_0 &= 1.98990542, \text{ for }  \lambda=10.12, \delta=1.20,  \\
    \rho_0 &= 2.93290917, \text{ for }  \lambda=10.12,  \delta=0.70,
\end{align*}
for the initial conditions of the type (iii), we obtain circular retrograde orbits. The photons fall towards the disk if the radius is smaller than this value, and they escape to infinity if it is bigger, as observed in the examples in Fig.~\ref{Ph_Sph_xy}, for which we chose initial radii $\tilde{\rho}_0=\rho_0\pm 0.10$ in both cases. The photons can orbit the disk multiple times before falling or escaping, depending on how close the initial $\rho_0$ is to the last photon orbit. The radii for the orbits shown in the figure can be computed explicitly by imposing the conditions that the spatial projection be a circle in the plane with $z=0$.
In contrast to Kerr space-times, every photon with initial conditions of this type but with positive $p^\phi$ (prograde orbits) will escape to infinity since we would need an initial $\rho_0<1$ to obtain a closed orbit, which is on the surface of the disk. Thus, a prograde circular orbit cannot be observed in counter-rotating disk space-times.

Similarly to what happens in Kerr space-times, photons with initial conditions of the type (iv) trace more complicated trajectories. However, depending on the value of the parameters $\lambda$ and $\delta$ for the disk space-time, one can find a radius $\rho_0$ for which the photon orbit is on the surface of an ellipsoid.

\begin{figure} [htb]
    \centering
    \includegraphics[width=0.45\textwidth]{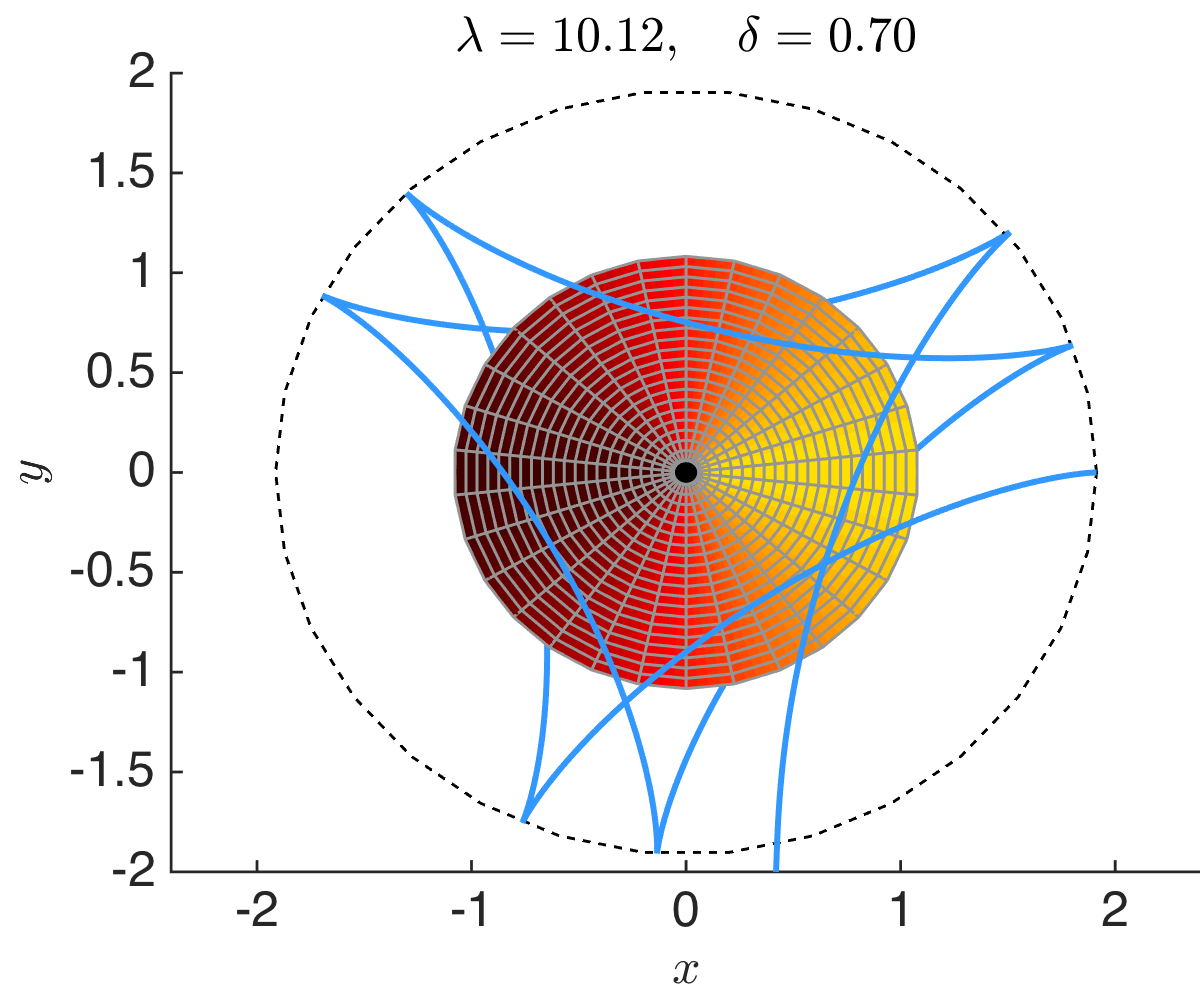} 
    \includegraphics[width=0.45\textwidth]{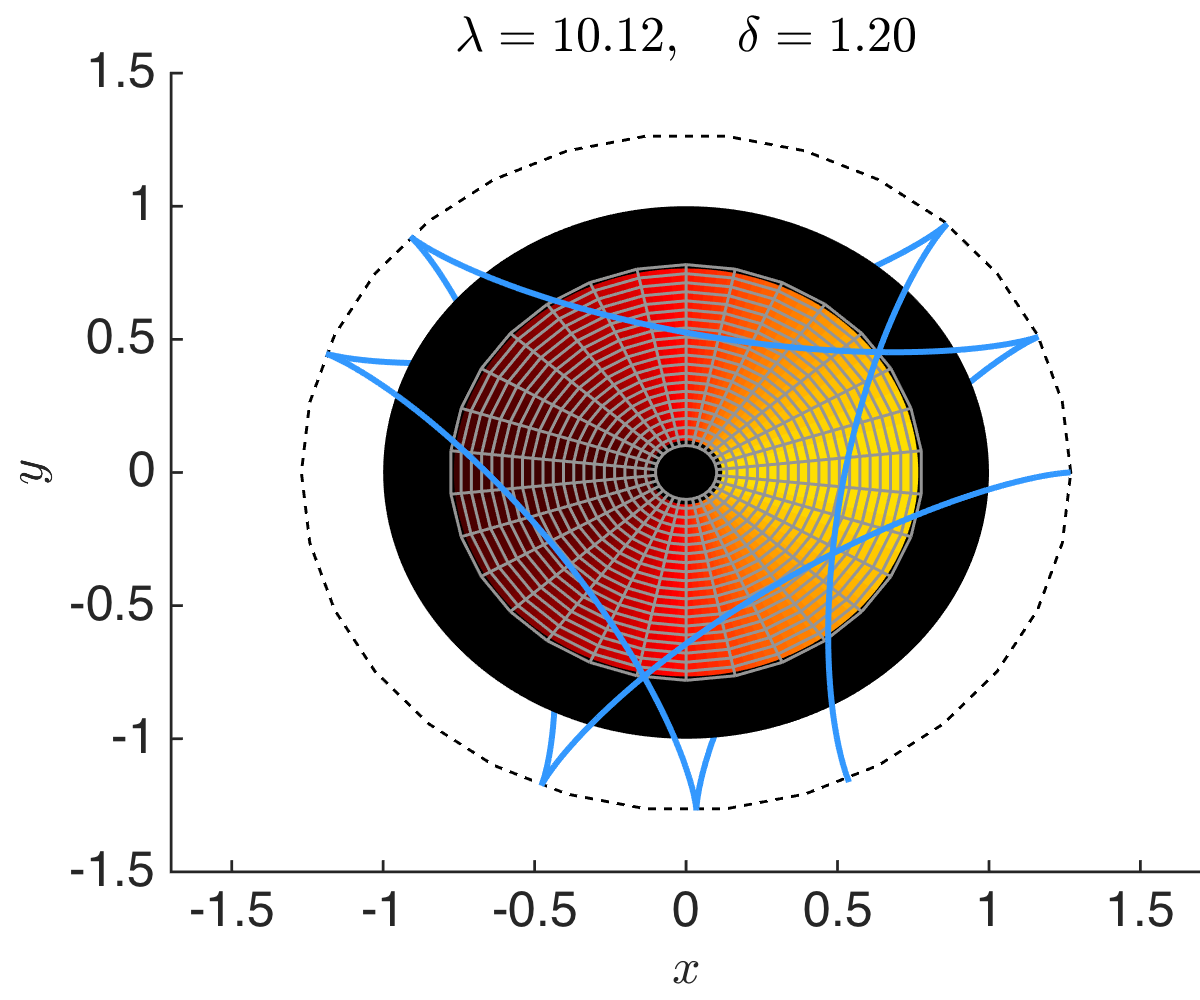}  
    \includegraphics[width=0.45\textwidth]{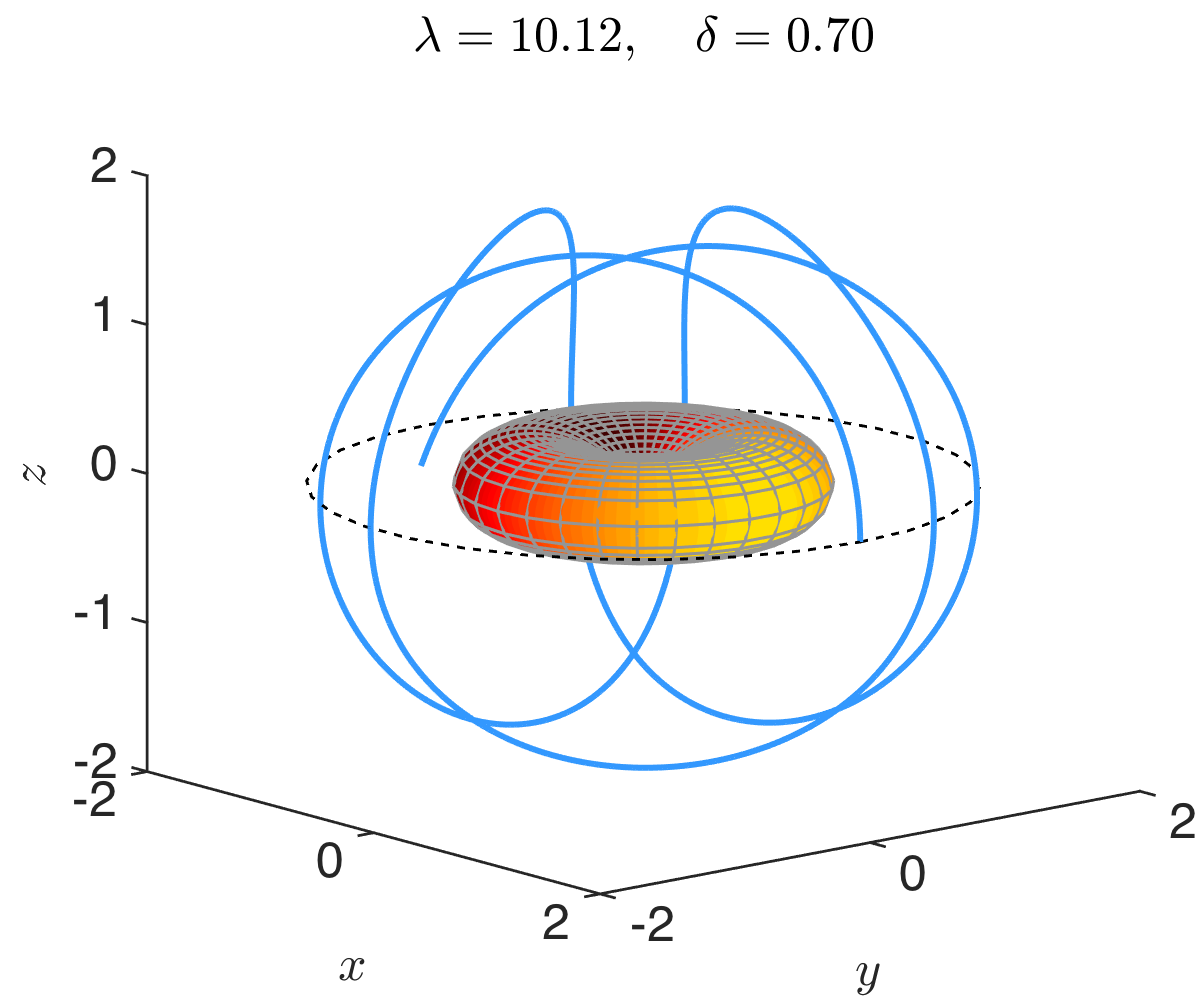} 
    \includegraphics[width=0.45\textwidth]{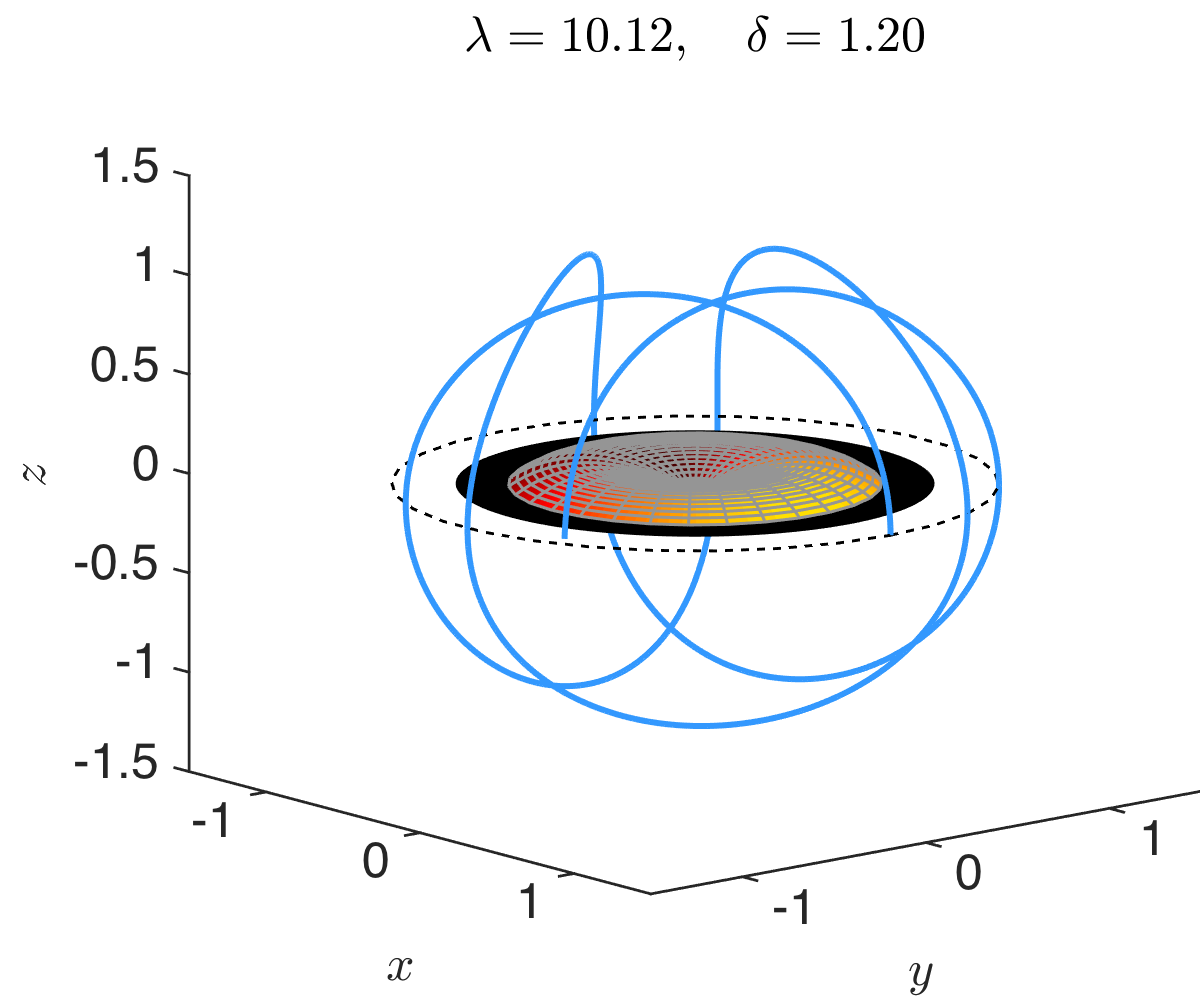} 
    \caption{Photons with orbits on ellipsoids. The dashed lines represent the $xy$ section of such ellipsoids.} \label{Ph_Sph_xyz}
\end{figure}
The initial radii, in dependence of the $\lambda$, $\delta$ parameters, for which the photon orbits are on an ellipsoid are the following:
\begin{align*}
    \rho_0 &= 1.27005031, \text{ for } \lambda=10.12, \delta=1.20, \\
    \rho_0 &= 1.91319267, \text{ for } \lambda=10.12, \delta=0.70.
\end{align*}
In contrast to the previous case, these radii cannot be easily obtained explicitly. They were determined by trial and error. Analogously to initial conditions of the type (iii), photons whose initial conditions are close to the conditions to obtain orbits as in Fig.~\ref{Ph_Sph_xyz} will follow similar trajectories until eventually falling towards the disk or escaping to infinity. 



\subsection{Shadow of the disk}

\begin{figure} [htb]
    \centering
    \includegraphics[width=6cm]{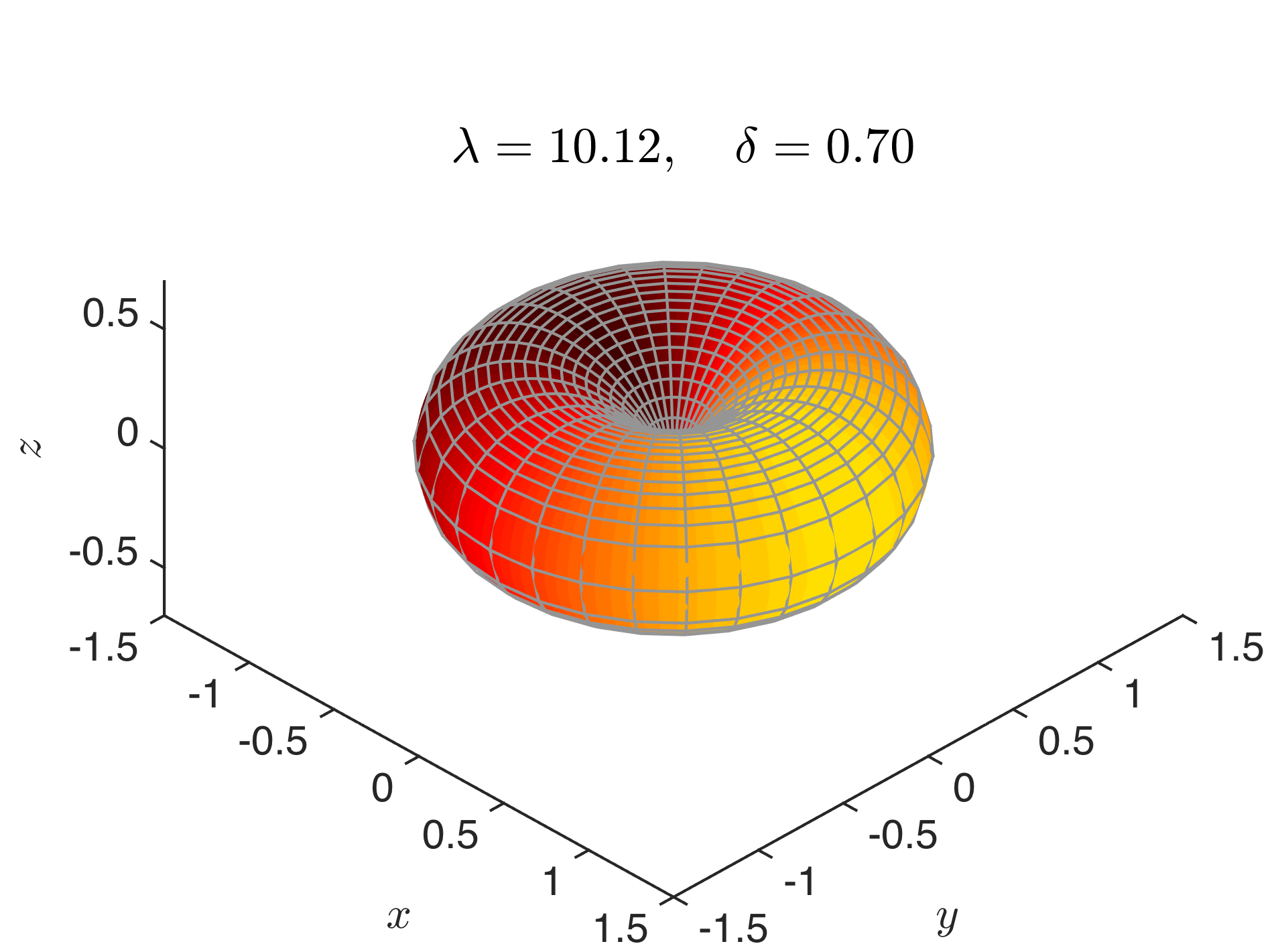} \includegraphics[width=6cm]{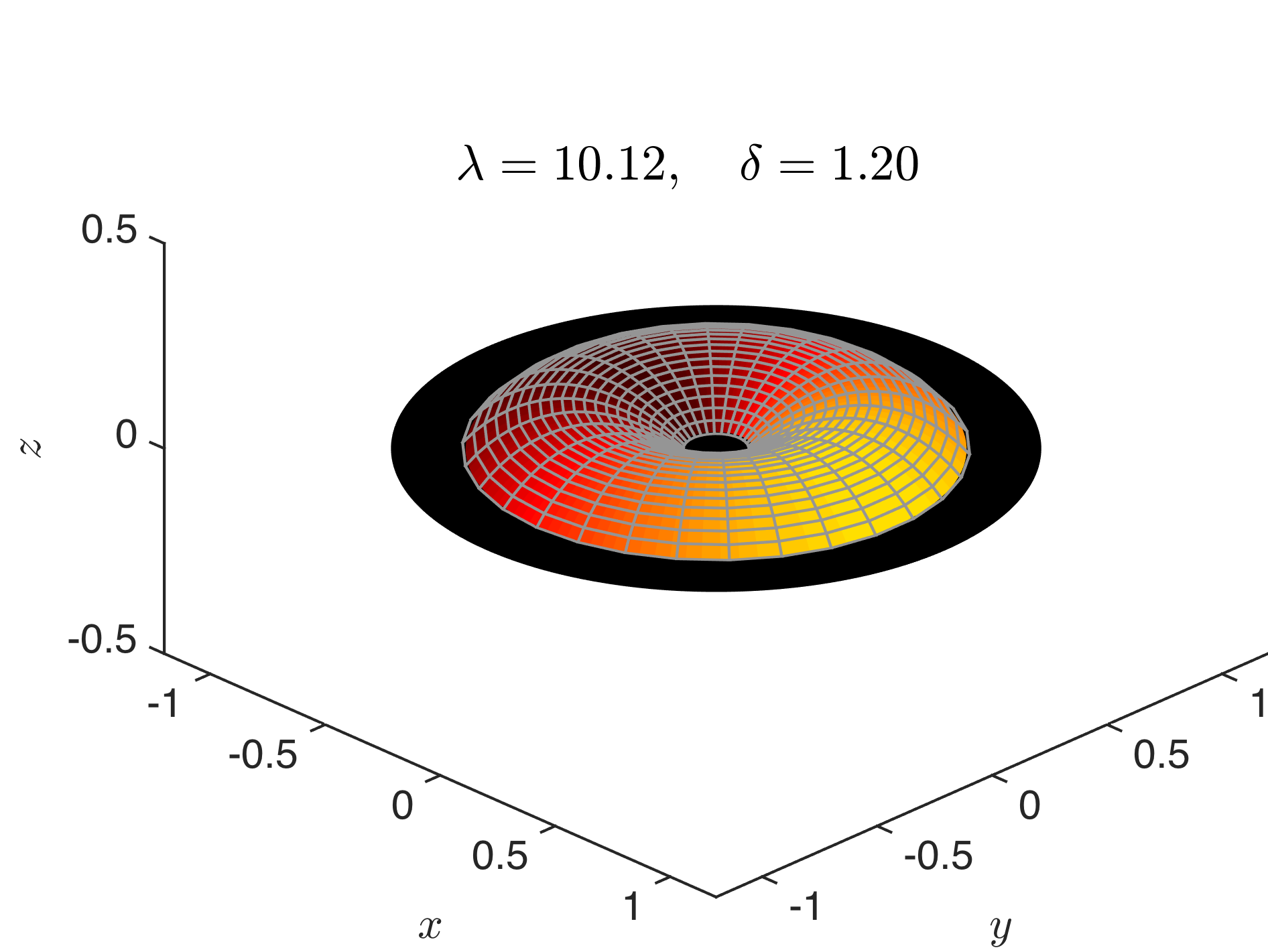}
    \caption{3D representation of the ergospheres.} \label{fig:3D_disk}
\end{figure}

Besides the photon orbits resulting from initial conditions of the type (iii) and (iv) in the 
previous subsection, there are other photon orbits on the surface of 
ellipsoids, i.e., \textit{photon spheres} using the terminology from 
Kerr space-times. The upper bound of the semi-axes of these ellipsoids is the radius $\rho_0$ of the photon with circular orbit shown in Fig.~\ref{Ph_Sph_xy}. A photon crossing one of these ellipsoids from the outside is trapped by the gravitational influence of the disk. Thus, in the simulation of a picture using the ray tracing technique, some photons will fall into these surfaces, implying that these surfaces will act as a shadow and will prevent light from a luminous background from reaching the observer. Nonetheless, if we assume the disk to be opaque and to emit light, it will be seen by an observer even if the disk is inside the photon spheres since, in this case, the photon will cross these surfaces from the inside with an outwards velocity. Fig.~\ref{fig:3D_disk} shows disks together with the ergospheres, which have a doughnut-like shape corresponding to space-times with different parameters.

We simulate the visualisation in both space-times using a camera with the properties mentioned in Section \ref{sec:RT}, a focal length $f_L=0.15$ for the disk with $\delta=0.70$ and a focal length $f_L=0.20$ for $\delta=1.20$. The reason is that given a fixed focal length, the apparent image of the disk with parameter $\delta=1.20$ will appear smaller due to its weaker gravitational pull, thus we need to increase the focal length $f_L$ to perceive a bigger image.
We use the same resolution $I_V \times I_H=300\times 300$ for all the simulations. In order to visualise the effect of the rotation, we also perform the simulations positioning the camera at various inclination angles $\alpha$ with respect to the $\zeta$-axis.
For visualisation purposes, we add artificial coloring to the shadow depending on the part of the ergosphere approached by the infalling photon corresponding to a given pixel (the coloring choice is shown in Fig.~\ref{fig:3D_disk}), and we color the upper face of the disk in black and the lower face in dark blue in order to increase the contrast with the shadow. 
\begin{figure} [htb]
  \centering
  \small $\lambda=10.12$, $\delta=0.70$ \hspace{2cm} $\lambda=10.12$, $\delta=1.20$ \par
  \fbox{\includegraphics[width=4.5cm]{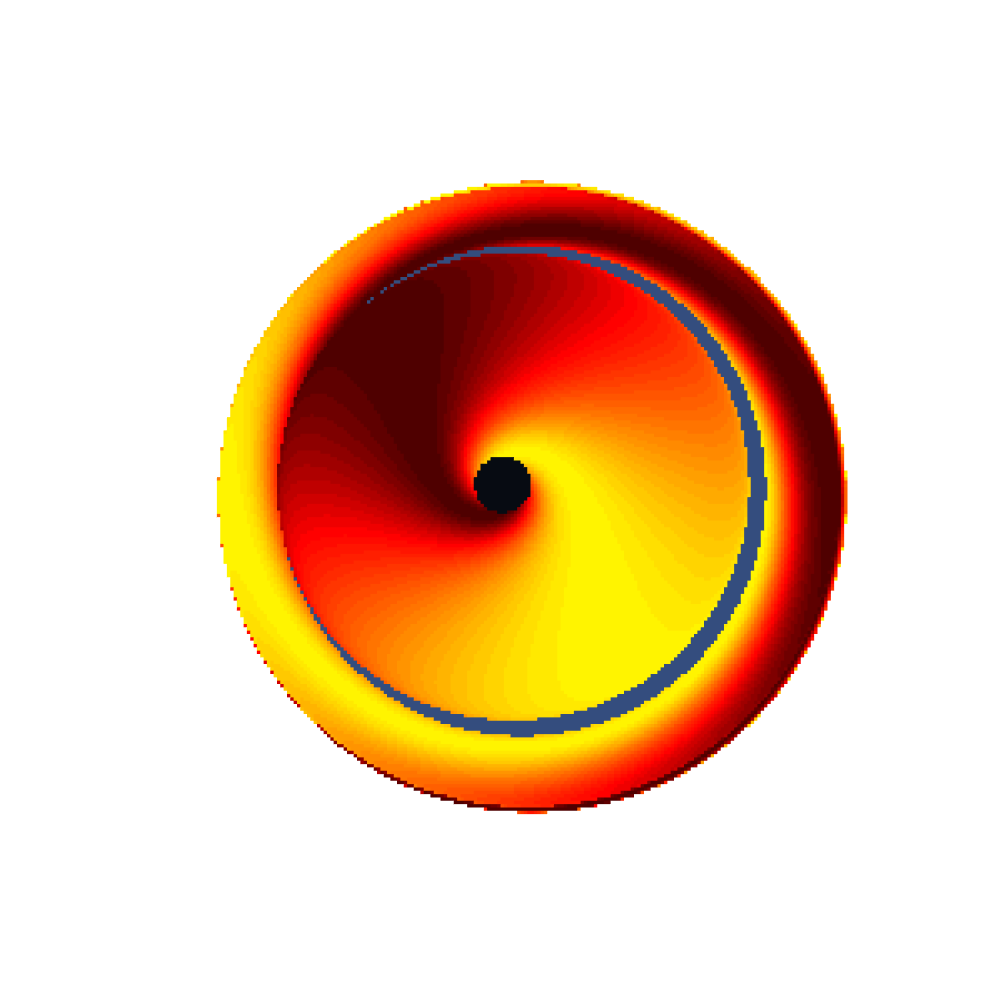} } \fbox{\includegraphics[width=4.5cm]{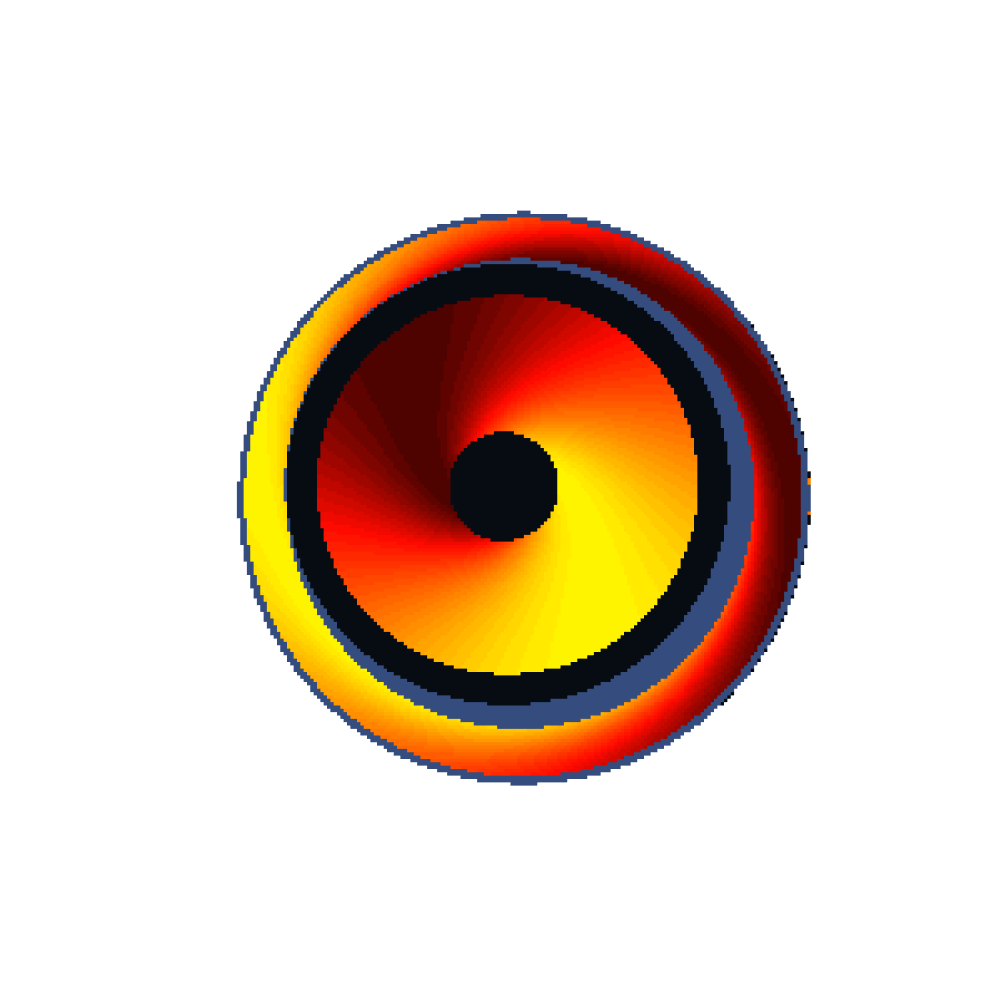} }
  \fbox{\includegraphics[width=4.5cm]{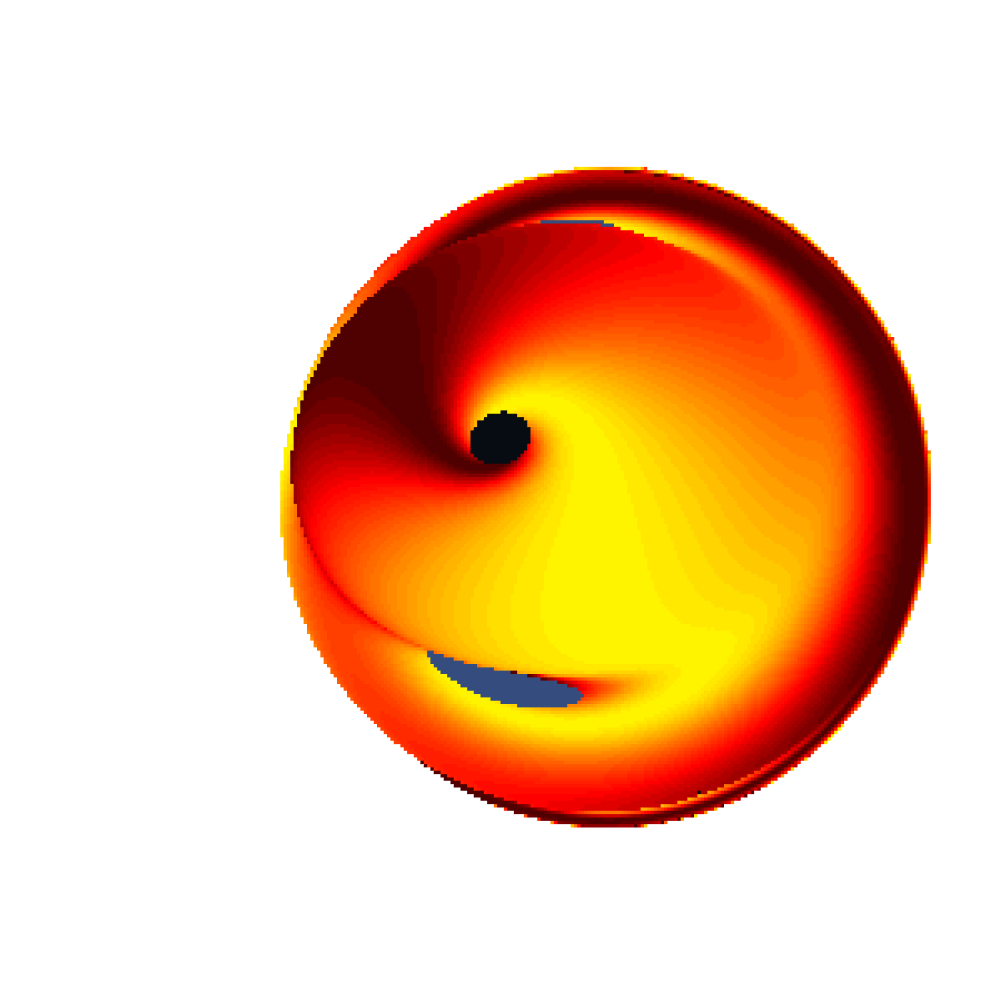} } \fbox{\includegraphics[width=4.5cm]{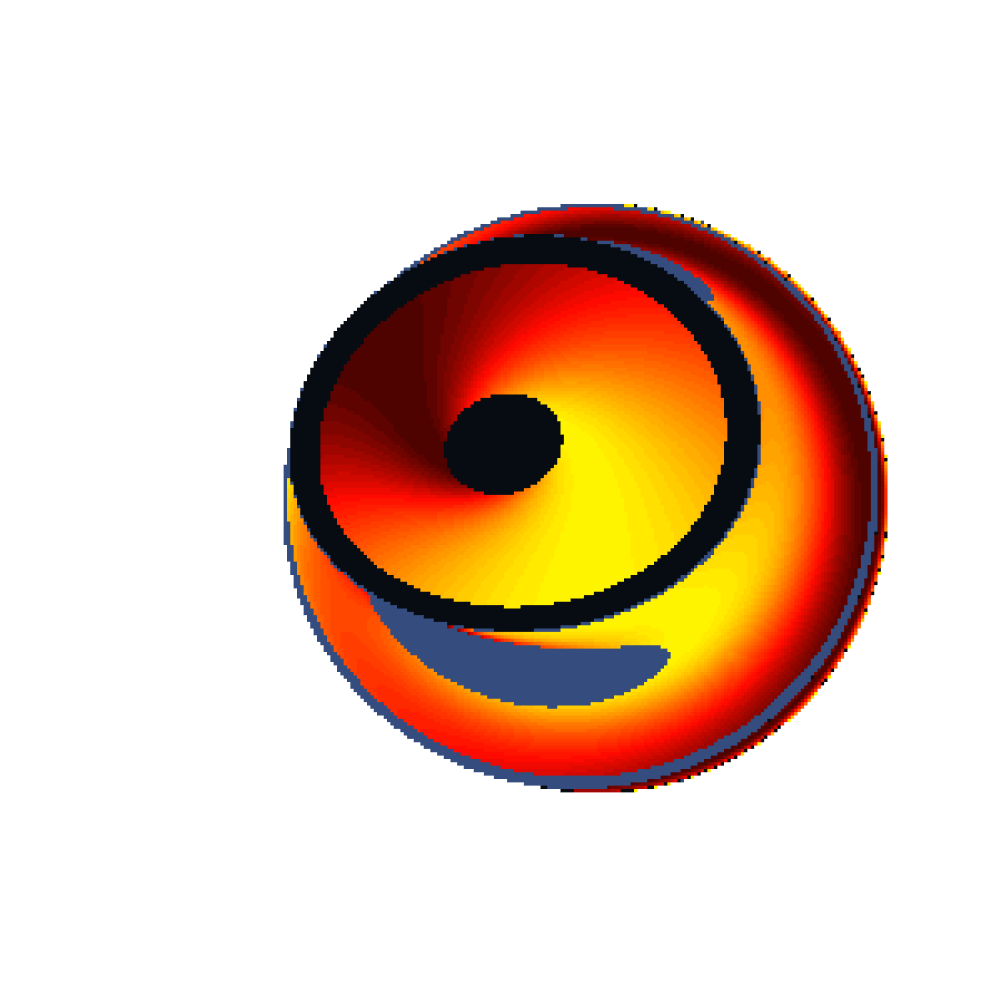} }
  \fbox{\includegraphics[width=4.5cm]{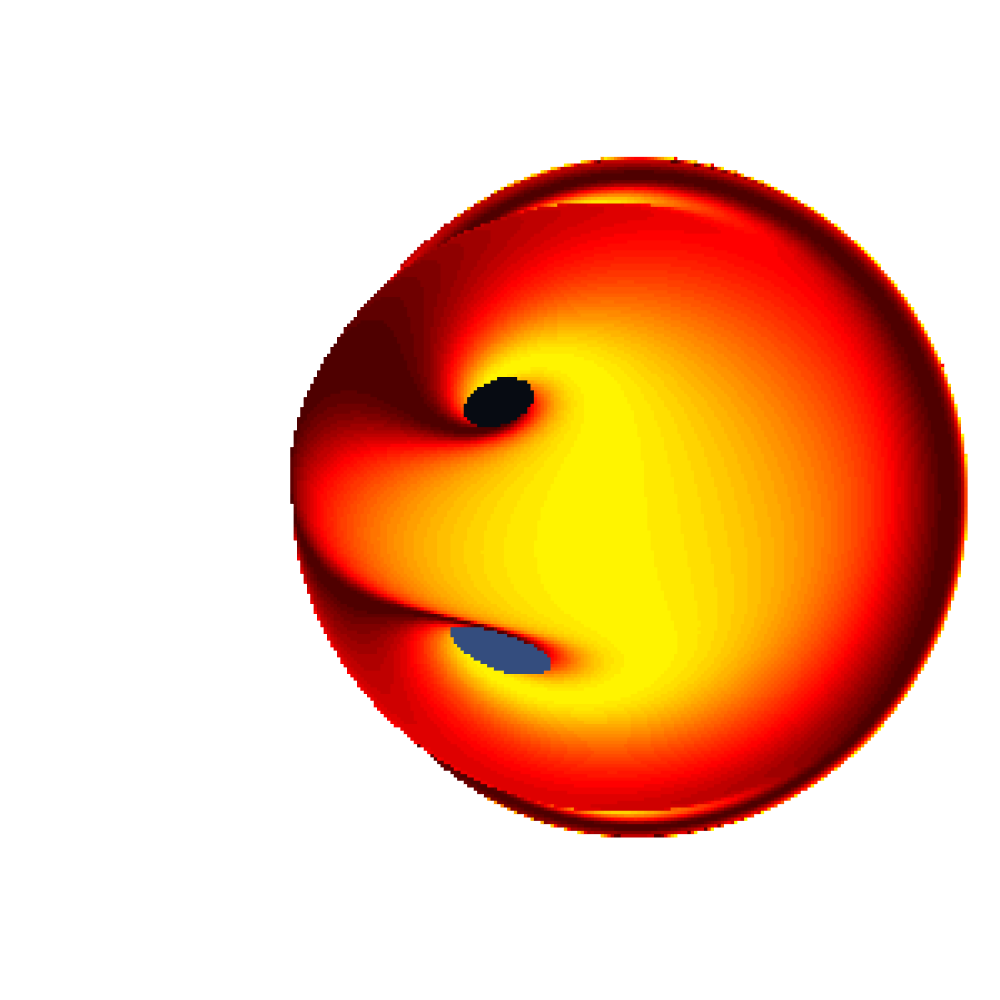} } \fbox{\includegraphics[width=4.5cm]{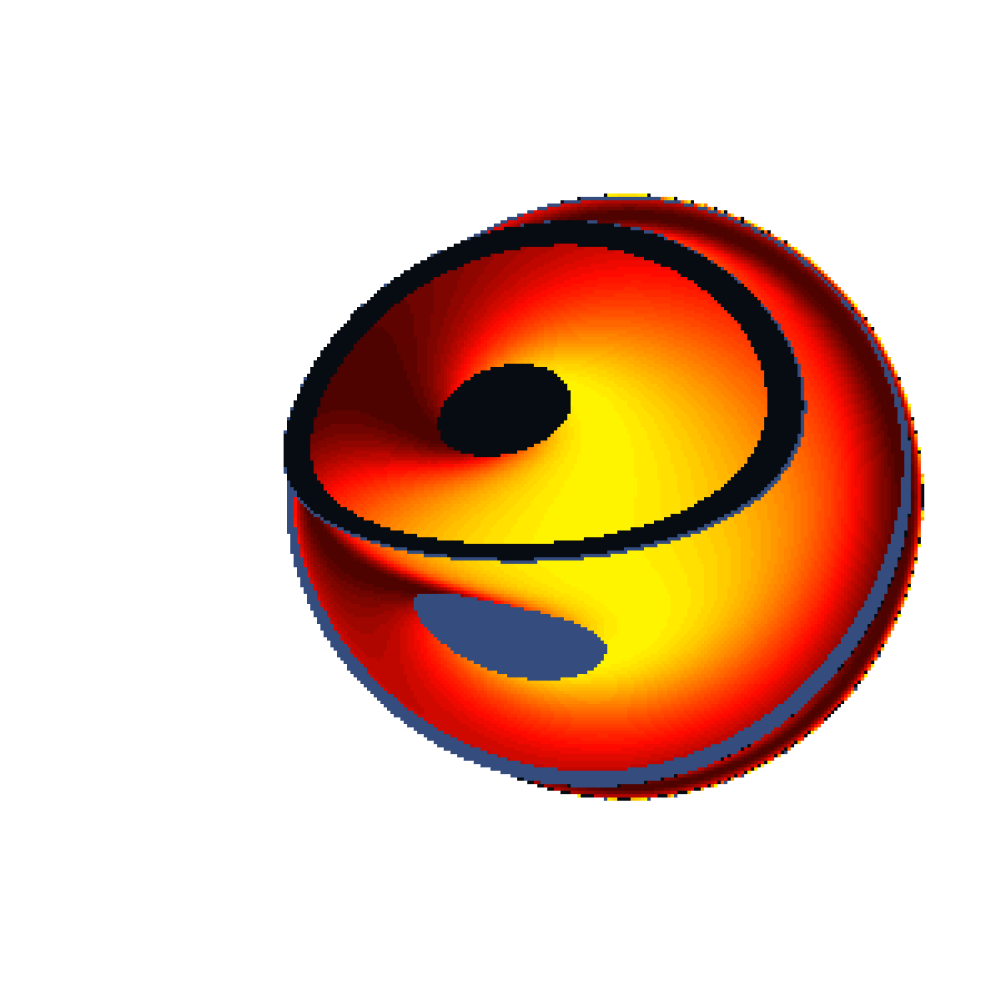} }
  \fbox{\includegraphics[width=4.5cm]{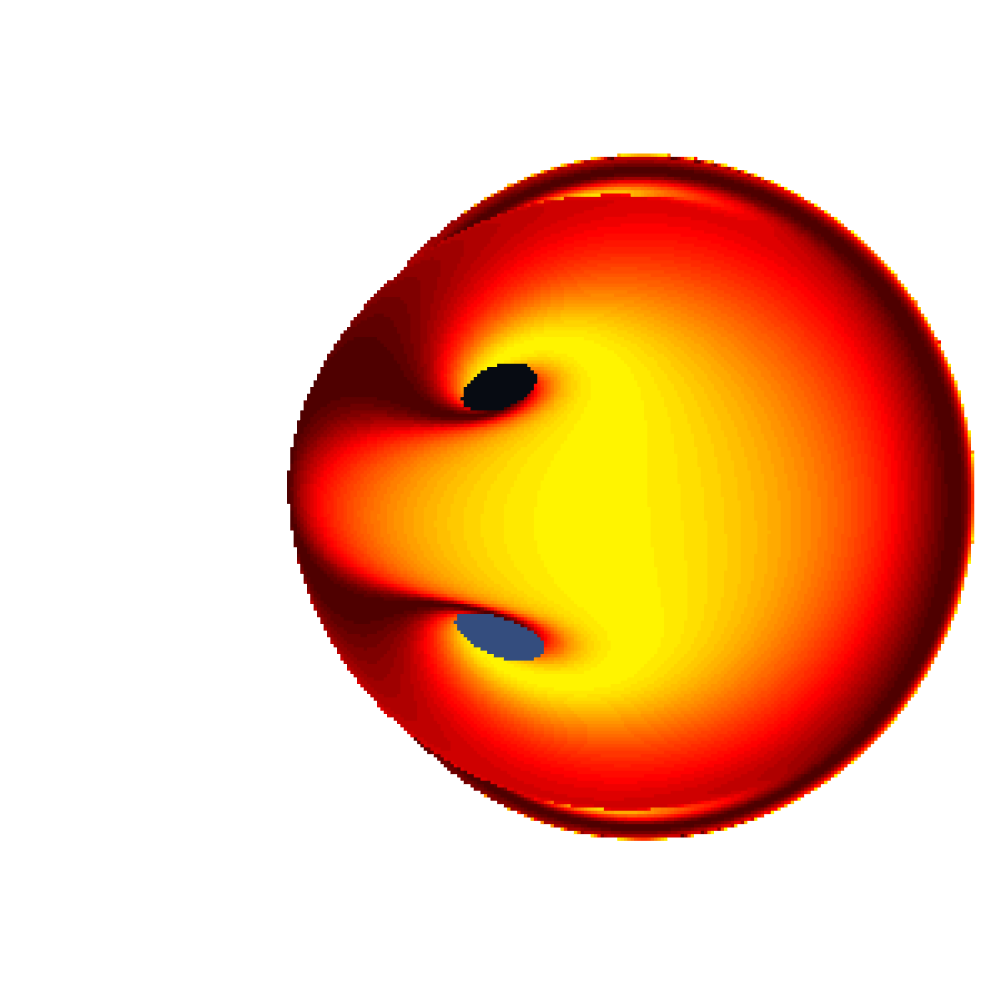} } \fbox{\includegraphics[width=4.5cm]{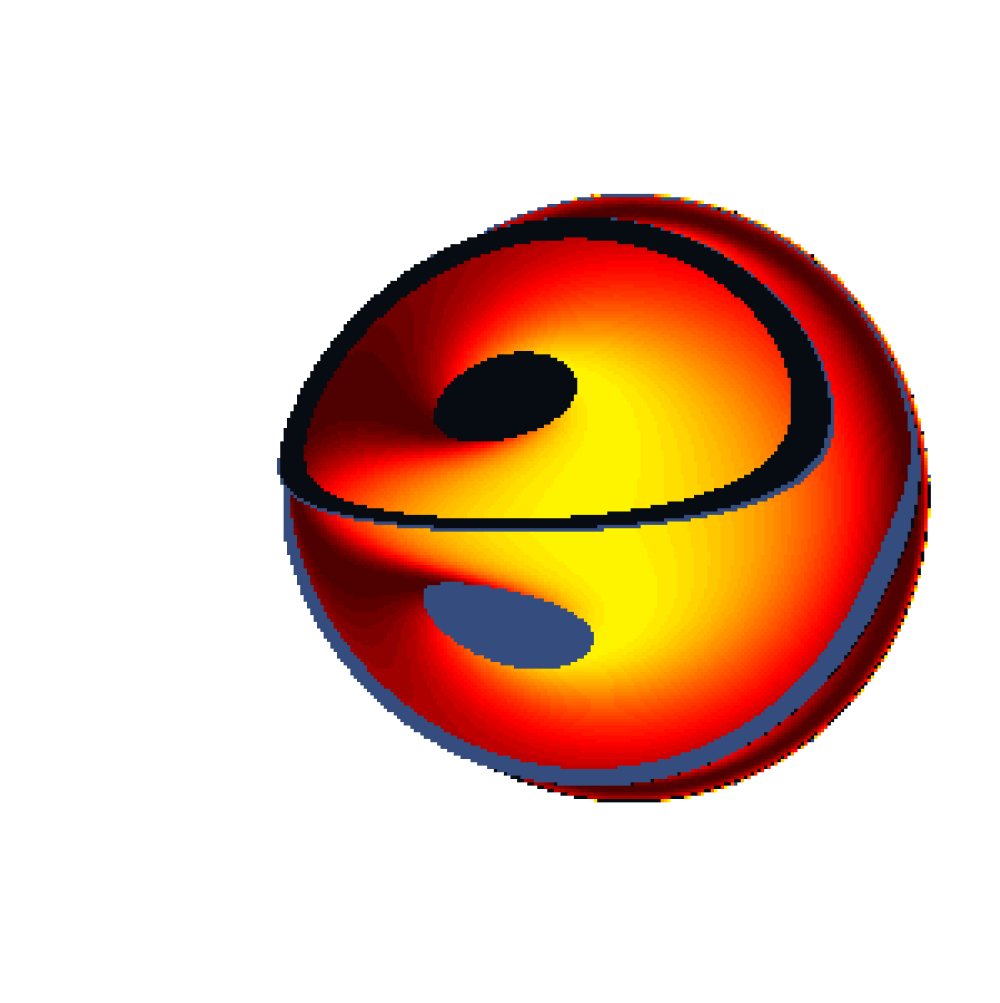} }
  
  \caption{Visualisation of counter-rotating dust disks for two different parameters. From top to bottom, the inclinations angles of the camera are $\alpha=10^\circ$, $\alpha=45^\circ$, $\alpha=70^\circ$ and $\alpha=80^\circ$.}  \label{fig:rt-cr-disk}
\end{figure}

It can be observed that the luminous background is covered by a quasi-circular shadow, analogous to the observed phenomenon in Kerr space-times. However, some portions of the dust disk are still observed. The doughnut-shaped ergosphere of the first space-time (with $\lambda=10.12$ and $\delta=0.70$) covers the disk except on the centre of the doughnut, whereas the outer radius of the ergosphere in the other space-time (with $\lambda=10.12$ and $\delta=1.20$) is smaller than the radius of the disk $\rho_{0}=1$, as observed in Fig.~\ref{fig:3D_disk}, which allows the visualisation of a greater portion of the disk.

The lower part of the disk is observed even if the camera is positioned above the equatorial plane; this is due to the bending of light in the vicinity of the disk, analogous to the behavior of light around a black hole. This bending can be observed better in Fig.~\ref{fig:parallel_i_ii} for individual photons. 

Fig.~\ref{fig:3D_disk} shows that the only observable part of the disk with parameters $\lambda=10.12$ and $\delta=0.70$ is at the centre of the ergosphere, but both the upper and the lower part can be seen by the observer. The left-hand side of Fig.~\ref{fig:rt-cr-disk} shows the visible parts of the disk, and as observed, the lower part is highly distorted as the angle $\alpha$ of the camera is closer to the $\zeta$-axis, as opposed to the upper part that looks more distorted when the angle $\alpha$ is closer to the equatorial plane.

The second example, the space-time with $\lambda=10.12$ and $\delta=1.20$, exhibits a smaller shadow as seen in Fig.~\ref{fig:rt-cr-disk}, which is related to the size of the ergosphere. From these figures, we could also infer that this disk has a smaller angular momentum compared to the first example, which agrees with the explicit values given in Table \ref{table}. Due to the smaller size of the ergosphere, the outer part of the disk is seen by the observer, but it gets distorted as the camera inclination gets closer to the equatorial plane, and the lower part becomes visible.

To determine the accuracy of these simulations, we computed the value 
of the Lagrangians $\Lag(\vx^{(N)})$ corresponding to each pixel in 
the same way we did in Subsection \ref{subsec:total_errors} for Kerr black holes. The value of the Lagrangians outside the shadows were of the same order of magnitude as those shown in Fig.~\ref{fig:lagrangian_kerr}, i.e., given tolerances $\texttt{AbsTol}=\texttt{RelTol}=10^{-8}$ for the \texttt{ode45} routine, we obtained $|\Lag(\vx^{(N)})|\leq 10^{-7}$ for the pixels that escaped to infinity.

\subsection{Coloring pattern for the celestial sphere}
In order to analyse the influence of a gravitational counter-rotating dust disk on the image of the background space, we choose the artificial coloring given by the Infrared Sky picture in the 2MASS catalog \cite{catalog} for the light-rays that escape to infinity and reach the ``celestial sphere'' around the origin. This coloring depends on the angles at which the light rays reach the celestial sphere. Coloring patterns of this type have been used in Schwarzschild and Kerr black holes simulations, see \cite{riazuelo, interstellar}. In these simulations we eliminate the coloring pattern used for the shadows in Fig.~\ref{fig:rt-cr-disk}. The main difference with respect to black hole shadows is that we assume that the opaque disks (which are inside the photon spheres) emit light, which is the red coloring covering the shadow in Fig. \ref{fig:rt-cr-disk}.

\begin{figure} [htb]
    \centering
    \small $\lambda=10.12$, $\delta = 0.70$  \hspace{2cm} $\lambda=10.12$, $\delta = 1.20$ \par
    \includegraphics[width=5cm]{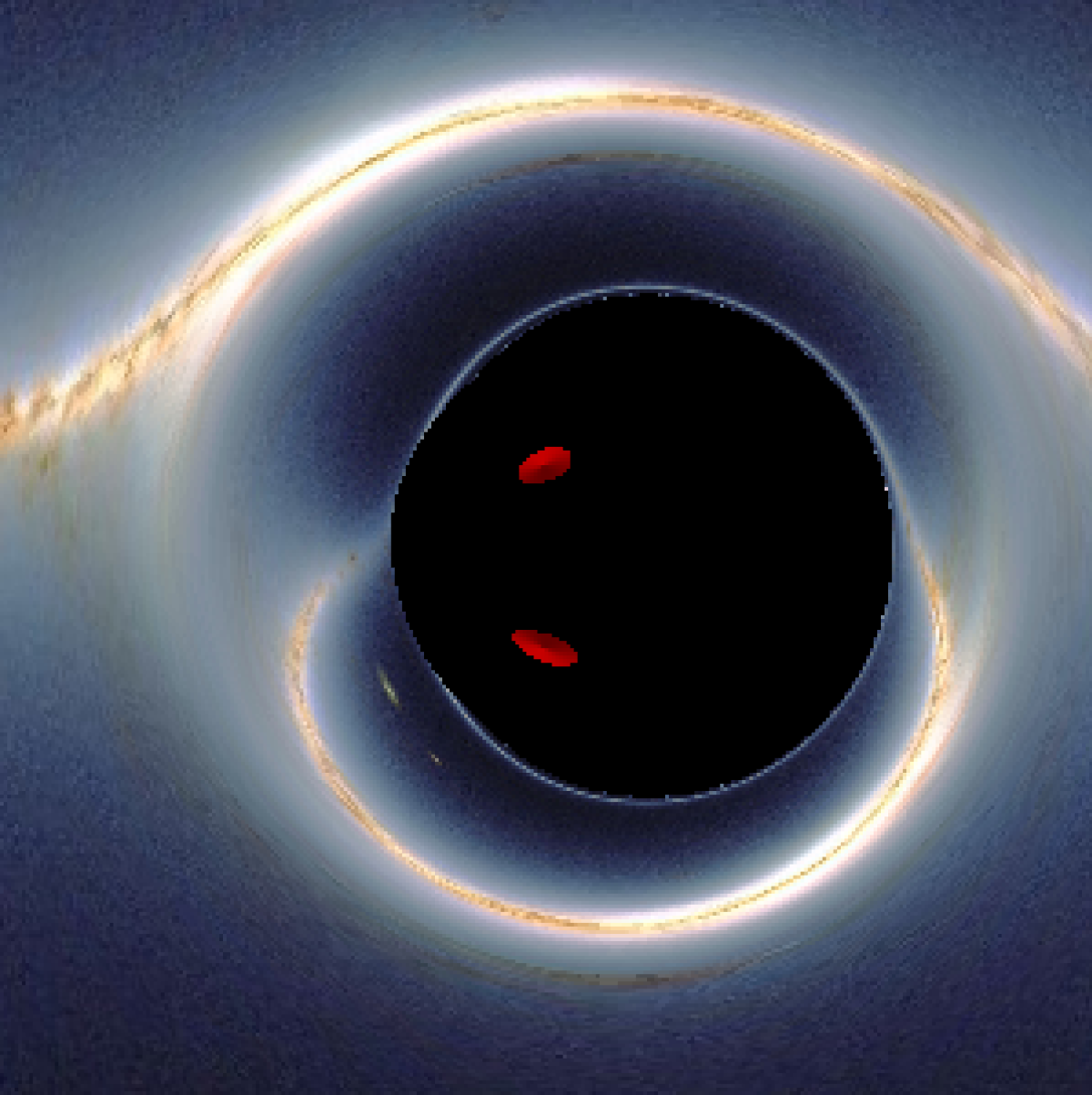}
    \includegraphics[width=5cm]{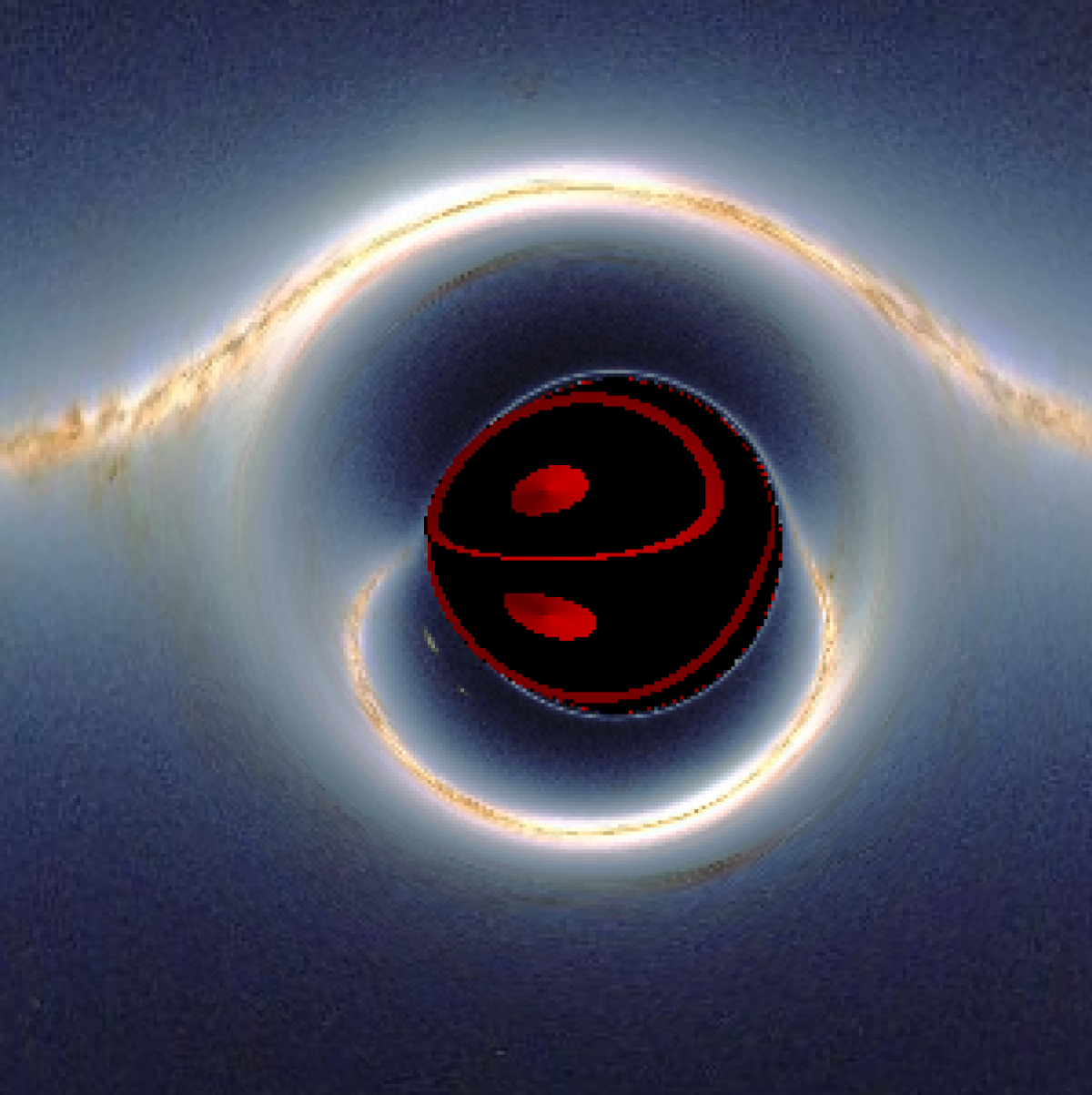}
    
    \caption{Counter-rotating disks and their shadows at the inclination angle $\alpha=80^\circ$.} \label{fig:shadow_CR_disk}
\end{figure}

If we assumed these gravitating disks not to emit light, then there would be full shadows analogous to those observed in Kerr space-times. In either case, the Einstein ring is observed and therefore a secondary image of the background space is also visible. Fig.~\ref{fig:shadow_CR_disk} shows simulations made with a camera with a focal length $f_L=0.10$ and resolutions of $I_H\times I_V=300\times 300$ pixels.

Regarding the size of the shadow (and hence the Einstein ring), it is noteworthy to mention that it depends on the mass of the counter-rotating disk. The shift of the shadow from the centre of the screen indicates the angular momentum of the disk. Recall that the greater the mass and angular momentum, the more the light rays will be bent by the counter-rotating dust disk as seen in Fig. \ref{fig:parallel_i_ii}. Therefore, given a fixed angular aperture for the camera, the shadow will be bigger the greater the mass of the disk is, as shown in Fig.~\ref{fig:shadow_CR_disk}. This behavior is also observed in Kerr space-times. The reason the shadows have the same size in Fig.~\ref{fig:shadow_kerr} is due to the fact that the distances are normalised to the mass of the black hole. In the case of a counter-rotating disk space-time, the distances are normalised by the radius of the disk.

\section{Outlook}\label{sec.out}

We have shown in this paper that geodesics can be computed in any stationary axisymmetric vacuum space-time, given the values of the metric functions on 2-dimensional Chebyshev grids. Physical properties, such as the ergo-regions and shadows of gravitating sources which are well documented for the Kerr black hole could also be studied for a family of counter-rotating dust disks. Since the solutions can be parametrised by a parameter varying between a Newtonian and an ultrarelativistic limit, where the exterior of the disk can be interpreted as the extreme Kerr solution, relativistic effects such as frame-dragging can be conveniently explored for this example. The second parameter appearing in the solution controls the transition from a static to a disk consisting only of one component of dust. The related effect on the light rays has been studied in the previous sections.

In future work, we would like to study higher genus algebro-geometric solutions to the Ernst equation. To do this, it is only necessary to perform the explicit computations of the Abel maps $\omega(\infty^\pm)-\omega(\xi)$ and the period matrices $\mathbb{B}_\xi$ on the Chebyshev grids, as it was done in genus 2 in Section \ref{sec:cr-disk}. Then, it suffices to apply the methods outlined in Sections \ref{sec:geodesics} and \ref{sec:RT} to obtain visualisations in these space-times. Other solutions that can be studied are the elliptic solutions presented in \cite{Ko91}, whose ergospheres are toroidal surfaces, and black holes with infinite gravitating disks, as discussed in \cite{prblack}. It would be also interesting to study geodesics in stationary axisymmetric solutions to the Einstein-Maxwell equations as in \cite{Har}.

In this paper we have only shown the trajectory of the photons, but the outcome of the computations in Section \ref{sec:geodesics} include the four-momenta at every point. Therefore, phenomena such as the slowing down of particles or the Doppler effect can also be studied. We are interested in identifying phenomena that are general to any stationary axisymmetric space-time and those that are characteristic to black holes. One could also study the optical effect of the counter-rotating disk if we assume it to be transparent.

\appendix
\section{Camera in a flat space-time} \label{append_camera}
In this Appendix we show the initial conditions of the photon associated to each pixel $(i,j)$. More precisely, we show the direction of the light rays hitting the screen of a virtual camera in a Minkowski space-time in Weyl coordinates. 

Let us consider a camera with the following features.
\begin{itemize}
    \item Size: horizontal width $d_H$ and vertical height $d_V$.
    \item Resolution: $I_H$ pixels wide and $I_V$ pixels height.
    \item Focal length $f_L$.
\end{itemize}
The position of the camera with respect to the origin is described by the following:
\begin{itemize}
    \item Object distance $R_c$.
    \item Inclination angle $\alpha$ with respect to the $z$-axis.
\end{itemize}

If the angle with respect to the axis is zero, then the position of each pixel $(i,j)$ is given by
\begin{equation} \label{xyz_zero}
    \begin{split}
    \tilde{x}_{ij} &= d_V \lc \frac{1}{2} - \frac{(i-1)}{(I_V-1)} \rc, \\
    \tilde{y}_{ij} &= d_H \lc \frac{1}{2} - \frac{(j-1)}{(I_H-1)} \rc , \\
    \tilde{z}_{ij} &= R_c + f_L.
    \end{split}
\end{equation}
If the camera is positioned at an angle $\alpha$ with respect to the $z$-axis, then the position of each pixel $(i,j)$ is simply the rotation of $\vec{v}_0:=(\tilde{x},\tilde{y},\tilde{z})^\trans$ by an angle $\alpha$ about the $y$-axis in counterclockwise direction, i.e.,
\begin{align*}
    x_{ij} &= \cos \alpha \cdot \tilde{x}_{ij} + \sin\alpha \cdot \tilde{z}_{ij},\\
    y_{ij} &= \tilde{y}_{ij},\\
    z_{ij} &= -\sin\alpha \cdot \tilde{x}_{ij} + \cos\alpha \cdot \tilde{z}_{ij}.
\end{align*}
The light ray reaching the pixel $(i,j)$ comes from the aperture in a straight line, thus the direction of the incoming light ray is given by the vector that goes from the aperture to the position of the pixel $(i,j)$. The aperture is located at $(R_c \sin\alpha,0,R_c \cos\alpha)$. Thus,
\begin{align*}
    p^x_{ij} &= ( x_{ij} - \sin\alpha \cdot R_c)/r_c,\\
    p^y_{ij} &=  y_{ij}/r_c,\\
    p^z_{ij} &= (z_{ij} - \cos\alpha \cdot R_c)/r_c,
\end{align*}
where $r_c=\sqrt{f_L^2+\tilde{x}_{ij}^2+\tilde{y}_{ij}^2}$ is the normalising constant such that $|\vec{p}_{ij}|^2=1$.

The transformation rules from Cartesian and cylindrical coordinates are
\begin{align*}
    \rho^2 &= x^2 + y^2,\\
    \tan\phi &= y/x,\\ 
    \zeta &= z,
\end{align*}
from which the following is deduced
\begin{align*}
    p^\rho_{ij} &= \cos\phi_{ij} \cdot p^x_{ij} + \sin\phi_{ij} \cdot p^y_{ij},\\
    p^\zeta_{ij} &= p^z_{ij},\\
    p^\phi_{ij} &= \frac{1}{\rho_{ij}} \lp \cos\phi_{ij} \cdot p^y_{ij} - \sin\phi_{ij} \cdot p^x_{ij} \rp.
\end{align*}
Therefore, the event and four-momentum associated to the pixel $(i,j)$ in Minkowski's space-time are
\begin{align*}
    \vx &= (t_0, \rho_{ij},\zeta_{ij},\phi_{ij}), \\
    \vp &= (1, p^\rho_{ij},p^\zeta_{ij},p^\phi_{ij}).
\end{align*}

\end{document}